\documentclass[prd,twocolumn,reprint,preprintnumbers,nofootinbib,superscriptaddress,floatfix]{revtex4-2}
\input{header_DSMC_new.tex}
\usepackage{orcidlink}
\newcommand{\orcid}[1]{\,\orcidlink{#1}}

\newif\ifshowauthorcomments
\showauthorcommentstrue

\DeclareRobustCommand{\resp}[1]{#1}

\begin{document}
\preprint{CERN-TH-2026-235}
\preprint{TU-1328}
\title{Neutrino Direct Simulation Monte Carlo: Accurate modeling of out-of-equilibrium decaying cosmological relics}

\author{Kensuke~Akita\orcid{0000-0002-5919-9347}}
\email{kensuke.akita.e2@tohoku.ac.jp}
\affiliation{Department of Physics, Tohoku University, Sendai, Miyagi 980-8578, Japan}
\affiliation{Department of Physics, The University of Tokyo, Bunkyo-ku, Tokyo 113-0033, Japan}
\author{Miguel Escudero\orcid{0000-0002-4487-8742}}
\email{miguel.escudero@cern.ch}
\affiliation{Theoretical Physics Department, CERN, 1211 Geneva 23, Switzerland}
\author{Oleksii~Ihnatenko\orcid{0009-0004-6596-6137}}
\email{oleksii.ihnatenko@knu.ua}
\affiliation{Taras Shevchenko National University of Kyiv, 64 Volodymyrska Street, Kyiv 01601, Ukraine}
\author{Maksym~Ovchynnikov}
\email{maksym.ovchynnikov@cern.ch}
\affiliation{Theoretical Physics Department, CERN, 1211 Geneva 23, Switzerland}

\begin{abstract}
We consider nonstandard cosmologies in which decaying relics inject energy at MeV temperatures, driving neutrinos out of equilibrium.
The subsequent neutrino evolution shapes predictions for the cosmic radiation density and primordial light element abundances.
Earlier work introduced Neutrino Direct Simulation Monte Carlo ($\nu$DSMC), which describes neutrino transport through Monte Carlo sampling of particle interactions.
Here, we extend this method by jointly evolving the relic population, neutrino spectra, electromagnetic plasma, and cosmic expansion, and coupling this evolution to Big Bang nucleosynthesis.
The framework includes direct and inverse decays of the relic, neutrino flavor conversion, quantum statistics, and pion production by energetic neutrinos.
To reveal the physical impact of these effects and independently test $\nu$DSMC, we compare several approaches to neutrino evolution across a range of nonstandard cosmological scenarios.
We find excellent agreement with independent solutions of quantum kinetic or quasi-classical Boltzmann equations.
Compared with these calculations, $\nu$DSMC includes a broader range of physical processes and is substantially faster for strongly nonthermal neutrino populations.
This makes broad, robust parameter scans for nonstandard cosmologies practical.
We show that when departures from thermal equilibrium are large, approximations that average over neutrino momenta can give qualitatively incorrect predictions, including the wrong sign of the change in the effective number of neutrino species and substantially incorrect primordial nuclear abundances.
We conclude that the extended $\nu$DSMC framework enables reliable cosmological tests of new physics by following neutrino thermalization and its observable consequences.
\end{abstract}

\maketitle

\section{Introduction}
\label{sec:introduction}

Neutrino decoupling, at cosmic times of order $1\,\s$, is among the earliest stages of cosmic evolution accessible to established cosmological observations.
Big Bang nucleosynthesis (BBN) and the cosmic microwave background (CMB) are sensitive to the expansion rate and particle content of the Universe during this era, while a future detection of the cosmic neutrino background would provide a complementary test.
These observations can test the Standard Cosmological Scenario and probe non-standard physics, including \resp{lepton flavor asymmetries}~\cite{Froustey:2021azz,Froustey:2024mgf,Domcke:2025lzg,Domcke:2025jiy}, non-standard neutrino interactions~\cite{Archidiacono:2013dua,Forastieri:2015paa,deSalas:2021aeh,Du:2021idh,Barenboim:2025okj,Gariazzo:2026nfb}, dark radiation, unstable relics~\cite{Pospelov:2010cw,Fradette:2018hhl,Gelmini:2020ekg,Sabti:2020yrt,Boyarsky:2020dzc,Boyarsky:2021yoh,Mastrototaro:2021wzl,Rasmussen:2021kbf,Akita:2024nam,Akita:2024ork,Bianco:2025boy,Jung:2025dyo,EscuderoAbenza:2025tsi,Escudero:2026mgw,Bianco:2026dvc}, late reheating~\cite{Kawasaki:2000en,Hannestad:2004px,Ichikawa:2005vw,Hasegawa:2019jsa,Barbieri:2025moq}, and evaporating primordial black holes~\cite{Acharya:2020jbv,Keith:2020jww,Sanchis:2025awq}.
For decaying relics and non-standard neutrino interactions, cosmological probes are complementary to searches at accelerator experiments~\cite{Beacham:2019nyx,Ovchynnikov:2024rfu,deBlas:2025gyz}.

\begin{figure*}[t!]
\centering
\includegraphics[width=\textwidth]{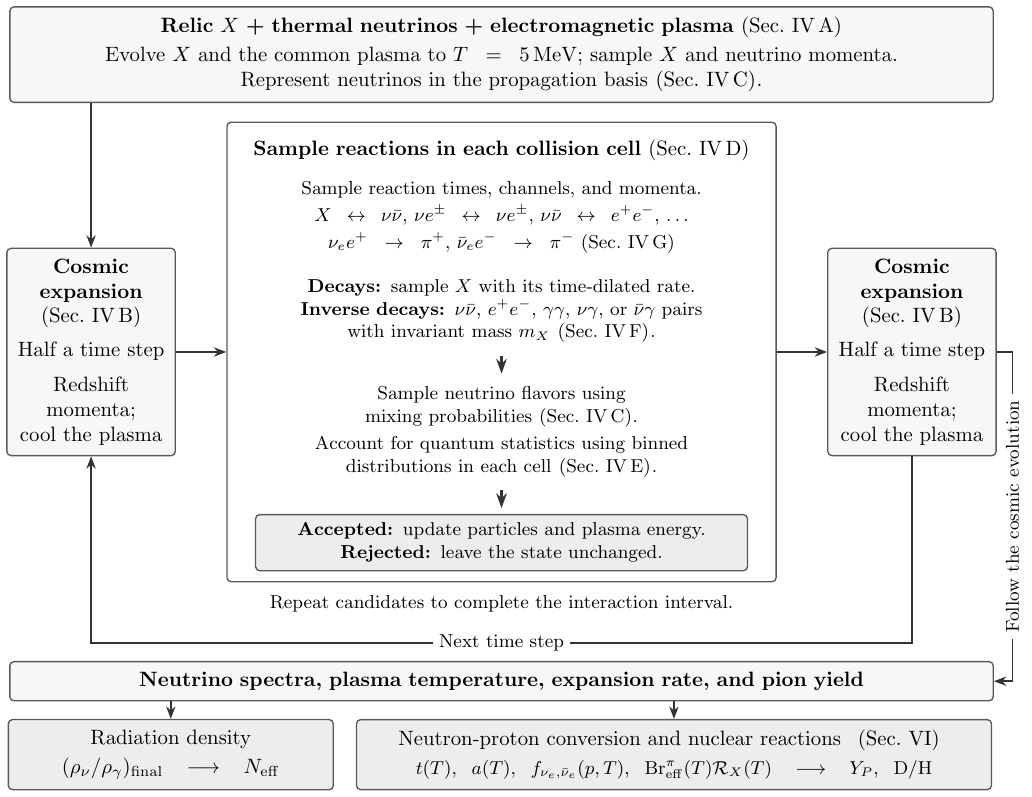}
\caption{\resp{Overview of \nudsmc, described in Sec.~\ref{sec:new-algorithm}.
Here, $T\equiv T_\gamma$ is the electromagnetic plasma temperature.
Above $T=5\,\mev$, neutrinos and the electromagnetic plasma share a common temperature.
Below this temperature, sampled neutrinos and relics $X$ evolve through the repeated expansion and reaction steps shown here.
The propagation basis is used when neutrino oscillations are enabled.
The final ratio of neutrino and photon energy densities, $(\rho_\nu/\rho_\gamma)_{\rm final}$, determines $N_{\rm eff}$.
The BBN inputs are cosmic time $t(T)$, scale factor $a(T)$, electron-neutrino and antineutrino occupations $f_{\nu_e,\bar\nu_e}(p,T)$ at physical momentum $p$, and the charged-pion yield per relic decay ${\rm Br}_{\rm eff}^{\pi}(T)$ [Eq.~\eqref{eq:brPiEff}].
Multiplying this yield by the relic decay rate per physical volume $\mathcal R_X(T)$ [Eq.~\eqref{eq:accepted-decay-density}] gives the pion production rate entering BBN.
The outputs are the helium-4 mass fraction $Y_P$ and the ratio of deuterium to hydrogen nuclei, D/H.
A case study applying this framework to heavy neutral leptons (HNLs) is discussed in Ref.~\cite{Akita:2026gee}.}}
\label{fig:nudsmc-overview}
\end{figure*}

CMB and BBN observations are becoming increasingly precise.
For example, in 2025, the effective number of neutrino species was measured with approximately $60\%$ greater precision by combining ACT, SPT, and Planck data~\cite{AtacamaCosmologyTelescope:2025nti,SPT-3G:2025bzu,Planck:2018vyg}, and the primordial helium abundance was measured with a precision of just $0.5\%$ in 2026~\cite{Yeh:2026lbt}.
Joint analyses with BAO and CMB data further show that BBN provides strong constraints on $N_{\rm eff}$ and that nuclear reaction networks and uncertainty treatments affect the inferred cosmological parameters~\cite{Akharman:2026izy}.
Using these data to test non-standard cosmologies requires predicting the full evolution of the neutrino momentum distributions together with the electromagnetic plasma and cosmic expansion.
During BBN, the electron-neutrino and antineutrino spectra determine the weak rates that interconvert neutrons and protons, while the total energy density controls the expansion rate.
The distributions left after decoupling also provide input for later calculations of cosmological perturbations and simulations of structure formation~\cite{Lesgourgues:2006nd}.

The final neutrino energy density relative to photons is commonly summarized by the effective number of neutrino species:
\begin{equation}
N_{\rm eff}\equiv\frac{8}{7}\left(\frac{11}{4}\right)^{4/3}
\left(\frac{\rho_\nu}{\rho_\gamma}\right)_{\rm final}.
\label{eq:neff-definition}
\end{equation}
Here, $\rho_\nu$ includes all active neutrinos and antineutrinos, and $\rho_\gamma$ is the photon energy density.
The ratio is evaluated after electron-positron annihilation and energy injection have ended, while the neutrinos remain relativistic.
This ratio does not specify the neutrino momentum distributions, their flavor content, or the earlier evolution of the plasma.
An accurate value of $N_{\rm eff}$ alone is generally insufficient for predicting primordial nuclear abundances.

Calculating this evolution requires a consistent treatment of neutrino-electron and neutrino-neutrino interactions, finite-temperature QED corrections to the electromagnetic equation of state and electron dispersion, and neutrino flavor conversion.
Most existing methods begin by analytically reducing the collision integrals in the neutrino evolution equations~\cite{Hannestad:1995rs}.
They then solve either quantum kinetic equations for the neutrino density matrix~\cite{Froustey:2019owm,Gariazzo:2019gyi,Akita:2020szl} or quasi-classical Boltzmann equations for the neutrino distribution functions.
The latter are treated on momentum grids~\cite{Dolgov:2000jw,Sabti:2020yrt,Boyarsky:2021yoh,Rasmussen:2021kbf,Akita:2024ork}, after integration over momentum~\cite{Escudero:2018mvt,EscuderoAbenza:2020cmq,Cielo:2023bqp,Escudero:2025kej}, or through finite-dimensional parametrizations of the distribution functions~\cite{Mangano:2001iu,Spinner:2026ytq}.
For the Standard Cosmological Scenario, these approaches consistently predict $\neff\approx3.044$~\cite{Escudero:2025kej}.
Beyond the standard scenario, however, their predictions can differ quantitatively and even qualitatively.
For late decaying relics, for example, different treatments have yielded opposite signs for $\Delta\neff$~\cite{Boyarsky:2021yoh,Ovchynnikov:2024xyd}.
Momentum-resolved calculations can also become computationally demanding when energetic neutrinos are injected~\cite{Sabti:2020yrt,Akita:2024ork}, while processes whose collision integrals do not admit a simple analytic reduction, such as hadronic decays, generally require process-specific treatment.

Refs.~\cite{Ovchynnikov:2024rfu,Ovchynnikov:2024xyd} introduced a complementary framework based on Neutrino Direct Simulation Monte Carlo (\nudsmc).
In this approach, sampled interactions of Monte Carlo neutrinos with the electromagnetic plasma and with particles injected by decaying relics replace direct evaluation of the neutrino Boltzmann collision kernels as the Universe expands.
The framework adapts the Direct Simulation Monte Carlo method originally developed to solve the non-relativistic Boltzmann equation for rarefied gases~\cite{roohi2016collision,stefanov2019basic}.
Refs.~\cite{Ovchynnikov:2024rfu,Ovchynnikov:2024xyd} incorporated an electromagnetic plasma bath, cosmological expansion, and neutrino oscillations.
Ref.~\cite{Ihnatenko:2025kew} subsequently refined the framework for a precision calculation of standard neutrino decoupling, obtaining $\neff\approx3.044$.

The computational cost of \nudsmc depends only weakly on the energy of the injected neutrinos, and the method does not require an analytic reduction of the collision integral.
The distributions of reaction products can be calculated in advance and tabulated.
Applications to high-energy neutrino injection at MeV temperatures resolved a discrepancy between existing treatments, demonstrating that injections with $E_\nu\gtrsim50\,\mev$ can decrease $\neff$~\cite{Ovchynnikov:2024rfu,Ovchynnikov:2024xyd}.

The original \nudsmc formulation did not include all ingredients required for a consistent treatment of neutrino evolution in non-standard cosmologies.
It neither evolved unstable relics explicitly nor included the additional reactions they induce, such as inverse decays of Standard Model particles into the relic species and the production of metastable particles such as charged pions.
Its treatments of flavor conversion and quantum statistics also required refinement.

Here, we extend \nudsmc to evolve relic decays together with the neutrino distributions, electromagnetic plasma, and cosmic expansion (see Fig.~\ref{fig:nudsmc-overview}).
We include direct and inverse decays, refine the treatments of flavor conversion and quantum statistics, and calculate the charged-pion source entering BBN.
Equilibrium tests and comparisons with independent calculations check the predicted radiation density, neutrino spectra, and primordial abundances.
The comparisons measure the computational cost and identify when a thermal description remains reliable and when persistent spectral distortions change the predicted radiation density and primordial abundances.

We have released the \textsc{Mathematica}~\cite{Mathematica} implementation of \nudsmc on \href{https://github.com/maksymovchynnikov/nuDSMC}{GitHub}~\cite{nuDSMC:repository}.
The faster \textsc{C++} implementation will be publicly released upon publication of this work.

The paper is organized as follows.
Sec.~\ref{sec:general} formulates the coupled evolution of an unstable relic, the electromagnetic plasma, and the neutrino distributions in an expanding Universe.
Sec.~\ref{sec:old-algorithm} reviews the original \nudsmc formulation and its limitations, while Sec.~\ref{sec:new-algorithm} presents the extended framework.
Sections \ref{sec:cross-checks} and \ref{sec:bbn-framework} describe the validation tests and the use of calculated neutrino spectra and plasma evolution in BBN, respectively.
Sec.~\ref{sec:case-studies} compares \nudsmc with other state-of-the-art treatments across the four cosmological scenarios.
Sec.~\ref{sec:performance-limitations} reports the computational performance and current physical limitations.
Finally, Sec.~\ref{sec:conclusions} summarizes the results and future applications.

\section{General formulation for decaying relics}
\label{sec:general}
The main focus of this work is the evolution of an unstable relic species $X$ around neutrino decoupling.
We characterize $X$ by its mass \mX, lifetime \tX, spin, decay modes, and phase-space distribution.
\resp{We denote the temperature of the electromagnetic plasma by $T\equiv T_\gamma$; photons, electrons, and positrons share this temperature throughout the evolution.}

We specify the initial condition for $X$ at a temperature \Tini, chosen below the QCD crossover but well above the neutrino decoupling scale:
\begin{align}
    Y_X(\Tini)
    &\equiv
    \left.\frac{n_X}{s}\right|_{\Tini}\,,
    \label{eq:initial-condition-X-1}
    \\
    n_X(\Tini)
    &=
    g_X\int\frac{\dd^3\mathbf p_X}{(2\pi)^3}\,
    f_X(p_X,\Tini)\,,
    \label{eq:initial-condition-X-2}
\end{align}
where $s$ is the entropy density of the primordial plasma and $g_X$ is the number of polarization states of $X$.
Throughout this paper, $f_X$ is the dimensionless occupation per polarization state; we assume equal occupations of these states.
The distribution summed over polarizations is therefore $g_Xf_X$, while the statistical factor for each final $X$ state is $1+f_X$ for a boson and $1-f_X$ for a fermion.
The distribution $f_X$ may be thermal, if $X$ was previously in equilibrium, or non-thermal, as in misalignment production, freeze-in, or decays of a heavier relic.
We assume that the $X$ momenta are isotropically distributed.

The evolution of $X$ in the plasma is split into two regimes separated by \Tspl.
We define \Tspl as the temperature above which neutrinos are treated as being in perfect thermal equilibrium with the electromagnetic plasma.
In practice, we use $\Tspl=5\mev$.
For $\Tspl<T\lesssim\Tini$, and for temperatures well below the QCD crossover, the Standard Model plasma is described as a single fluid with temperature $T$, energy density $\rho_{\rm pl}(T)$, and pressure $P_{\rm pl}(T)$.
The interactions of $X$ with this plasma may include decays, inverse decays, scatterings, and other processes.
Schematically,
\begin{align}
&\left(\partial_t-Hp_X\partial_{p_X}\right)f_X(p_X,t)
=
\mathcal C_X^{\rm pl}[f_X;T](p_X,t)\,,
\label{eq:evolution-above-split-a}
\\[0.8em]
&\frac{\dot a}{a}
=H(t) =
    \sqrt{
    \frac{8\pi}{3m_{\rm Pl}^{2}}
    \left(\rho_{\rm pl}+\rho_X\right)}\,,
\label{eq:evolution-above-split-b}
\\[0.8em]
&\frac{d\rho_{\rm pl}}{dt}
+
3H(\rho_{\rm pl}+P_{\rm pl})
=
\mathcal Q_{\rm pl}^{X}(t)\,.
\label{eq:evolution-above-split-c}
\end{align}
where
\begin{align}
    &\rho_X
    =
    g_X\int\frac{\dd^3\mathbf p_X}{(2\pi)^3}\,
    E_X f_X(p_X,t)\,.
\end{align}
The collision operator $\mathcal C_X^{\rm pl}$ may be written as
\begin{equation}
    \mathcal C_X^{\rm pl}
    =
    \mathcal C_X^{1\leftrightarrow2,{\rm pl}}
    +
    \mathcal C_X^{2\leftrightarrow2,{\rm pl}}
    +\dots\,,
\end{equation}
where $1\leftrightarrow2$ and $2\leftrightarrow2$ denote reaction classes that include both forward and inverse directions.
Energy conservation fixes the plasma source term $\mathcal Q_{\rm pl}^{X}$:
\begin{equation}
    \mathcal Q_{\rm pl}^{X}
    =
    -g_X
    \int\frac{\dd^3\mathbf p_X}{(2\pi)^3}\,
    E_X\,
    \mathcal C_X^{\rm pl}[f_X;T](p_X,t)\,.
\end{equation}
For $T\lesssim\Tspl$, neutrinos begin to decouple from the electromagnetic bath.
The electromagnetic bath remains a thermal fluid with temperature $T$, while the neutrino spectra must be evolved explicitly.
A fully general description is provided by quantum kinetic equations for the neutrino flavor density matrix; see, e.g., Refs.~\cite{Akita:2020szl,Froustey:2020mcq}.
In the absence of a sizeable neutrino-antineutrino asymmetry, we use adiabatic propagation with averaged oscillation phases, following the Adiabatic Transfer of Averaged Oscillations (ATAO) of Ref.~\cite{Froustey:2020mcq}.
We combine this propagation approximation with the flavor-projected collision prescription described in Sec.~\ref{sec:oscillations}.
In this adiabatic, oscillation-phase-averaged description, the propagation basis at each neutrino energy and plasma temperature is the eigenbasis of the in-medium Hamiltonian,
\begin{multline}
    \left(U^{\rm m}(E,T)\right)^\dagger
    \Omega_\nu(E,T)
    U^{\rm m}(E,T)
    \\ =
    {\rm diag}(\lambda_1,\lambda_2,\lambda_3)\,.
\end{multline}
We denote the neutrino and antineutrino occupation numbers in this basis by $f_i(p,t)$ and $\bar f_i(p,t)$, and define
\begin{equation}
    \Pi_{\alpha i}(E,T)
    =
    |U^{\rm m}_{\alpha i}(E,T)|^2\,.
\end{equation}
These projectors are calculated from the Hamiltonian at the relevant neutrino energies and plasma temperature; they are not independent evolution variables.
The evolution below \Tspl is then described by
\begin{widetext}
\begin{align}
\left(\partial_t-Hp\partial_p\right)f_i(p,t)
&=
\mathcal C_i^{2\leftrightarrow2}[f,\bar f;T](p,t)
+
\mathcal C_i^{1\leftrightarrow2}[f,\bar f,f_X;T](p,t)+\dots \,,
\nonumber\\
\left(\partial_t-Hp\partial_p\right)\bar f_i(p,t)
&=
\bar{\mathcal C}_i^{2\leftrightarrow2}[f,\bar f;T](p,t)
+
\bar{\mathcal C}_i^{1\leftrightarrow2}[f,\bar f,f_X;T](p,t)+\dots\,,
\nonumber\\
\left(\partial_t-Hp_X\partial_{p_X}\right)f_X(p_X,t)
&=
\mathcal C_X^{1\leftrightarrow2}[f,\bar f,f_X;T](p_X,t)
+
\mathcal C_X^{2\leftrightarrow2}[f,\bar f,f_X;T](p_X,t)+\dots \,,
\nonumber\\
\frac{d\rho_{\rm EM}}{dt}
+
3H(\rho_{\rm EM}+P_{\rm EM})
&=
\mathcal Q_{\rm EM}(t)\,,
\nonumber\\
\frac{\dot a}{a}
=
H
&=
\sqrt{
\frac{8\pi}{3m_{\rm Pl}^{2}}
\left(
\rho_{\rm EM}+\rho_\nu+\rho_X
\right)}\,.
\label{eq:evolution-below-split}
\end{align}
\end{widetext}

\section{Earlier versions of \nudsmcheading}
\label{sec:old-algorithm}

Earlier studies used the version of the \nudsmc algorithm summarized in this section~\cite{Ovchynnikov:2024rfu,Ovchynnikov:2024xyd,Ihnatenko:2025kew}.
Refs.~\cite{Ovchynnikov:2024rfu,Ovchynnikov:2024xyd} presented simplified prototype implementations, and Ref.~\cite{Ihnatenko:2025kew} refined the approach for a precision calculation of $\neff$ in the Standard Cosmological Scenario.
The discussion below mainly follows Ref.~\cite{Ihnatenko:2025kew}.

\subsection{The \nudsmcheading algorithm in earlier studies}

The calculation divides the evolution of the Universe into timesteps $\Delta t$ that are short enough to resolve both cosmological expansion and neutrino thermalization.
It follows a pool of $N_\nu$ Monte Carlo neutrinos $\nu_i$, each characterized by a state label $\alpha_i$ (for example, a flavor or propagation eigenstate) and an energy $E_i$.
The electromagnetic bath is described by its temperature \resp{$T$}.
To suppress Monte Carlo noise, typical calculations use $N_\nu\gtrsim10^5$ particles, with the largest runs reaching $N_\nu\simeq10^8$.
The initial volume $V\equiv V_{\rm ini}$ is specified at $T_{\rm split}$.

Let $O_{\rm exp}$ and $O_{\rm int}$ denote cosmological expansion and particle interactions, respectively.
Earlier implementations applied these operators during each timestep in the order
\begin{align}
O_{\rm exp}\to O_{\rm int}\,.
\label{eq:splitting-old}
\end{align}
That is, the system was expanded first and the interactions were simulated afterward.
The operator $O_{\rm exp}$ updated the volume $V$, the neutrino energies, and the \resp{electromagnetic plasma} energy density according to
\begin{equation}
\Delta \ln(a) \approx H\times \Delta t\,.
\label{eq:expansion-old}
\end{equation}
The interaction operator used the No-Time-Counter (NTC) scheme, one of the simplest DSMC algorithms~\cite{roohi2016collision,stefanov2019basic}.
Refs.~\cite{Ovchynnikov:2024rfu,Ovchynnikov:2024xyd,Ihnatenko:2025kew} adapted this scheme to include neutrino oscillations, the electromagnetic bath, and quantum-statistical factors.

After expansion, the system was divided into $N_{\rm cells}$ computational cells.
Each cell had volume $V_{\rm cell}=V/N_{\rm cells}$, approximately $N_{\nu,{\rm cell}}=N_\nu/N_{\rm cells}$ neutrinos, and \resp{electromagnetic plasma} energy $E_{\rm EM,cell}=\rho_{\rm EM}V_{\rm cell}$.
The corresponding energy density determined a unique cell temperature, $T_{\rm cell}=T(\rho_{\rm EM,cell})$, with $\rho_{\rm EM,cell}=E_{\rm EM,cell}/V_{\rm cell}$.
Because the physical system is homogeneous and isotropic, these cells had no physical boundaries; they served only to organize collisions and improve performance.
The earlier implementations required $N_{\nu,{\rm cell}}\gtrsim10^3$, so that the neutrino population in each cell could be treated as a statistical ensemble, for example when defining an effective neutrino temperature.

At the beginning of each interval $\Delta t$, the algorithm estimated an upper bound on the number of binary interactions involving the Monte Carlo neutrinos $\{\alpha_i,E_i\}$ and the $e^\pm$ particles:
\begin{equation}
N_{\text{pairs}}=\sum_{I=1}^{3}N_{\text{pairs}}^{(I)}\,.
\end{equation}
Here, $N_{\rm pairs}^{(I)}$ is the proposed upper bound for process class $I$:
\begin{align}
&\text{(I)}: \quad\nu_{\alpha}\nu_{\beta}\!\to\! \nu_{\alpha}\nu_{\beta}/\nu_{\alpha} \bar{\nu}_{\beta}\!\to\!Y\,,
\label{eq:proc-scatt-1}
\\
&\text{(II)}: \quad \nu e^\pm\!\to\!\nu e^\pm \,,
\label{eq:proc-scatt-2}
\\
&\text{(III)}: \quad e^+e^-\!\to\!\nu\bar\nu\,,
\label{eq:proc-scatt-3}
\end{align}
where
\begin{equation}
Y = \begin{cases}\nu_{\alpha}\bar{\nu}_{\beta}, \quad\alpha \neq \beta, \\ \nu_{\gamma}\bar{\nu}_{\gamma} \ \text{or} \ e^{+}e^{-}, \quad\alpha =
\beta \end{cases}\,.
\end{equation}
Explicitly,
\begin{equation}
N_{\rm pairs}^{(I)} = \frac{\Delta t}{V_{\rm cell}}\times \frac{N_{\rm I,cell}(N_{\rm I,cell}-1)}{2}\times (\sigma v)_{I}^{\rm max}\,.
\label{eq:sampling-pairs-proposal}
\end{equation}
Here, $N_{\rm I,cell}$ is the number of particles in the cell that participate in process class $I$; depending on the class, these particles are neutrinos or electrons.
The quantity $(\sigma v)_I^{\rm max}=\max_Y[(\sigma v)_I^Y]$ bounds the cross section times relative velocity over all allowed final states $Y$.
Its value depends on the \resp{weak interaction process and on the prescription for sampling particles}, as discussed below.

Neutrinos were drawn from the Monte Carlo pool in each cell, whereas $e^\pm$ were sampled from a Fermi-Dirac distribution with the temperature-dependent mass
\begin{equation}
m_{e,{\rm th}}^2(T) = m_{e}^2+\delta m_{e,(2)}^2(T)
\label{eq:thermal-electron-mass}
\end{equation}
at temperature $T$.
Here, $m_e$ is the vacuum electron mass, and $\delta m_{e,(2)}^2(T)$ is the leading $\mathcal{O}(e^2)$ finite-temperature electron self-energy correction used in Ref.~\cite{Akita:2020szl}.
The separate $\mathcal{O}(e^3)$ contribution included in the plasma thermodynamics is the ring-resummed plasmon correction to the electromagnetic equation of state; it does not modify the electron dispersion relation at this order.

A sampled pair was \emph{pre-accepted} with probability
\begin{equation}
P_{\rm acc}^{(I)} = (\sigma v)_{\rm pair}/(\sigma v)^{(I)}_{\max}\,,
\label{eq:pre-acceptance}
\end{equation}
where $(\sigma v)_{\rm pair}$ is the cross section times relative velocity for that pair.
After pre-acceptance, the final state $Y$ was selected with weight $(\sigma v)_Y^{(I)}/(\sigma v)_{\rm max}^{(I)}$.
Its kinematics were generated in the center-of-momentum frame, using a randomly sampled incoming direction, and then boosted to the laboratory frame of the initial pair.

The scattering event was finally accepted with the Pauli-blocking factor
\begin{equation}
P_{\rm acc}^{\rm blocking} = (1-f_{Y_1}(E_{Y_{1}}))(1-f_{Y_2}(E_{Y_{2}}))\,,
\label{eq:acceptance-Pauli}
\end{equation}
where $Y_1$ and $Y_2$ are the final-state particles.
The total acceptance probability is therefore $P_{\rm acc}^{(I)}P_{\rm acc}^{\rm blocking}$.

After each interaction, the algorithm updated both the neutrino pool and the \resp{electromagnetic plasma} energy $E_{\rm EM,cell}$.
For example, it removed two neutrinos after $\nu\bar\nu\to e^+e^-$, added two after the inverse process, and replaced their state information after scattering.
The updated plasma energy fixed the cell energy density and temperature, and hence $m_{e,{\rm th}}(T)$, through $\rho_{\rm EM,cell}=E_{\rm EM,cell}/V_{\rm cell}$.

The proposal in Eq.~\eqref{eq:sampling-pairs-proposal} depends on the scaling of the \resp{weak interaction cross section}.
For two particles with four-momenta $p_i=(E_i,\mathbf p_i)$, \resp{the scaling in the massless limit is}
\begin{equation}
\sigma v \simeq C\times E_1E_2f(\alpha)\,,
\label{eq:weak-cross-section-scaling}
\end{equation}
where $C=G_F^2\dots$ is a dimensional constant, $\alpha$ is the angle between the incoming momenta, ${\cos\alpha=\mathbf p_1\cdot\mathbf p_2/(|\mathbf p_1||\mathbf p_2|)}$, and $f(\alpha)=(1-\cos\alpha)^2$.
The explicit \resp{upper bound} in Eq.~\eqref{eq:sampling-pairs-proposal} nevertheless depends on the particle sampling prescription.

Refs.~\cite{Ovchynnikov:2024rfu,Ovchynnikov:2024xyd} sampled particles uniformly, independently of their energy and type, as in standard DSMC implementations.
In that case, $\max[(\sigma v)_I]$ is the largest possible cross section for any pair in the cell.
This prescription is inefficient for weak interactions: Eq.~\eqref{eq:weak-cross-section-scaling} gives $N_{\rm pairs}^{(I)}\propto E_{\rm max,cell,(I)}^2$, while a typical pair is accepted with probability $P_{\rm acc,int}^{(I)}\propto E_1E_2/E_{\rm max,(I)}^2\ll1$.
Most sampled pairs therefore do not interact.

Ref.~\cite{Ihnatenko:2025kew} instead sampled each particle with weight $\omega_{i,j}\propto E_{i,j}$.
In this scheme, \resp{the proposed upper bound} in Eq.~\eqref{eq:sampling-pairs-proposal} becomes
\begin{multline}
N_{\rm pairs}^{\rm actual}=\frac{\Delta t}{V_{\rm cell}}\sum(\sigma v)_{\rm pair}\\\ <\frac{\Delta t}{V_{\rm cell}}C^{(I)}_{\rm max}\sum_{i,j} E^{\textbf{I},1}_{i}E^{\textbf{I},1}_{j}=\frac{\Delta t}{V_{\rm cell}}C_{\rm max}^{(I)}\frac{E_{\rm total}^{\rm (I),1}E_{\rm total}^{\rm (I),2}}{N_{\rm I, cell}}\,,
\end{multline}
where the constant $C^{(I)}_{\max}$ defines the \resp{upper bound} for process class $I$, and $E_{\rm total}^{(I),1/2}$ denotes the total energy of the two particle species participating in that class.
The resulting number of proposed pairs is
\begin{equation}
    N_{\rm pairs}^{(I)} \simeq \frac{\Delta t}{V} \,C_{\max}^{(I)}\times E_{\rm total}^{(I),1}\times E_{\rm total}^{(I),2}\,.
\end{equation}
The corresponding acceptance factor is
\begin{equation}
    P_{\rm acc}^{(I)} = \frac{C^{(I)}}{C_{\max}^{(I)}}\times \frac{g(E_{1},E_{2},\alpha)}{E_{1}E_{2}f(\alpha)}\,,
\end{equation}
where $g(E_1,E_2,\alpha)$ contains the exact dependence on energy and angle, including finite particle masses.
This factor is typically $\mathcal O(1)$, except in non-relativistic regions of phase space.

For the quantum-statistical acceptance in Eq.~\eqref{eq:acceptance-Pauli}, the earlier implementation estimated the final-state occupation numbers from an effective temperature for each species:
\begin{align}
    f_{Y_{1}}(E_{Y_{1}}) \approx 1/(\exp(E_{Y_{1}}/T_{Y_{1}})+1)\,,
\end{align}
The temperature $T_{Y_1}$ was inferred from the energy density $\rho_{Y_1}(T_{Y_1})$ under the assumption of \resp{an equilibrium energy distribution}.
This treatment is accurate for electromagnetic particles at the temperatures of interest but only approximate for neutrinos.

Finally, neutrino oscillations were represented by globally reassigning the flavor labels in the neutrino pool at the beginning of every timestep.
Before each call to the interaction routine, the labels were drawn from \resp{transition probabilities averaged over oscillations}:
\begin{equation}
\{\alpha_{i},E_{i}\} \to \{\alpha'_{i},E_{i}\}\,,
\end{equation}
where $\alpha_i'$ was sampled using the in-medium PMNS matrix $U_{i\alpha}(T)$:
\begin{equation}
    P_{\alpha\to \alpha'}(T,E_{i}) = \sum_{j}|U_{\alpha j}(E_{i},T)|^{2}|U_{j\alpha'}(E_{i},T)|^{2}\,.
\end{equation}
The in-medium PMNS matrix was evaluated following Ref.~\cite{Sabti:2020yrt}.

\subsection{Limitations}

The formulation above recovered the standard result $\neff\approx3.044$~\cite{Ihnatenko:2025kew}, but it has several physical and computational limitations in non-standard cosmologies.

\emph{First}, the expansion and interaction operators $O_{\rm exp}$ and $O_{\rm int}$ in Eq.~\eqref{eq:splitting-old} do not commute.
In particular, the amount of energy transferred from neutrinos to the EM sector differs depending on whether interactions occur before or after redshifting.
The ordering in Eq.~\eqref{eq:splitting-old} therefore introduces an $\mathcal O(c\Delta t)$ error, whose coefficient $c$ depends on the properties of $X$, the timestep prescription, and the physical scenario.

\emph{Second}, as its name suggests, the NTC method samples collisions over each time step according to their expected rates and assigns no explicit continuous collision time to individual interactions.
In the large-$N_{\rm cell}$ and small-$\Delta t$ limit, this procedure recovers the Poissonian structure of the collision process~\cite{stefanov2019basic}.
This limit precludes the large timesteps that would improve performance.
With decaying particles, continuous injection during the timestep would require splitting the interaction operator $O_{\rm int}$ into two operators, $O_{\rm int} \to \{O_{\rm 2\leftrightarrow 2}, O_{\rm 1\leftrightarrow 2}\}$, handling the $2\leftrightarrow2$ scatterings and $1\leftrightarrow 2$ processes separately.
This would introduce an additional $\mathcal O(\Delta t)$ error beyond the splitting between $O_{\rm exp}$ and $O_{\rm int}$.
Physically, the injected neutrinos thermalize differently depending on their injection time within the timestep.

\emph{Third}, estimating neutrino Pauli blocking from an effective temperature is reliable only when the neutrino energy distribution in each cell closely follows a Fermi-Dirac shape.
This assumption is well justified in the Standard Cosmological Scenario.
It fails, however, when substantial energy is exchanged between neutrinos and the electromagnetic sector or when $X$ decays produce non-thermal neutrino spectra.

\emph{Fourth}, the oscillation prescription has an intrinsic drawback.
At the start of every timestep, it redraws the flavor of each neutrino from the \resp{transition probabilities averaged over oscillations}.
A calculation with smaller timesteps therefore applies more independent flavor updates over the same physical time, although the physical oscillation and collision rates have not changed.
The resulting flavor composition depends on the numerical timestep and on any other algorithmic choice that changes how often the interaction kernel is called.

The prescription also discards the evolution of each neutrino between successive interactions.
The in-medium Hamiltonian changes as the plasma temperature $T$ falls, but the old update selects the next flavor only from probabilities evaluated at the current $T$.
Physical neutrinos propagate between weak interactions through an expanding, slowly varying medium.
This motivates \resp{evolving flavor in the propagation basis} between physical interactions.

This timestep dependence is especially severe for injections of non-thermal, high-energy neutrinos.
Since weak interaction rates grow with neutrino energy, a small population of energetic injected neutrinos can transiently raise the maximal interaction rate used to set the adaptive timestep.
The algorithm then uses smaller timesteps while these neutrinos are present and returns to larger timesteps after they have downscattered.
Because the old prescription redraws each flavor once per timestep, \resp{this change in timestep} also changes the number of artificial flavor updates and hence the predicted flavor composition.

\emph{Fifth}, the relic population and its energetic decay products open reaction channels that are negligible in the standard MeV plasma.
In addition to $X$ decays, $X\to{\rm SM}$, the inverse processes ${\rm SM}\to X$ can be relevant.
Scatterings of non-thermal neutrinos and electromagnetic particles can also produce heavier Standard Model species, including pions and muons.
Their non-thermal abundances may greatly exceed the corresponding Boltzmann-suppressed equilibrium yields, thereby modifying both \resp{energy exchange between neutrinos and the electromagnetic plasma} and the neutron-to-proton ratio that sets the BBN initial conditions.

\section{Extended \nudsmcheading framework}
\label{sec:new-algorithm}
The extended \nudsmc algorithm addresses the limitations identified in Sec.~\ref{sec:old-algorithm} and adds the direct and inverse decays of $X$, together with charged-pion production.
We focus on $1\leftrightarrow2$ decays and write the combined interaction operator schematically as $O_{\rm int}=O_{2\leftrightarrow2+1\leftrightarrow2}$.
Extensions to other decay topologies and to non-standard neutrino interactions mediated by $X$ are left for future work.

Section~\ref{sec:Tini-Tsplit} describes the evolution to the onset of neutrino decoupling in the presence of decaying $X$ particles.
Section~\ref{sec:expansion-interaction-splitting} introduces the improved splitting of expansion and interactions, and Sec.~\ref{sec:oscillations} describes neutrino oscillations.
Sections~\ref{sec:MCF} and \ref{sec:Pauli-blockings} present the collision kernel that handles decays and $2\leftrightarrow2$ scatterings simultaneously and the local occupation numbers used for Pauli blocking and Bose stimulation.
The implementations of $1\leftrightarrow2$ decays and charged-pion production are given in Secs.~\ref{sec:1-to-2} and \ref{sec:charged-pion}, respectively.

\subsection{Evolution to the onset of neutrino decoupling}
\label{sec:Tini-Tsplit}

As discussed in Sec.~\ref{sec:general}, neutrinos and electromagnetic particles can be treated as a single fluid of temperature $T$ until neutrino decoupling begins at $T_{\rm split}\simeq5\,\mev$.
The $X$ population may already affect the expansion rate above this temperature, which in turn changes its abundance and momentum distribution at $T_{\rm split}$.
The pre-decoupling evolution of the $X$ distribution and of the expansion rate must therefore be solved consistently.

In \nudsmc, the initial conditions are specified at $T_{\rm ini}$, with the default choice $T_{\rm ini}=20\,\mev$, where $X$ is assumed to contribute negligibly to the expansion rate.
As in Eqs.~\eqref{eq:initial-condition-X-1} and \eqref{eq:initial-condition-X-2}, the $X$ population is defined by its abundance $Y_X(T_{\rm ini})$ and a tabulated momentum distribution $f_X(p_X,T_{\rm ini})$.\footnote{These quantities are physically related.
For convenience, the implementation permits an arbitrary normalization of the distribution function.}
Built-in choices include Fermi-Dirac and Bose-Einstein distributions, appropriate when $X$ was once thermalized and subsequently decoupled at $T_{\rm dec}>T_{\rm ini}$, and a Dirac-delta distribution in which all $X$ particles have the same momentum $p$, including $p=0$.
The initial distribution can also be specified by a table.

Before neutrino decoupling, the $X$ distribution and the expansion rate are evolved according to Eqs.~\eqref{eq:evolution-above-split-a}--\eqref{eq:evolution-above-split-c}, with the $X$ population discretized in fine momentum bins.
At $T_{\rm split}$, this stage provides the initial time $t(T_{\rm split})$, volume $V(T_{\rm split})$, and distribution $f_{X}(p_{X},T_{\rm split})$ for the momentum-resolved neutrino evolution.
Monte Carlo $X$ particles are then sampled from this distribution to reproduce the $X$ number density.
The number of sampled Monte Carlo neutrino particles of each flavor is fixed by $\rho_{\nu_{\alpha}}(T_{\rm split}) \approx N_{\nu_{\alpha}}\times 3.15T_{\rm split}/V(T_{\rm split})$.

\subsection{Initial conditions and \resp{splitting of expansion and interactions}}
\label{sec:expansion-interaction-splitting}

The update in Eq.~\eqref{eq:splitting-old} first expands the Universe through a full timestep and only then simulates the particle interactions.
We replace this asymmetric ordering by Strang splitting: the system expands for half a timestep, undergoes all simulated $1\leftrightarrow2$ and $2\leftrightarrow2$ reactions, and then expands for the remaining half.
This reduces the \resp{error from splitting expansion and interactions} to $\mathcal{O}((\Delta t)^{2})$:
\begin{equation}
    \left(O_{\rm exp}(\Delta t/2)\right) \to O_{1\leftrightarrow 2 + 2\leftrightarrow 2} \to \left( O_{\rm exp}(\Delta t/2)\right)\,.
    \label{eq:splitting-general}
\end{equation}

During each half-step, the expansion operator $O_{\rm exp}$ updates the scale factor and redshifts all particle momenta and spatial scales.
For the interval $\delta t=\Delta t/2$, the scale factor is evolved beyond the first-order relation in Eq.~\eqref{eq:expansion-old} by including the change in the Hubble rate:
\begin{equation}
    \Delta \ln(a) \approx H\times \delta t + \dot{H}\times \frac{(\delta t)^{2}}{2}\,,
    \label{eq:a-change}
\end{equation}
The quantity $\dot H$ follows from the Friedmann equations:
\begin{equation}
\dot{H} = -\frac{4\pi}{m_{\text{Pl}}^{2}}\sum_{y = X,\nu,\text{EM}}(P_{y}+\rho_{y}),
\end{equation}
where $P_y$ and $\rho_y$ are the pressure and energy density of species $y$.
The momenta of the $X$ particles and neutrinos redshift as $p_i\mapsto p_i\xi$, where $\xi=\exp[-\Delta\ln(a)]$.
The \resp{electromagnetic plasma} energy density is updated according to
\begin{equation}
   \rho_{\text{EM}} \to \rho_{\text{EM}}\times\exp\left[-\frac{3\Delta \ln(a)(\rho_{\text{EM}}+P_{\text{EM}})}{\rho_{\text{EM}}}\right]\,.
\end{equation}
The \resp{electromagnetic plasma} temperature $T$ is then obtained by inverting $\rho_{\rm EM}=\rho_{\rm EM}(T)$.
Finally, the system volume is updated as
\begin{equation}
V \to V\times \exp\left[3\Delta \ln(a)\right]\,.
\end{equation}

\subsection{Neutrino oscillations}
\label{sec:oscillations}

Neutrino oscillations redistribute the injected population among flavors with different weak interaction rates.
We follow adiabatic propagation with averaged oscillation phases, as in ATAO~\cite{Froustey:2020mcq}, and evaluate Standard Model collisions in the propagation basis, including interference between flavor amplitudes.

\begin{samepage}
\subsubsection{Propagation between interactions}

We work in the regime of negligible lepton asymmetries.
The only flavor-nonuniversal refractive term included in the Hamiltonian is the \resp{thermal charged-lepton contribution, which is the same for neutrinos and antineutrinos.
We take the Dirac phase in the vacuum mixing matrix to be $\delta_{\rm CP}=0$.}\footnote{\label{fn:cp-phase}\resp{Below $5\,\mev$, the refractive and collision terms from the electromagnetic plasma treat $\nu_\mu$ and $\nu_\tau$ identically.
In standard neutrino decoupling, this symmetry makes the effect of $\delta_{\rm CP}$ on $N_{\rm eff}$ and the electron-neutrino spectrum negligible~\cite{Froustey:2020mcq}.
Ref.~\cite{Froustey:2021azz} also establishes negligible effects on these observables for the primordial asymmetries studied there.
The primary neutrino injection considered here has equal branching fractions into the three flavors and preserves this symmetry.
Decay channels that distinguish $\nu_\mu$ from $\nu_\tau$ would require a separate assessment.}}

\end{samepage}

The flavor-universal part of the refractive potential is omitted, since it contributes only an overall phase and does not affect flavor conversion.
The oscillation Hamiltonian in flavor space is therefore
\begin{multline}
    \Omega_\nu(E_{\nu},T)
    =
    \frac{1}{2E_{\nu}}\,
    U_{\rm PMNS}\,M^2\,U_{\rm PMNS}^\dagger
    \\ +
    \mathrm{diag}(V_e-V_x,0,0),
    \label{eq:Omega-nu}
\end{multline}
where
\begin{equation}
    M^2
    \equiv
    \mathrm{diag}(m_1^2,m_2^2,m_3^2)
\end{equation}
is the vacuum neutrino mass-squared matrix and $U_{\rm PMNS}$ is the vacuum leptonic mixing matrix.

For $x=\mu,\tau$, the thermal charged-lepton contribution is
\begin{equation}
    V_e-V_x
    =
    -\frac{2\sqrt{2}\,G_F\,E_{\nu}}{m_W^2}
    \left(
        \rho_{e^-}+P_{e^-}+\rho_{e^+}+P_{e^+}
    \right),
    \label{eq:Vex-thermal}
\end{equation}
see, e.g., Ref.~\cite{Froustey:2020mcq}.
In the ultra-relativistic limit, $\rho_{e^-}+P_{e^-}+\rho_{e^+}+P_{e^+}\simeq 7\pi^2T^4/45$, so that
\begin{equation}
    V_e-V_x
    \simeq
    -\frac{14\sqrt{2}\pi^2}{45}\,
    \frac{G_F\,E_{\nu}\,T^4}{m_W^2}.
    \label{eq:Vex-ur}
\end{equation}
It is often useful to rewrite the same matter effect in mass-squared units,
\begin{equation}
    A_{\rm m}(E_{\nu},T)
    \equiv
    2E_{\nu}(V_e-V_x),
    \label{eq:a-def}
\end{equation}
or, using Eq.~\eqref{eq:Vex-ur},
\begin{equation}
    A_{\rm m}(E_{\nu},T)
    \simeq
    -\frac{28\sqrt{2}\pi^2}{45}\,
    \frac{G_F\,E_{\nu}^2\,T^4}{m_W^2}.
    \label{eq:athermal}
\end{equation}
Equivalently, the matter basis may be obtained by diagonalizing
\begin{equation}
    2E_\nu\Omega_\nu
    =
    U_{\rm PMNS}M^2U_{\rm PMNS}^\dagger
    +
    \mathrm{diag}\!\bigl(A_{\rm m}(E_\nu,T),0,0\bigr).
    \label{eq:matter-masssq}
\end{equation}

The approximation in Eq.~\eqref{eq:Omega-nu} neglects the flavor-nonuniversal part of neutrino self-refraction.
In the full quantum-kinetic equations, the $\nu$-$\nu$ forward-scattering potential contains both a CP-odd asymmetry piece proportional to $(n_{\nu}-n_{\bar{\nu}})_{\alpha\beta}$ and a CP-even finite-temperature piece proportional to $(\rho_{\nu}+\rho_{\bar{\nu}})_{\alpha\beta}$.
In the scenarios considered here, we assume no sizeable primordial lepton asymmetry and take the residual traceless flavor structure of the neutrino self-potential to remain subleading.
We therefore include only the charged-lepton thermal contribution in Eq.~\eqref{eq:Vex-thermal}.
\resp{With the assumptions on lepton asymmetries and the CP phase stated above, neutrinos and antineutrinos have the same matter projectors and use the same flavor probabilities in the reaction rates.}

At each local value of $(E_{\nu},T)$, we define the instantaneous propagation basis as the basis that diagonalizes $\Omega_\nu$,
\begin{multline}
    \bigl(U^{\rm m}(E_{\nu},T)\bigr)^\dagger
    \Omega_\nu(E_{\nu},T)\,
    U^{\rm m}(E_{\nu},T)
    \\ =
    \mathrm{diag}\!\bigl(\lambda_1,\lambda_2,\lambda_3\bigr),
    \label{eq:Um-diag}
\end{multline}
where $U^{\rm m}(E_{\nu},T)$ is the matter mixing matrix and $\lambda_i(E_\nu,T)$ are the corresponding in-medium eigenvalues.
The \resp{labels of propagation eigenstates} $i=1,2,3$ are assigned by continuous adiabatic continuation to the vacuum mass eigenstates in the limit $V_e-V_x\to0$.

The flavor and propagation bases are related by
\begin{equation}
    |\nu_\alpha\rangle
    =
    \sum_{i=1}^{3}
    U^{{\rm m}*}_{\alpha i}(E_{\nu},T)\,
    |\nu_i^{\rm prop}\rangle,
\end{equation}
so that the local projector probabilities are
\begin{equation}
    \Pi_{\alpha i}(E_{\nu},T)
    \equiv
    \bigl|U^{\rm m}_{\alpha i}(E_{\nu},T)\bigr|^2.
    \label{eq:projector}
\end{equation}
They give the probability that a neutrino in propagation eigenstate $i$ is measured as flavor $\alpha$ by a local weak interaction.

During free streaming, each Monte Carlo neutrino keeps its propagation label as its momentum redshifts and the plasma cools.
This update assumes adiabatic evolution of the instantaneous eigenstates and neglects transitions between propagation branches.
We also average the relative oscillation phases and evolve only the occupations $f_i(E,T)$ in the propagation basis.
The corresponding density matrix in the flavor basis is
\begin{equation}
    \rho_{\alpha\beta}(E,T)
    =
    \sum_j
    U^{\rm m}_{\alpha j}(E,T)\,
    f_j(E,T)\,
    U^{{\rm m}\,*}_{\beta j}(E,T).
    \label{eq:rho-reconstructed}
\end{equation}
It generally contains off-diagonal entries even after the propagation phases have been averaged.
At a fixed background, the resulting flavor transition probabilities are
\begin{equation}
    P_{\alpha\beta}^{\rm avg}(E_{\nu},T)
    =
    \sum_{i=1}^{3}
    \Pi_{\alpha i}(E_{\nu},T)\,
    \Pi_{\beta i}(E_{\nu},T).
    \label{eq:oscillation-avg}
\end{equation}
Along an evolving trajectory, the projectors at production and at the next interaction are evaluated at their respective energies and temperatures.

To compute $\Pi_{\alpha i}(E_{\nu},T)$ efficiently and with controlled accuracy over a wide range of $(E_{\nu},T)$, we use the compact three-flavor matter parametrization of Ref.~\cite{Denton:2018hal}.
In that approximation the medium modifies the effective $1$-$3$ and $1$-$2$ mixing through $\widetilde\theta_{13}$ and $\widetilde\theta_{12}$, while $\theta_{23}$ is kept at its vacuum value at leading order.
\resp{We use the CP phase specified above.}
The effective angles are obtained from Ref.~\cite{Denton:2018hal}:
\begin{widetext}
\begin{align}
\cos 2\widetilde\theta_{13}
   &=
    \frac{
    \cos 2\theta_{13} - A_{\rm m}(E_{\nu},T)/\Delta m^2_{ee}
    }
    {
    \sqrt{
    \bigl(\cos 2\theta_{13}
    - A_{\rm m}(E_{\nu},T)/\Delta m^2_{ee}\bigr)^2
    + \sin^2 2\theta_{13}}
    },
    \label{eq:DMP-theta13}
    \\
\cos 2\widetilde\theta_{12}
    &=
    \frac{
    \cos 2\theta_{12} - A_{\rm m}^{(0)}(E_{\nu},T)/\Delta m^2_{21}
    }
    {
    \sqrt{
    \bigl(\cos 2\theta_{12}
    - A_{\rm m}^{(0)}(E_{\nu},T)/\Delta m^2_{21}\bigr)^2
    + \sin^2 2\theta_{12}
    \cos^2(\widetilde\theta_{13}-\theta_{13})}
    } .
    \label{eq:DMP-theta12}
\end{align}
\end{widetext}
Here,
\begin{equation}
    \Delta m^2_{ee}
    =
    \cos^2\theta_{12}\,\Delta m^2_{31}
    +
    \sin^2\theta_{12}\,\Delta m^2_{32},
    \label{eq:Dm-ee}
\end{equation}
and
\begin{multline}
    A_{\rm m}^{(0)}(E_{\nu},T)
    =
    A_{\rm m}(E_{\nu},T)\cos^2\widetilde\theta_{13}
    \\ +
    \Delta m^2_{ee}
    \sin^2(\widetilde\theta_{13}-\theta_{13}) .
    \label{eq:a0-DMP}
\end{multline}
Ref.~\cite{Denton:2018hal} gives these formulae for the standard matter potential.
In the present cosmological setting, we apply the same parametrization using the local thermal refractive potential in Eq.~\eqref{eq:athermal}.
The dependence on the mass ordering enters through the vacuum spectrum, i.e. through the signed choices of $\Delta m^2_{31}$ and $\Delta m^2_{32}$, and hence through $\Delta m^2_{ee}$.

\subsubsection{Collisions in the propagation basis}

After averaging the propagation phases, the density matrix in the flavor basis has the off-diagonal elements in Eq.~\ref{eq:rho-reconstructed}, which contribute to weak collision rates (Appendix~\ref{app:osc-coherence}).
We evaluate Standard Model collision amplitudes in the local propagation basis, adding the flavor contributions before squaring.
For a reaction $r$, the incoming kets and outgoing bras give
\begin{multline}
    \mathcal M^{\rm m}_{r,\boldsymbol j;\boldsymbol i}
    =\sum_{\boldsymbol\alpha,\boldsymbol\beta}
    \mathcal M^{\rm f}_{r,\boldsymbol\beta;\boldsymbol\alpha}
    \\
    \times\prod_{a\in{\rm in}}U^{\rm m}_{\alpha_a i_a}(E_a,T)
    \prod_{b\in{\rm out}}U^{{\rm m}*}_{\beta_b j_b}(E'_b,T),
    \label{eq:coherent-matter-amplitude}
\end{multline}
where $\boldsymbol i$ and $\boldsymbol j$ are the incoming and outgoing propagation labels, and $\boldsymbol\alpha$ and $\boldsymbol\beta$ are the corresponding flavor labels.
The incoming neutrino factors are the overlaps $\langle\nu_\alpha|\nu_i^{\rm prop}\rangle=U^{\rm m}_{\alpha i}$; outgoing neutrinos contribute the conjugate overlaps.
The sums and products run over all neutrino and antineutrino legs, with these conjugations reversed for antineutrinos.
The matter mixing matrices are real for the CP phase used here and are evaluated at each particle's energy and the local temperature.
The spin-summed $|\mathcal M^{\rm m}_r|^2$ determines the differential collision probabilities, including interference between flavor amplitudes.
We sample the outgoing propagation labels jointly with the collision kinematics.

The model specifies relic decays through incoherent flavor channels with the stated partial widths.
A neutrino daughter produced in flavor $\alpha$ is initialized with
\begin{equation}
    \mathbb P(\nu_\alpha\to\nu_j;E,T)
    =\Pi_{\alpha j}(E,T)
    =\bigl|U^{\rm m}_{\alpha j}(E,T)\bigr|^2,
    \label{eq:measure-prop}
\end{equation}
with the same rule for antineutrinos under the stated CP assumptions.

\subsubsection{Pauli blocking}

The final-state propagation labels sampled from Eq.~\ref{eq:coherent-matter-amplitude} determine the Pauli factors.
For outgoing neutrinos and antineutrinos, the event acceptance contains
\begin{equation}
    \Pi_{\rm Pauli}
    =\prod_{b\in{\rm out},\,\nu,\bar\nu}
    \left[1-f_{j_b}(E'_b,T)\right].
    \label{eq:coherent-pauli}
\end{equation}
Each occupation is estimated from the current population in the collision cell, as described in Sec.~\ref{sec:Pauli-blockings}.

\subsubsection{Relation to quantum kinetic evolution}

For an initially dominant scalar at rest with $m_X=12.5\,\mev$ and $\tau_X\simeq0.5\,\s$, we take branching fractions $1/2$, $3/8$, and $1/8$ into $e^-e^+$, $\nu_e\bar\nu_e$, and $\nu_\mu\bar\nu_\mu$, respectively.
With inverse decays disabled, including collision interference lowers the final $N_{\rm eff}$ by about $0.01$.

The propagation approximation requires adiabatic evolution of the matter eigenstates and rapid averaging of the relevant relative phases between collisions.
The oscillation frequencies are the splittings of the full in-medium Hamiltonian,
\begin{equation}
    \omega_{ij}(E,T)
    =
    |\lambda_i(E,T)-\lambda_j(E,T)| .
    \label{eq:omegaij-splitting}
\end{equation}
For a reaction class $r$, the ratio of oscillation frequency to interaction rate is
\begin{equation}
    R_{\rm osc}^{(r)}(E,T)
    =
    \min_{(i,j)\in{\cal P}_r}
    \frac{\omega_{ij}(E,T)}
    {\Gamma_r^{\rm eff}(E,T)} ,
    \label{eq:Rosc-phase-averaging}
\end{equation}
where $\Gamma_r^{\rm eff}$ is the physical interaction rate and ${\cal P}_r$ contains the pairs of propagation eigenstates whose relative phases affect that reaction.
Phase averaging requires $R_{\rm osc}^{(r)}\gg1$ for the populations contributing appreciably to the observable.
Around standard decoupling, the thermal rates and expansion time give the scale separation estimated in Appendix~\ref{app:osc-conditions}.
For relic decays, the time-dependent neutrino distributions, temperature, and expansion rate determine these ratios.
The comparisons with QKE in Sec.~\ref{sec:case-studies} assess the resulting radiation density and spectra in the cosmological benchmarks.

\subsection{Majorant Collision Frequency scheme}
\label{sec:MCF}
The interaction operator $O_{1\leftrightarrow 2 + 2\leftrightarrow 2}$ in Eq.~\eqref{eq:splitting-general} must describe many interactions within one macro-step without erasing the physical ordering of individual events.
We therefore replace the NTC collision scheme with an advanced version of the Majorant Collision Frequency (MCF) scheme~\cite{venkattraman2012comparative,roohi2016collision}.
At a fixed cell state, MCF constructs an upper bound $\Gamma_{\max}$ on the total physical interaction rate and draws the time to the next candidate from the exponential distribution,
\begin{equation}
\delta\tau=-\frac{\ln\xi}{\Gamma_{\max}}\,.
\end{equation}
It then selects an interaction class according to its contribution to $\Gamma_{\max}$ and accepts the candidate with the ratio of the exact rate to its upper bound.
A rejected candidate leaves the state unchanged.
After an accepted event, the particle populations and all majorant rates are updated before the next waiting time is drawn.
Candidates are therefore Poisson distributed between changes of state, and their sequence follows the continuous-time Markov jump process describing physical collisions.
In particular, this Poissonian event structure is preserved for a finite number of Monte Carlo particles per cell even when $\Gamma_{\max}\Delta t\gg1$.
Thus, this formulation is especially attractive for our application: $\Delta t$ need not resolve the mean collision time; its remaining accuracy constraint comes from the \resp{splitting of expansion and interactions} discussed above.

For the Standard Model $2\leftrightarrow2$ reactions, the total collision majorant is the sum over reaction classes,
\begin{equation}
\Gamma_{\max}^{2\leftrightarrow2}
=\sum_{I=1}^{3}\Gamma_{\max}^{(I)}\,.
\end{equation}
For each of the two incoming particle populations $a$, define the energy represented in the cell by $\mathcal E_{I,a}=\sum_{i\in a}E_i$ for the unit-weight Monte Carlo neutrinos used here, or by the corresponding thermal energy moment times $V_{\rm cell}$ for an \resp{electromagnetic plasma} population.
The channel majorant then has the common form
\begin{equation}
\Gamma_{\max}^{(I)}
=\frac{C_{\max}^{(I)}}{V_{\rm cell}}\,
 \mathcal E_{I,1}\mathcal E_{I,2}\,.
\end{equation}
The factor $C_{\max}^{(I)}$ includes the identical-particle symmetry and channel multiplicities and bounds the remaining angular, flavor, and finite-mass dependence of the weak rate.
We preserve the energy-weighted selection introduced in the NTC scheme~\cite{Ihnatenko:2025kew}: an incoming neutrino is proposed with probability $p_i=E_i/\mathcal E_{I,a}$.
Energetic neutrinos, which have larger weak interaction rates, are consequently tested more often.
This is an importance-sampling probability.
The physical neutrino weights remain unchanged, and \resp{accepting candidates with the ratio of the exact rate to its upper bound restores the physical rate}.

Our extension places direct and inverse decays of the unstable $X$ particles on the same continuous event clock as the $2\leftrightarrow2$ reactions.
The cellwise total majorant becomes
\begin{equation}
\Gamma_{\max}=\Gamma_{\max}^{2\leftrightarrow2}
              +\Gamma_{\max}^{1\leftrightarrow 2}\,,
\label{eq:proposal-rate}
\end{equation}
where $\Gamma_{\max}^{1\leftrightarrow 2}$ bounds both $X$ decays and inverse decays, as detailed in Sec.~\ref{sec:1-to-2}.
A decay product is thus injected at its sampled time and can interact during the remainder of the macro-step, without an additional splitting between decay and scattering operators.

\resp{The division into collision cells is also adaptive.}
At each macro-step, $N_{\rm cells}$ is chosen to maintain an approximately fixed target number of importance-sampled $2\leftrightarrow2$ collision proposals per cell.
As the physical interaction rates decrease, the ensemble is repartitioned into fewer, larger cells, increasing both $V_{\rm cell}$ and the number of Monte Carlo neutrino particles per cell.
This coarsening prevents overpartitioning from driving the expected number of accepted interactions below unity in each cell, where finite Monte Carlo fluctuations could obscure residual interactions that still produce appreciable thermalization globally.
Since the cells are computational partitions of a homogeneous system, repartitioning preserves the total physical volume and the global particle and energy content.

\subsection{Local occupation numbers and \resp{factors accounting for quantum statistics}}
\label{sec:Pauli-blockings}

Each final-state fermion contributes a Pauli-blocking factor $1-f_y(E_y)$ to the physical event rate, while each final-state boson contributes a Bose-stimulation factor $1+f_y(E_y)$.
Electrons, positrons, and photons remain thermal, so their occupation numbers follow from the local plasma temperature.
Neutrinos and $X$ particles can instead have nonthermal momentum distributions.
Their occupation numbers are therefore evaluated from the momentum-resolved populations in each collision cell.
For neutrinos, this evaluation is performed in the propagation basis defined in Sec.~\ref{sec:oscillations}.

For an isotropic population of species $s$, let $N_{s,b}$ be the number of particles represented in the momentum bin $[p_{b,-},p_{b,+}]$ of a cell with physical volume $V_{\rm cell}$.
The mean occupation per internal state in that bin is
\begin{equation}
    \widehat f_{s,b}=\frac{6\pi^2 N_{s,b}}{g_sV_{\rm cell}(p_{b,+}^3-p_{b,-}^3)},
    \label{eq:cell-occupation-reconstruction}
\end{equation}
where $g_s$ counts the equally populated internal states, and a weighted Monte Carlo sample replaces $N_{s,b}$ by the sum of particle weights.
For a propagation label $i$, the code counts neutrinos and antineutrinos together; their equal populations and one helicity state each give $\widehat f_{\nu_i,b}=\widehat f_{\bar\nu_i,b}=3\pi^2N_{i,b}/[V_{\rm cell}(p_{b,+}^3-p_{b,-}^3)]$.
For $X$, the normalization uses $g_X$ polarization states.

The reference implementation uses twelve uniform bins over $0\leq p\leq3T$, with width $T/4$, followed by logarithmic bins extending to the largest neutrino momentum in the cell.
At each macro-step, the number of logarithmic bins is the nearest integer to $\max\{10,20[1+\log_{10}(N_\nu^{\rm MC}/(3\times10^5))]\}$, where $N_\nu^{\rm MC}$ is the total number of Monte Carlo neutrinos and antineutrinos across all cells.
Thus, their spacing is $\Delta\ln p=\ln[p_{\max}/(3T)]/N_{\log}$.
The resolved $X$ distribution uses the same hybrid binning and $N_{\log}$, with a momentum range extended to cover inverse-decay products as well as existing particles.
Occupations at a reaction momentum are obtained by linear interpolation between the bin-center estimates; the particle momenta themselves remain continuous.

Each accepted reaction changes particle momenta or creates or removes particles in its collision cell.
We therefore update the occupation estimates after every accepted reaction, so subsequent reactions use the current distributions when evaluating Pauli blocking and Bose enhancement.
The adaptive cell partition described above maintains sufficient local Monte Carlo statistics for these estimates.
The special zero-momentum $X$ component is treated separately in Sec.~\ref{sec:zero-mode}.

\subsection{$1\leftrightarrow 2$ interactions with $X$ particles}
\label{sec:1-to-2}

We consider two classes of $X$ particles: bosons decaying into $\nu_\alpha\bar\nu_\alpha$, $e^+e^-$, or $\gamma\gamma$, and fermions decaying into $\nu_\alpha\gamma$ and its charge-conjugate mode.
The first class includes Majorons, ALPs, dark photons, and other spin-0 or spin-1 species~\cite{Chikashige:1980ui,EscuderoAbenza:2025tsi,Ilten:2018crw}; the second includes HNLs coupled through a dipole portal~\cite{Magill:2018jla,Ovchynnikov:2023wgg}.
Other $X$ species can be implemented analogously.

In vacuum, decays of $X$ can be incorporated into \nudsmc by adding the direct $1\to2$ decay rate to the total majorant in Eq.~\eqref{eq:proposal-rate}, simulating each selected decay, and inserting its products into the corresponding DSMC populations.
A finite-temperature plasma additionally requires quantum-statistical factors, thermal masses, and inverse decays.
Thermal-bath corrections become small for $m_X\gtrsim50\,\mev$ at MeV temperatures; inverse reactions involving nonthermal neutrinos depend on their actual distributions.

First consider the quantum-statistical and thermal-mass corrections.
For a channel $X\to1+2$, the vacuum rest-frame partial width is $\Gamma_{X\to12}\equiv{\rm Br}_{12}/\tau_X$.
We denote by $\Gamma^{\rm kin}_{X\to12}(T)$ the rest-frame partial width with the adopted thermal-mass prescription, before Pauli blocking or Bose enhancement.
The decay rate per unit plasma time for a parent of energy $E_X$ is
\begin{equation}
    \Gamma^{\rm med}_{X\to12}(E_X,T)
    = \frac{m_X}{E_X}\,\Gamma^{\rm kin}_{X\to12}(T)
    \left\langle\Pi_Q\right\rangle_{\rm dec}\,,
    \label{eq:Xs-thermal-width}
\end{equation}
The factor $m_X/E_X$ accounts for time dilation.
The brackets denote an average over the decay distribution, with daughter energies and occupations evaluated in the plasma frame; Appendix~\ref{app:thermal-1-2} gives the channel-dependent widths, statistical averages, and inverse rates for bosonic and fermionic relics.
For a given set of daughter energies, the quantum-statistical factor is
\begin{multline}
\Pi_{Q}(\{y(E_{y})\}) \\ = \prod_{y \in \rm bosons}(1+f_{y}(E_{y}))\times \prod_{y \in \rm fermions}(1-f_{y}(E_{y}))\,.
\label{eq:quantum-statistics-direct-decay}
\end{multline}
This is the product of the Pauli-blocking and Bose-stimulation factors introduced in Sec.~\ref{sec:Pauli-blockings}, evaluated for the decay daughters.
Depending on the $X$ mass and decay kinematics, these factors can make the in-medium decay rate differ substantially from its vacuum value.
In \nudsmc the decay rate is evaluated from the local occupation factors and sampled stochastically; in an evolving plasma it is therefore not a fixed vacuum exponential.

Second, for sufficiently light particles with $m_X\simeq3T$, direct decay $X\to\{y\}$ at MeV temperatures is accompanied by the inverse process $\{y\}\to X$.
The $X$ population may then enter thermal equilibrium with the plasma.
For bosonic $X$, Bose stimulation enhances the inverse process by $1+f_X(E_X)$; for fermionic $X$, Pauli blocking gives the suppression factor $1-f_X(E_X)$.

\subsubsection{Describing direct and inverse decays}

A direct decay converts a relic into two daughter particles, $X\to1+2$; an inverse decay is the reverse reaction, $1+2\to X$.
Inverse decays require different sampling from non-resonant $2\leftrightarrow2$ collisions.
For a narrow $X$, the incoming energies and their relative angle must place the produced particle on shell.
Choosing that angle at random would therefore almost always miss the resonance.
After the angle is integrated analytically, the inverse-decay rate for massless incoming particles has the energy dependence
\begin{equation}
(\sigma v)_{1+2\to X}(E_1,E_2)
\propto \frac{1}{\tau_X}
\frac{\theta(4E_1E_2-m_X^2)}{E_1^2E_2^2}\,.
\end{equation}
The step function enforces $4E_1E_2\geq m_X^2$.
Because this energy dependence differs from that of weak scattering, each inverse decay is treated as a separate MCF reaction channel.
The full sampling uses the corresponding thermal-mass kinematics described in Appendix~\ref{app:thermal-1-2}.
The nonuniform kernel also leaves a characteristic transient shape.
Relative to a Fermi-Dirac spectrum with the same neutrino number and energy, the neutrino distribution develops a broad intermediate-energy shoulder bracketed by depleted low- and high-energy tails.
Energy-selective inverse removal and boosted two-body decays together generate this pattern.
The fixed-volume test in Appendix~\ref{app:thermal-1-2-checks} displays it directly.

Within each computational cell, the $X$ and Standard Model populations are evolved together.
Direct decays, inverse decays, and ordinary $2\leftrightarrow2$ reactions contribute to the common event rate in Eq.~\eqref{eq:proposal-rate}.
For a direct decay, the MCF algorithm draws a candidate $X$ and decay channel using an upper bound on the decay rate.
The factor $m_X/E_X$ accounting for time dilation enters an acceptance test.
The daughter momenta are generated in the $X$ rest frame and boosted to the plasma frame.
Their local occupations determine the statistical factor in Eq.~\eqref{eq:quantum-statistics-direct-decay}.
For fermionic daughters, this factor is at most unity and gives the Pauli acceptance probability.
For bosonic daughters, an upper bound on their stimulation factor is included in the proposal rate, and the acceptance probability contains the ratio of the actual factor to that bound.
Accepted neutrino daughters enter the momentum-resolved neutrino population, while each electromagnetic daughter transfers its energy to the thermal plasma.

For an inverse decay, incoming neutrinos are drawn from the simulated momentum distributions, while electrons, positrons, and photons are drawn from their thermal distributions.
The sampling enforces the mass threshold and includes the Bose-stimulation or Pauli-blocking factor of the produced $X$.
When oscillations are included, the interaction probability uses the energy-dependent projection of each propagation state onto the flavor coupled to $X$.
An accepted event removes the sampled incoming particles, or the corresponding energy for particles drawn from the thermal plasma, and inserts an $X$ with energy $E_X=E_1+E_2$.

After an accepted event, the affected populations and rate bounds are updated before the next candidate is drawn.
An $X$ created by an inverse decay can therefore decay during the remainder of the same interaction step, and its daughters can undergo subsequent scatterings.
Tests of vacuum decay, detailed balance, and thermal equilibrium are described in Appendix~\ref{app:thermal-1-2-checks}.

\subsubsection{Zero mode processes}
\label{sec:zero-mode}

The local reconstruction in Sec.~\ref{sec:Pauli-blockings} applies to resolved moving particles.
For the late reheating scenarios, however, a typical bosonic $X$ population carries most of the energy density of the Universe, with $\rho_X \gg \rho_{\rm plasma}$ at early times, and a large part may occupy the zero-momentum state.
The $1\leftrightarrow 2$ exchange between this population and its decay products is special.
On the one hand, the rest-frame $X$ population is a macroscopically occupied single state.
On the other hand, the exact inverse process into a rest-frame $X$ requires decay products with $E_1=E_2=m_X/2$ and opposite momenta.
This phase-space point has zero measure in the continuous Monte Carlo pool and must therefore be treated separately.

To estimate quantum-statistical factors near $p_X=0$, we assign a finite momentum volume to the zero mode.
The total physical volume $V_{\rm sys}=\sum_c V_c$ fixes the momentum resolution shared by the collision cells, whose individual volumes are $V_c$.
Writing $L=V_{\rm sys}^{1/3}$, the momentum spacing is
\begin{equation}
    p_{\rm FV} = \frac{2\pi}{L}.
\end{equation}
All cells use this momentum interval and evaluate occupations from their local volumes and populations.
Expansion changes the shared momentum scale; repartitioning at fixed $V_{\rm sys}$ changes the number of states represented in each cell.
We represent the neighborhood of zero momentum by $p_X<p_0$, where
\begin{equation}
    p_0 \equiv \frac{p_{\rm FV}}{2}.
\end{equation}
Only particles with $p_X=0$ are stored collectively, with occupation per $X$ polarization state
\begin{equation}
    N_0 = \frac{N_{X,{\rm rest}}}{g_X}\,,
    \label{eq:zero-mode-occupancy}
\end{equation}
Here, $N_{X,{\rm rest}}$ counts the at-rest particles in $V_{\rm sys}$ and $g_X$ is the number of $X$ polarizations.
The rates below describe a homogeneous population in this volume.
For cell $c$, with $N_{X,0,c}$ at-rest particles, we use $N_0=V_{\rm sys}N_{X,0,c}/(g_XV_c)$ and multiply the rates by $V_c/V_{\rm sys}$.
The occupations and temperature are evaluated in that cell.
Particles with $p_X>0$ are tracked with their individual momenta.
Inverse decays with $0<p_X<p_0$ include stimulation by the at-rest population, normalized to one momentum state, and by moving $X$ particles, normalized to the phase-space volume of the sphere $p_X<p_0$.
Production at $p_X\geq p_0$ uses the resolved momentum distribution.

For the thermal EM channels, ${X\leftrightarrow e^+e^-}$ and ${X\leftrightarrow\gamma\gamma}$, the zero-mode dynamics is local in the cell because the daughters are drawn from an equilibrium bath.
Schematically,
\begin{align}
    R_{X_0\to y\bar y}
        &=
        \Gamma_{X\to y\bar y}\,
        {\cal F}^{\rm kin}_{y}\,
        g_X N_0\,
        \left(1\mp f_y(m_X/2)\right)^2 ,
        \\
    R_{y\bar y\to X_0}
        &=
        \Gamma_{X\to y\bar y}\,
        {\cal F}^{\rm kin}_{y}\,
        g_X(1+N_0)\,
        f_y^2(m_X/2) .
\end{align}
The upper sign convention is the usual one: Pauli blocking for $e^\pm$ gives $(1-f_e)^2$, while Bose enhancement for photons gives $(1+f_\gamma)^2$.
The factor ${\cal F}^{\rm kin}_{y}=\Gamma^{\rm kin}_{X\to y\bar y}(T)/\Gamma_{X\to y\bar y}$ is the thermal-mass correction to the rest-frame partial width defined above.
These events transfer the energy $m_X$ directly between the at-rest $X$ population and the EM bath.
The cell temperature is then recalculated from its updated electromagnetic energy density.

A rest-frame $X$ decay produces two back-to-back neutrinos with the fixed energy
\begin{equation}
    E_* = \frac{m_X}{2}.
\end{equation}
This monoenergetic contribution is the decay line.
The reverse reaction requires the same energy and opposite momenta, which form a zero-measure configuration in a smooth Monte Carlo sample.
During each interaction step, $N_{{\rm line},i}$ counts the combined neutrino and antineutrino population in the decay line and propagation state $i$.
The CP-symmetric calculation assigns the same occupation to $\nu_i$ and $\bar\nu_i$; without oscillations, the index $i$ labels flavor.

The same finite-volume boundary $p_0$ defines how this delta-function contribution is represented in energy:
\begin{equation}
    E_* - p_0 < E < E_* + p_0 ,
\end{equation}
with the lower edge clipped at zero.
Its width is thus the finite-volume momentum resolution $2p_0=p_{\rm FV}$; it is unrelated to the decay width $\Gamma_X$.
In the calculations below, this interval is divided into two equal-width energy bins and the exact daughters occupy the lower, redshifting bin.
Denoting its edges by $E_{{\rm line},-}$ and $E_{{\rm line},+}$, the occupation of either neutrinos or antineutrinos is
\begin{equation}
    f_{{\rm line},i}
    =
    N_{{\rm line},i}\,
    \frac{3\pi^2}{
    V_{\rm cell}
    \left[
        E_{{\rm line},+}^3-E_{{\rm line},-}^3
    \right]} .
    \label{eq:line-occupancy}
\end{equation}
For the symmetric interval relevant here, $E_{{\rm line},-}=E_*-p_0$ and $E_{{\rm line},+}=E_*$.
The factor $3\pi^2$ accounts for the combined neutrino and antineutrino count: each species separately has mean number $N_{{\rm line},i}/2$ and one helicity state.
The prefactor in Eq.~\eqref{eq:line-occupancy} is the occupation increment $\Delta f_{\rm line}$ per particle added to the combined count.
Thus, $G_{\rm line}=1/\Delta f_{\rm line}$ is the total number of one-particle states represented by this bin for $\nu_i$ and $\bar\nu_i$ together.

The occupation entering a flavor-$\alpha$ reaction is obtained with the projectors of Sec.~\ref{sec:oscillations},
\begin{equation}
f_{{\rm line},\alpha}
=\sum_i\Pi_{\alpha i}(E_*,T)f_{{\rm line},i}.
\label{eq:line-flavor-occupation}
\end{equation}
Pauli blocking also includes individually represented neutrinos already lying in the same energy bin.
We denote the total occupation, including these particles and the decay line, by $f_{{\rm bin},i}$, and its flavor projection by $f_{{\rm bin},\alpha}$.
For exchange with the represented decay-line particles, the continuum rates are
\begin{align}
    R_{X_0\to\nu_\alpha\bar\nu_\alpha}
        &=
        \frac{{\rm Br}_\alpha}{\tau_X}\,
        g_X N_0\,
        \left(1-f_{{\rm bin},\alpha}\right)^2 ,
        \label{eq:zero-mode-direct}
        \\
    R_{\nu_\alpha\bar\nu_\alpha\to X_0}
        &=
        \frac{{\rm Br}_\alpha}{\tau_X}\,
        g_X(1+N_0)\,
        f_{{\rm line},\alpha}^2 .
        \label{eq:zero-mode-inverse}
\end{align}
The implementation sums over the propagation labels $i,j$ of the two particles with weights $\Pi_{\alpha i}\Pi_{\alpha j}$.
Within one propagation bin, the two entries are sampled without replacement from the combined CP-symmetric population.
After the first entry is selected, the available particle or vacancy count decreases by one, corresponding to an occupation increment $\Delta f_{\rm line}$.
This gives $(1-f_{{\rm bin},i})\max(1-f_{{\rm bin},i}-\Delta f_{\rm line},0)$ for decay and $f_{{\rm line},i}\max(f_{{\rm line},i}-\Delta f_{\rm line},0)$ for inverse decay, as derived in Appendix~\ref{app:line-pair-counting}.
Particles in different propagation bins use the product of the two respective occupation factors.
When the bin contains many one-particle states, these finite-count factors approach Eqs.~\eqref{eq:zero-mode-direct} and \eqref{eq:zero-mode-inverse}.
For exchange between $X$ particles at rest and an isolated decay line, $f_{{\rm bin},\alpha}=f_{{\rm line},\alpha}$, and detailed balance gives
\begin{equation}
    \frac{N_0}{1+N_0}
    =
    \left(\frac{f_{{\rm line},\alpha}}{1-f_{{\rm line},\alpha}}\right)^2 .
\end{equation}
Expansion continually redshifts neutrinos through $E_*$, opening states for new decay daughters.
When both expansion and neutrino inverse decays are present, we combine the $N_0$-enhanced forward and inverse contributions into the net transfer through this energy interval.
We estimate the incoming occupation $f_{\rm in}$ from the upper bin, before adding new daughters, and solve for the outgoing occupation $f_{\rm out}$ below $E_*$.
For an uncoupled flavor channel with $\Gamma_\alpha={\rm Br}_\alpha/\tau_X$, the partner energies $E$ and $2E_*-E$ give the statistical factor $1-f(E)-f(2E_*-E)$.
Taking the background quantities constant across the narrow interval, its steady balance with redshift gives
\begin{align}
 f_{\rm out}-f_{\rm in}
 &=\kappa\Gamma_\alpha(1-f_{\rm in}-f_{\rm out})\nonumber\\
 &=\frac{\kappa\Gamma_\alpha}{1+\kappa\Gamma_\alpha}
       (1-2f_{\rm in}),
 \label{eq:cold-boundary-current}\\
 \kappa&=\frac{2\pi^2 n_{X,0}}{H E_*^3},
 \qquad n_{X,0}=\frac{N_{X,0,c}}{V_c},
 \nonumber
\end{align}
where $H$ is the Hubble rate and $n_{X,0}$ is the local physical number density of at-rest particles.
For strong exchange, $\kappa\Gamma_\alpha\gg1$, Eq.~\ref{eq:cold-boundary-current} gives $f_{\rm out}\to1-f_{\rm in}$: an empty incoming distribution allows the outgoing states to fill up to unit occupation.
The sign of $f_{\rm out}-f_{\rm in}$ determines whether the cold population decays or grows by absorbing neutrinos.
This net transfer includes stimulation by at-rest $X$ in inverse reactions involving both the represented daughters and individually sampled neutrinos.
Spontaneous inverse reactions and stimulation by moving $X$ are sampled separately.
Without expansion we use the microscopic rates above.
With expansion present and neutrino inverse reactions disabled, we solve the decay balance across the same narrow interval with Pauli blocking for both daughters.

With oscillations, the coupling $\sum_\alpha\Gamma_\alpha\Pi_{\alpha i}\Pi_{\alpha j}$ connects the incoming and outgoing occupations of the propagation states.
When the total current is outward, states with negative individual currents lose neutrinos through conversions at $E_*$, and $X$ decays produce the net increase in pairs.
For a total inward current, we sample the finite-volume decay and inverse processes in both directions.
Each batch is limited by the available incoming particles and the number of unoccupied final neutrino states; dividing its event frequency by the batch size preserves the reaction rate.

The decay-line counter is a temporary representation of physical neutrinos in this finite-volume bin.
It enters the $2\leftrightarrow2$ collision rates, and an accepted collision can consume a decay-line neutrino and create ordinary finite-energy particles.
The inverse channel
\begin{equation}
    \nu_{\rm line}+\nu_{\rm finite}\to X_{\rm resolved},
\end{equation}
consumes one decay-line neutrino and one ordinary Monte Carlo neutrino to produce a moving, resolved $X$.
Together, these interactions transfer neutrinos out of the monoenergetic contribution and account for its collisional broadening and inverse removal.
A neutrino produced at time $t_{\rm birth}$ has comoving momentum $q=a(t_{\rm birth})E_*$.
Expansion redshifts its energy as $E(t)=q/a(t)$.
The surviving $X$ particles remain at rest.
The collective counts and individual particles describe the same physical populations at different stages of the interaction step.

Thermal inverse reactions are exponentially suppressed when $m_X\gtrsim T_{\rm cell}\ln(1/\epsilon)$, with $\epsilon=10^{-4}$ setting the thermal cutoff.
Exchange with the nonthermal decay line remains active at larger $m_X/T_{\rm cell}$ when neutrino inverse reactions and expansion are included.

\subsection{Production of pions in neutrino thermalization}
\label{sec:charged-pion}

If the particles injected by $X$ decay are sufficiently energetic, their thermalization can produce Standard Model species beyond the $\{\nu,e^\pm,\gamma\}$ sector.
At invariant masses $\sqrt{s}\lesssim4\pi f_\pi\simeq1\,\gev$, the relevant production channels are exclusive and include
\begin{align}
   &\nu_{e}+ e^{+} \to \pi^{+}\,, \quad
    \bar{\nu}_{e}+ e^{-} \to \pi^{-}\,, \label{eq:heavy-prod-processes-1}\\
   &\nu_{e}+e^{+}\to \nu_{\mu}+\mu^{+}\,, \quad
    \nu_{\alpha}+\bar{\nu}_{\alpha}\to \pi^{+}+\pi^{-}\,, \label{eq:heavy-prod-processes-2} \\
   &e^{+}+e^{-}\to \pi^{0}+\gamma\,, \quad
    e^{+}+e^{-}\to \pi^{+}+\pi^{-}\,, \label{eq:heavy-prod-processes-3}
\end{align}
together with the corresponding charge-conjugated channels.
The equilibrium populations of pions, muons, kaons, and heavier hadrons are Boltzmann-suppressed at MeV temperatures, but energetic injected particles can generate much larger non-thermal populations.

The exclusive channels~\eqref{eq:heavy-prod-processes-1}-\eqref{eq:heavy-prod-processes-3} dominate when injected particles scatter on the low-energy background.
Quark showers require substantially larger invariant masses and become relevant only for extremely energetic injection or for injected abundances large enough to make self-scattering important.
For a particle of energy $E$ colliding with a thermal particle, $\sqrt{s}\simeq\sqrt{6ET}$, where the thermal-particle energy has been estimated as $3.15T$.
At $T\simeq1\,\mev$, the condition $\sqrt{s}\gtrsim1\,\gev$ therefore requires $E\gtrsim1.7\times10^2\,\gev$.

These particles affect both the plasma evolution and the light-element abundances.
Channels with purely electromagnetic final states, such as $\nu_\alpha\bar\nu_\alpha\to\pi^0\to\gamma\gamma$, accelerate energy transfer from the neutrino sector to the electromagnetic sector.
Metastable particles such as $\mu^\pm$, $\pi^\pm$, $K^\pm$, and $K_L$ have a richer evolution.
Depending on the plasma temperature, their lifetimes may be comparable to or longer than their interaction times.
They may therefore decay, self-annihilate into lighter metastable or electromagnetic species, lose energy electromagnetically, or interact with nucleons~\cite{Akita:2024nam,Akita:2024ork}.

Charged mesons induce additional proton-neutron conversion processes, for example
\begin{equation}
    \pi^{+}+n\to p+\pi^{0}\,, \qquad
    \pi^{-}+p\to n+\pi^{0}/\gamma\,,
\end{equation}
and may also contribute to nuclear dissociation~\cite{Reno:1987qw,Kohri:2001jx,Kawasaki:2004qu,Pospelov:2010cw,Kawasaki:2017bqm,Hasegawa:2019jsa}.
Even a small charged-meson yield per decaying LLP can appreciably modify BBN dynamics~\cite{Boyarsky:2020dzc,EscuderoAbenza:2025tsi}.

For neutrinophilic $X$ decays, Ref.~\cite{Bianco:2025boy} studied cascade-induced production of metastable particles for $1\,\gev\leq\mX\leq10\,\mathrm{TeV}$ and $\tX\gtrsim10^4\,\s$.
That analysis used \textsc{Pythia8} to simulate hadron production through final-state radiation in $X$ decays and through subsequent scatterings of the neutrino daughters.
A recent study of electrophilic decays~\cite{Bianco:2026dvc} considered very heavy LLPs with $m_X>10\,\gev$ and lifetimes down to $\tX\simeq10^{-2}\s$ using similar machinery.
It found that final-state radiation from $X$ decays dominates meson production, while subsequent processes such as $\gamma+\gamma_{\rm bg}\to{\rm hadrons}$ provide a smaller contribution.

For the masses considered here, $m_X\lesssim10\,\gev$, the abundance of such secondary non-thermal particles is typically too small to appreciably affect energy exchange between the neutrino and electromagnetic sectors.
Injected neutrinos and electromagnetic particles predominantly interact with the thermal background, and the resulting invariant masses generally remain below the thresholds for processes~\eqref{eq:heavy-prod-processes-1}-\eqref{eq:heavy-prod-processes-3}.

We nevertheless track the charged-pion yield because of its impact on BBN.
We include the leading charged-pion production channel sourced by non-thermal neutrinos,
\begin{equation}
    \nu_{e}+e^{+}\to\pi^{+}\,,
    \qquad
    \bar{\nu}_{e}+e^{-}\to\pi^{-}\,.
    \label{eq:pion-producting-process}
\end{equation}
The rate is helicity-suppressed.
Among the charged-pion production processes listed above, however, this channel has the lowest invariant-mass threshold, $s=m_{\pi^\pm}^2$; pair-production channels such as $\nu_\alpha\bar\nu_\alpha\to\pi^+\pi^-$ and $e^+e^-\to\pi^+\pi^-$ require $s\geq4m_{\pi^\pm}^2$.
Consequently, this channel dominates for injected neutrino energies $E_\nu\lesssim1\,\gev$.

Motivated by Ref.~\cite{Bianco:2026dvc}, we omit secondary pion production from purely electrophilic cascades.
The dominant hadronic injection then comes from the decays
\begin{align}
&X\to (e^{+}e^{-})^{*} \to \gamma^{*}\to {\rm hadrons}\,, \label{eq:EM-induced-hadronic-decays-electron-1} \\ &X\to e^{+}e^{-}+\gamma^{*} \to e^{+}e^{-}+{\rm hadrons}\,,\label{eq:EM-induced-hadronic-decays-electron-2} \\ &X\to \gamma \gamma^{*}\to \gamma+{\rm hadrons}\,\,.
\label{eq:EM-induced-hadronic-decays-photon}
\end{align}
For moderate $X$ masses, the decays~\eqref{eq:EM-induced-hadronic-decays-electron-2} and \eqref{eq:EM-induced-hadronic-decays-photon} remain exclusive; a quark-shower description becomes appropriate only at much larger masses~\cite{EscuderoAbenza:2025tsi}.
The virtual photon favors an invariant mass near that of the final hadronic state, and intermediate vector-meson resonances further enhance this region.

We implement process~\eqref{eq:pion-producting-process} in \nudsmc as a deterministic source counter.
In each cell, the pion-production rate is evaluated using the same angle-averaged narrow-width kernel as for inverse $2\to1$ production of $X$.
The pions remain external to the DSMC particle pool, and their production rate is integrated throughout each timestep on a substep grid that is passed to the BBN calculation.
The crossed process $\nu_e e^+\to\pi^+$ fixes the normalization through the partial width $\Gamma_{\pi\to e\nu}={\rm Br}(\pi\to e\nu)/\tau_\pi$.
The total pion width is instead dominated by $\pi\to\mu\nu_\mu$.

The pion source is calculated without removing the incoming neutrino or charged lepton from the transport.
We describe it by the effective charged-pion yield per $X$ decay in a time interval around temperature $T$,
\begin{equation}
    {\rm Br}_{\rm eff}^{\pi}(T)
    \equiv
    \frac{\Delta N_{\pi^+}+\Delta N_{\pi^-}}
         {\Delta N_{X\to{\rm daughters}}}\,.
    \label{eq:brPiEff}
\end{equation}
The denominator counts all individually sampled forward decays, including decays of regenerated $X$, and adds the net outward conversion of at-rest particles in Eq.~\ref{eq:cold-boundary-current}.
The corresponding decay rate per physical volume is
\begin{equation}
    \mathcal R_X(T)\equiv
    \frac{\Delta N_{X\to{\rm daughters}}}
         {\int_{\Delta t} V(t')\,dt'}\,,
    \label{eq:accepted-decay-density}
\end{equation}
where $V(t')$ is the physical volume represented by the simulation and the integral covers the same time interval $\Delta t$ as Eq.~\eqref{eq:brPiEff}.
The total production rate of charged pions per physical volume is therefore ${\rm Br}_{\rm eff}^{\pi}(T)\mathcal R_X(T)$.
Numerically, we evaluate this product directly as $(\Delta N_{\pi^+}+\Delta N_{\pi^-})/\int_{\Delta t}V(t')\,dt'$, preserving pion production even in intervals with no $X$ decay, where the yield per decay is undefined.
For the small yields considered here, ${\rm Br}_{\rm eff}^{\pi}\ll1$, the energy transferred to pions gives a negligible correction to neutrino thermalization.

Finally, to exclude the ordinary equilibrium pion population, the counter is constructed from the positive excess of the DSMC neutrino distribution over the would-be thermal distribution at the electromagnetic temperature:
\begin{multline}
    \Gamma_{\pi}^{\rm cnt}(T)
    =
    \int dE_{\nu}\,
    \mathcal{K}_{\nu e\to\pi}(E_{\nu},T)
    \\ \times\left[
        f_{\nu_{e}}^{\rm DSMC}(E_{\nu})
        -
        f_{\nu}^{\rm FD}(E_{\nu},\resp{T})
    \right]_{+}\,,
\end{multline}
with the thermal $e^{\pm}$ distribution $f_e^{\rm FD}(E_e,\resp{T})$ included in the kernel $\mathcal{K}_{\nu e\to\pi}$:
\begin{multline}
\mathcal K_{\nu e\to\pi}
\\ =
\frac{m_\pi^3\Gamma_{\pi\to e\nu}}
{2\pi^2(m_\pi^2-m_e^2(T))}
\int_{E_e^{\rm min}}^\infty dE_e\, f_e^{\rm FD}(E_e,\resp{T})\,.
\end{multline}
The time dependence of ${\rm Br}_{\rm eff}^{\pi}\mathcal R_X$ determines the pion input to the BBN calculation in Sec.~\ref{sec:bbn-framework}.
The microscopic branching fraction of the primary LLP decay remains a separate input.

\section{Cross-checks}
\label{sec:cross-checks}

We validate the extended \nudsmc framework using two classes of benchmarks: controlled systems with simple analytic expectations and established scenarios for which independent state-of-the-art calculations agree.
The more complex comparisons in Sec.~\ref{sec:case-studies}, where existing calculations do not always agree, provide additional tests of neutrino oscillations and of the momentum-resolved evolution.

The tests are summarized below:
\begin{itemize}
\item In the Standard Cosmological Scenario, \nudsmc reproduces the established precision prediction $N_{\rm eff}\simeq3.044$ obtained with independent neutrino-decoupling methods~\cite{Froustey:2020mcq, Gariazzo:2019gyi,Akita:2020szl,Escudero:2025kej}.
It also agrees with the dedicated \nudsmc calculation of Ref.~\cite{Ihnatenko:2025kew}.

\item To check the decay probabilities and relativistic time dilation, we disable cosmic expansion, inverse decays, and all other interactions, and choose conditions in which quantum-statistical effects are negligible.
The energy $E_X$ of each surviving $X$ particle then remains constant.
For particles in a narrow interval around $E_X$, the expected number remaining at time $t$ is
\begin{equation}
N_X(E_X,t)=N_X(E_X,0)\exp\!\left[-\frac{t}{\gamma_X\tau_X}\right],
\qquad \gamma_X=\frac{E_X}{m_X},
\label{eq:decay-time-dilation-test}
\end{equation}
where $\tau_X$ is the lifetime in the rest frame.
Relativistic particles therefore decay more slowly by the Lorentz factor $\gamma_X$.
The simulated populations reproduce this exponential decrease for each tested decay mode (Appendix~\ref{app:thermal-1-2-checks}).

\item To test cosmological expansion in the presence of decaying relics, we turn off all interactions except relic decay and compare the evolution directly with the Friedmann equations.
The scale factor, neutrino energy density, and time evolution agree at the sub-percent level.

\item We test equilibration through $X\leftrightarrow\nu\bar\nu$ alone in a fixed volume, with equal branching fractions into the three neutrino flavors (Appendix~\ref{app:thermal-1-2-checks}).
Initially, the neutrinos and antineutrinos have thermal distributions with vanishing chemical potentials, and no $X$ particles are present.
Inverse decays produce $X$ particles, whose subsequent decays redistribute the neutrino momenta.
The reactions conserve energy and $N_\nu+2N_X$, where $N_\nu$ counts both neutrinos and antineutrinos.
These conservation laws, together with $\mu_X=2\mu_\nu$, determine the common equilibrium temperature and chemical potentials.
The simulated neutrino and $X$ distributions approach the predicted Fermi-Dirac and Bose-Einstein forms.

\item We separately test $X\leftrightarrow e^+e^-$ in a fixed volume, starting with a thermal electromagnetic plasma and no $X$ particles (Appendix~\ref{app:thermal-1-2-checks}).
Electromagnetic interactions maintain thermal electron, positron, and photon distributions as energy is exchanged with $X$.
Chemical equilibrium requires $\mu_X=\mu_{e^-}+\mu_{e^+}=0$, and energy conservation determines the final temperature.
The simulated $X$ distribution and plasma temperature agree with this prediction.

\item We repeat the neutrino equilibration test with the weak $2\leftrightarrow2$ reactions enabled, allowing energy exchange with the electromagnetic plasma (Appendix~\ref{app:thermal-1-2-checks}).
Cosmic expansion remains disabled.
The neutrinos, electromagnetic plasma, and $X$ particles reach a common temperature, with vanishing chemical potentials.
The distributions agree with the equilibrium prediction obtained from conservation of total energy.

\item We test the dynamical occupation-number estimates for neutrinos and $X$ particles in out-of-equilibrium calculations with all interactions enabled.
For comparison, we also calculate global occupation numbers from the full neutrino and $X$ populations at the beginning of each timestep and hold them fixed within every cell.
These global estimates reduce Monte Carlo noise, but they cease to represent the local state once interactions substantially change the cell populations.
The frozen-occupation approximation is therefore valid only under the restrictive timestep condition
\begin{equation}
    \Delta t \ll
    \min\!\left(\tau_X,\Gamma_{2\leftrightarrow2,\max}^{-1}\right)\,,
    \label{eq:timestep-limitation-frozen-occupancies}
\end{equation}
on top of the requirement $\Delta t \ll H^{-1}$.

When Eq.~\eqref{eq:timestep-limitation-frozen-occupancies} is satisfied and each cell contains $N_{\nu,\rm cell}\gtrsim10^3$ neutrinos, the dynamic and frozen prescriptions agree within the Monte Carlo noise.
\end{itemize}

\section{BBN framework}
\label{sec:bbn-framework}

To calculate light-element abundances in the presence of a relic $X$, we adapt the BBN framework of Ref.~\cite{EscuderoAbenza:2025tsi}.
The abundance evolution includes the expansion rate modified by $X$, nonthermal neutrino spectra, and conversion of nucleons by metastable hadrons.

All \nudsmc abundance calculations use the \textsc{primat} nuclear reaction rates~\cite{Pitrou:2018cgg}.
Reverse rates follow detailed balance.

\subsection{Cosmic expansion}

Relic decays modify the Hubble rate $H(T)$, the time-temperature relation $t(T)$, and the scale factor $a(T)$.
These changes affect the time available for weak and nuclear reactions, the redshift of neutrino momenta, and the dilution of baryons and relics.
When temperature is used as the evolution variable, the reaction terms are multiplied by $dt/dT$, calculated from the plasma temperature as a function of cosmic time.
The inputs describing $X$ are its mass, lifetime, initial abundance, and the multiplicities of metastable mesons.
The transport calculation determines $t(T)$, $a(T)$, the electron-neutrino and antineutrino distributions $f_{\nu_e/\bar\nu_e}(T,p)$, the relic number density $n_X(T)$, and the charged pion production rate ${\rm Br}_{\rm eff}^{\pi}(T)\mathcal R_X(T)$ used in BBN.
The two factors in this source are defined in Eq.~\ref{eq:brPiEff} and Eq.~\ref{eq:accepted-decay-density}.
Nuclear reactions and meson absorption depend on the baryon density $n_B(T)=\eta_B(T)n_\gamma(T)$.
Conservation of baryon number gives the baryon-to-photon ratio
\begin{equation}
\eta_{B}(T) = \eta_{B}(T_{\text{CMB}})\times \left( \frac{a(T_{\text{CMB}})T_{\text{CMB}}}{a(T)T}\right)^{3},
\end{equation}
where $\eta_B(T_{\rm CMB})=6.109\times10^{-10}$~\cite{Planck:2018vyg} is its value at recombination.

\subsection{Weak neutron-proton conversion}

Weak neutron-proton conversion sets the neutron fraction available for
primordial nucleosynthesis.
We calculate neutrino capture on neutrons, antineutrino capture on protons,
neutron decay, and their inverse reactions using the evolving electron-neutrino
and antineutrino spectra.
The rates include nucleon recoil and thermal motion, together with Pauli
blocking by electrons, positrons, and outgoing neutrinos.

At each time, let $f_\nu(E)$ and $f_{\bar\nu}(E)$ denote the dimensionless
occupations of $\nu_e$ and $\bar\nu_e$, with energy $E$ in the plasma frame.
After integrating over the other particles, the rates per target nucleon are
\begin{equation}
\begin{aligned}
\Gamma_{n\to p}={}&K\int_0^\infty dE\,\bigl[f_\nu\mathcal W_1
+(1-f_{\bar\nu})(\mathcal W_2+\mathcal W_3)\bigr],\\
\Gamma_{p\to n}={}&K\int_0^\infty dE\,\bigl[(1-f_\nu)\mathcal W_4
+f_{\bar\nu}(\mathcal W_5+\mathcal W_6)\bigr].
\end{aligned}
\label{eq:weak-conversion-kernels}
\end{equation}
Here, $K=(\tau_n\lambda m_e^5)^{-1}$ in natural units, with neutron lifetime
$\tau_n=878.4\,\mathrm{s}$.
The dimensionless vacuum neutron-decay integral $\lambda$ is calculated with
the physical electron mass $m_e$ and the neutron-decay matrix element and
radiative correction specified below.
The kernels $\mathcal W_i(E,T)$ depend on the electromagnetic temperature $T$.
Here, $i=1,5,3$ describes $\nu_e n\to e^-p$, $\bar\nu_e p\to e^+n$, and
$n\to pe^-\bar\nu_e$, respectively; their inverses carry indices $i=4,2,6$.
For Fermi-Dirac electrons and positrons with negligible chemical potential and
classical thermal nucleons, detailed balance gives
$\mathcal W_4=\kappa e^{-E/T}\mathcal W_1$,
$\mathcal W_2=\kappa^{-1}e^{-E/T}\mathcal W_5$, and
$\mathcal W_6=\kappa e^{E/T}\mathcal W_3$.
The equilibrium neutron-to-proton density ratio is
$\kappa(T)=m_n^2K_2(m_n/T)/[m_p^2K_2(m_p/T)]$, where $K_2$ is a modified
Bessel function.

We include nucleon recoil exactly in the phase space and tree-level
quasielastic matrix element, using the physical neutron and proton masses.
The nucleon charged weak current contains vector, axial-vector, weak-magnetism,
and induced-pseudoscalar form factors, which describe nucleon structure and
depend on the four-momentum transfer between the neutron and proton~\cite{StrumiaVissani2003}.
Crossing relates the scattering and decay processes, while averaging over
Maxwell-J\"uttner nucleon distributions includes their thermal motion.

For the electromagnetic corrections, we take nucleons to be static.
The capture kernels contain the photon-inclusive correction at
$\mathcal O(\alpha)$~\cite{FukugitaKubota2004}, evaluated at the electron or positron
energy in the proton rest frame, with the Coulomb factor for the electron channel.
Neutron decay uses the radiative function adopted in
\textsc{primat}~\cite{Pitrou:2018cgg}.
Both expressions include resummed leading logarithms.
We add the finite-temperature corrections of \textsc{primat} to the kernels,
using Born kinematics.
We incorporate the zero-momentum electron self-energy into the mass used for
phase space, thermal occupations, and the matrix element, and remove its
linearized counterpart from the perturbative contribution to avoid double
counting.
The remaining momentum-dependent correction to the electron dispersion
relation is included to first order in $\alpha$.
Radiative contributions proportional to inverse powers of the nucleon mass
lie beyond this prescription.

The weak conversion calculation includes electrons, positrons, electron
neutrinos, and individual nucleons.
It omits muon production, inelastic hadronic reactions, and neutrino
interactions with nuclei.
For comparison, the approximation of infinite nucleon mass evaluates Born
spectral integrals and multiplies each conversion direction by the
corresponding temperature-dependent thermal correction
factor~\cite{EscuderoAbenza:2025tsi}.
The two prescriptions also differ in their treatment of radiative effects.

\subsection{Conversion by metastable mesons}

Metastable mesons can convert neutrons and protons before decaying.
For $h=\pi^\pm,K^-,K_L$, \resp{the conversion rate induced by mesons} is
\begin{equation}
    \Gamma_{p\leftrightarrow n}^{h}(T)  = \sum_{j}n_{h,\text{inst}}\langle \sigma v\rangle^{h}_{\pn, j},
\end{equation}
Here, $j$ labels \resp{a \pn process induced by a meson}, and $n_{h,\rm inst}$ is the instantaneous density set by meson production, decay, and interactions~\cite{Akita:2024nam,Akita:2024ork}.
The quantity $\langle\sigma v\rangle_j^h$ is the thermally averaged conversion cross section.
At $T\gtrsim50\,\text{keV}$, charged mesons rapidly lose kinetic energy through elastic electromagnetic scattering, so this cross section is typically evaluated near threshold, $E_h\approx m_h$.

Meson evolution generally couples the plasma, nucleon, and meson sectors: meson-nucleon interactions redistribute energy, and their rates depend on the relative neutron and proton densities.
In the models considered here, however, the injected meson abundances are small enough that their feedback on the neutrino and electromagnetic sectors can be neglected.
When calculating the meson abundances, we approximate the neutron and proton target densities by $n_n=n_p=n_B/2$, where $n_B$ is the total baryon density~\cite{EscuderoAbenza:2025tsi}.

Using the yield normalization of Eq.~\eqref{eq:brPiEff} and the relic decay rate of Eq.~\eqref{eq:accepted-decay-density}, the production rate of meson species $h$ per physical volume is ${\rm Br}_{\rm eff}^{h}(T)\mathcal R_X(T)$.
If meson self-annihilation is omitted, the quasistatic estimate for its instantaneous density is
\begin{equation}
    n_{h,\rm inst}(T)\simeq
    \frac{{\rm Br}_{\rm eff}^{h}(T)\mathcal R_X(T)}
    {\tau_h^{-1}+\sum_{N=n,p}n_N
    \langle\sigma v\rangle_{h+N\to y}}\,.
    \label{eq:meson-instant-number-density}
\end{equation}
\resp{The numerical calculation of meson evolution} includes meson self-annihilation.
Its contribution is subdominant for the \resp{dilute population of secondary pions} in \setupname{IV}.
Models with sizable direct meson injection can receive an appreciable correction from this process~\cite{Akita:2024ork}.
For mesons produced directly in $X$ decay, ${\rm Br}_{\rm eff}^{h}$ is the mean multiplicity of $h$ per decay, including the branching fractions of the primary decay channels.
For the secondary pions considered here, the CP-symmetric neutrino and electron populations give equal yields of the two charges, ${\rm Br}_{\rm eff}^{\pi^+}={\rm Br}_{\rm eff}^{\pi^-}={\rm Br}_{\rm eff}^{\pi}/2$.
The numerator of Eq.~\eqref{eq:meson-instant-number-density} is evaluated from the pion counts as described in Sec.~\ref{sec:charged-pion}, including the time dependence generated by decays, inverse decays, and neutrino thermalization.

In Eq.~\eqref{eq:meson-instant-number-density}, $\tau_h$ is the meson lifetime.
The total meson-nucleon cross section $\langle\sigma v\rangle_{h+N\to N(N')+y}$ includes \pn conversion and quasi-elastic processes in which $h$ disappears into lighter particles such as $\pi^0$ or $\gamma$.
The target densities in this meson-loss rate are $n_N=n_B/2$, with $n_B=\eta_B(T)n_\gamma(T)$ calculated from baryon-number conservation and the evolving scale factor.
The resulting conversion rates per target nucleon enter the nuclear network, which multiplies them by the evolving neutron and proton abundances to obtain the conversion fluxes.

Finally, the code integrates the coupled abundances of $n,p,d,t,{}^3\text{He},{}^4\text{He},{}^7\text{Li}$, and ${}^7\text{Be}$.
The reference abundances and cosmological inputs used in the comparisons are specified in Sec.~\ref{sec:physics-setups}.

\section{Case studies}
\label{sec:case-studies}

In this section, we compare \nudsmc with several state-of-the-art approaches to neutrino evolution across complementary physical scenarios.
These scenarios reveal how weak $2\leftrightarrow2$ reactions, neutrino oscillations, the $1\leftrightarrow2$ decay and inverse-decay processes, and \resp{production of secondary particles} shape \neff, the neutrino momentum distributions, and primordial nuclear abundances.

\begin{table*}[t!]
\caption{Different approaches used in this study to trace the evolution of the plasma at MeV temperatures.
The columns identify the method, how it represents the neutrino population, its principal methodological limitations, and its role in the comparisons.
These method names are used throughout the figures and discussion.}
\label{tab:approaches}
\small
\begin{tabular}{@{}
    p{0.18\textwidth}|
    p{0.245\textwidth}|
    p{0.245\textwidth}|
    p{0.245\textwidth}@{}}
\hline
\textbf{Method} &
\textbf{Representation of the neutrino population} &
\textbf{Main limitations of the approach} &
\textbf{Why it is included}
\\
\hline
\raggedright\textbf{\textit{$\bm{\nu}$DSMC}} &
Monte Carlo particles carrying momenta and propagation eigenstate labels, or flavor labels without oscillations &
Quasi-classical treatment of neutrino evolution &
Extended and applied in this work
\\
\hline
\raggedright\textbf{\textit{Thermal}}~\cite{Escudero:2025kej,Escudero:2026thermalBSM} &
A neutrino temperature and chemical potential for each independent flavor
sector &
Momentum-dependent spectral distortions and flavor coherence are not resolved &
Tests whether thermal spectra reproduce $N_{\rm eff}$ and primordial helium and deuterium abundances
\\
\hline
\raggedright\textbf{\textit{Quasi-classical Boltzmann equations (QCBE)}}~\cite{Akita:2024nam} &
Diagonal neutrino distribution functions $f_{\nu_\alpha}(p)$ on a momentum grid &
Computational cost and quasi-classical treatment of neutrino evolution &
State-of-the-art momentum-resolved approach used for studies of unstable relics
\\
\hline
\raggedright\textbf{\textit{Quantum kinetic equations (QKE)}}~\cite{Barbieri:2025moq,Akita:2020szl} &
Neutrino density matrix $\rho_{\alpha\beta}(p)$ on a momentum grid &
Restricted reaction scope and high computational cost &
Maximally robust treatment of neutrino oscillations
\\
\hline
\end{tabular}
\end{table*}

\subsection{Methods we consider}
The comparisons have two roles.
The established case of late electromagnetic reheating compares \nudsmc with existing calculations of neutrino decoupling and flavor conversion.
The studies of neutrinophilic decays test whether the neutrino number and energy densities suffice to predict $N_{\rm eff}$ and the primordial abundances, or whether the full momentum distributions are required.
Additional comparisons isolate the effects of inverse decays and secondary charged pions.

The methods compared in the case studies represent the neutrino state at different levels of detail, as summarized in Tab.~\ref{tab:approaches}.

\thermalmethod is the momentum-averaged Boltzmann calculation introduced \resp{for applications beyond the Standard Model} in Refs.~\cite{Escudero:2018mvt,EscuderoAbenza:2020cmq} and recently refined in Refs.~\cite{Escudero:2025kej,Escudero:2026thermalBSM}.
It evolves the number and energy densities of each independent neutrino sector and reconstructs a Fermi-Dirac distribution with an effective temperature and chemical potential.
The common-$T_\nu$ \thermalmethod calculation uses one pair $(T_\nu,\mu_\nu)$ for all three flavors as an approximation to rapid flavor equilibration.
Its calculation without oscillations evolves separate pairs for $\nu_e$ and for a common $\nu_x$ sector representing the degenerate $\nu_\mu$ and $\nu_\tau$ distributions.
The adopted \thermalmethod implementation already incorporates both directions of $X\leftrightarrow\nu_\alpha\bar\nu_\alpha$.
For this comparison, $X$ is initially at rest.
The implementation tracks this component separately from \resp{the population of $X$ particles with nonzero momentum} regenerated by inverse decays.
For the initially at-rest component, it includes the Fermi-Dirac and Bose-Einstein occupation factors in the decay and inverse-decay rates.
For the regenerated component, it instead uses momentum-integrated Maxwell-Boltzmann rates.
In both cases, the source terms transfer equal and opposite particle number and energy between the $X$ and neutrino sectors.
For the abundance comparisons, the reconstructed neutrino spectra and calculated expansion rate enter the BBN evolution with the finite-mass weak rates of Sec.~\ref{sec:bbn-framework}.

\qcbemethod evolves the diagonal distributions $f_{\nu_e}(p)$, $f_{\nu_\mu}(p)$, and $f_{\nu_\tau}(p)$ with the weak $2\leftrightarrow2$ collision terms of Refs.~\cite{Akita:2024nam,Akita:2024ork}.
It resolves the momentum dependence and carries no off-diagonal density matrix.
\resp{In the comparison with direct decays}, it uses the same prescription as the other calculations, including Pauli blocking and excluding inverse decays, and is shown without oscillations.
The case studies compare its $N_{\rm eff}$ prediction and its electron-neutrino and antineutrino distribution functions, which enter the weak $p\leftrightarrow n$ conversion rates.
For the performance benchmarks in Sec.~\ref{sec:performance-limitations}, we also use the \qcbemethod spectra and expansion rate in the BBN calculation of Sec.~\ref{sec:bbn-framework} to determine the numerical accuracy required for $Y_P$ and D/H.

The \qkemethod approach solves quantum kinetic equations for the momentum-dependent neutrino density matrices.
Their diagonal entries describe the distributions of each flavor, while the off-diagonal entries describe flavor coherence.
This approach evolves oscillations and collisions together and provides an independent check of the approximate treatment of oscillations in \nudsmc.
We use two implementations for the physical scenarios summarized in Tab.~\ref{tab:physics-setups}.

For \setupname{I}, we use the late electromagnetic reheating results of Ref.~\cite{Barbieri:2025moq}, calculated with FortEPiaNO~\cite{deSalas:2016ztq,Gariazzo:2019gyi}.
The comparison is sensitive to the treatment of neutrino-neutrino collisions as well as to flavor conversion.
Appendix~\ref{app:setup-I-qke-comparison} quantifies this sensitivity by varying the off-diagonal collision damping in the public FortEPiaNO configuration.

For \setupname{II}, we use an independent code for neutrinophilic decays, developed here by extending the calculations of Refs.~\cite{Akita:2020szl,Akita:2022hlx}.
This implementation includes the full neutrino-neutrino and neutrino-electron collision terms at tree level.
We implement the $\phi\leftrightarrow\nu\bar{\nu}$ collision term at the density matrix level by analogy with $e^-e^+\leftrightarrow\nu\bar{\nu}$.\footnote{\resp{The collision term for neutrinophilic decays at the density matrix level may depend on the detailed processes, for example, $\phi\leftrightarrow\nu\bar{\nu}$ or $\phi\leftrightarrow\nu\nu$ (as for Majorons).
However, for some parameters, we have checked that the off-diagonal components in this collision term have only a minor effect on $N_{\rm eff}$ of $10^{-3}$, which is under the numerical precision, by dropping off these components.}}
With neutrino oscillations disabled, this implementation gives the same results as \qcbemethod.

No reference calculation includes momentum-dependent distortions, flavor conversion, inverse decays, quantum-statistical factors in the direct decays, and \resp{production of secondary pions} over the complete range considered here.
Each scenario therefore compares the calculations that address its physical question.

\subsection{Physical scenarios}
\label{sec:physics-setups}
BBN analyses commonly constrain the baryon density and effective number of neutrino species using primordial helium and deuterium measurements, often together with CMB data~\cite{Fields:2019pfx,Yeh:2020mgl}.
Interpreting these constraints as a change in radiation density with otherwise standard neutrino evolution does not generally describe energy injection during decoupling.
Ref.~\cite{Ganguly:2025mdi} illustrates this dependence on the physical origin of $\Delta N_{\rm eff}$ for electromagnetic entropy injection, including the associated changes in neutrino temperatures and weak conversion rates.
The scenarios below test how the evolving neutrino momentum distributions, energy exchange with the plasma, and production of secondary particles determine $N_{\rm eff}$, $Y_P$, and D/H in electromagnetic and neutrinophilic decays.

Tab.~\ref{tab:physics-setups} summarizes the physical content and purpose of the four scenarios.
Their parameters and calculation choices are defined below.

\begin{table*}[t!]
\caption{Compact summary of the four cosmological scenarios used to compare \nudsmc with other approaches.}
\label{tab:physics-setups}
\small
\begin{tabular}{@{}
    p{0.17\textwidth}|
    p{0.47\textwidth}|
    p{0.30\textwidth}@{}}
\hline
\textbf{Scenario} &
\textbf{Qualitative description} &
\textbf{Purpose}
\\
\hline
\textbf{\setupname{I}}\newline
{\footnotesize\itshape EM late reheating} &
A scalar reheats the electromagnetic plasma through $X\to e^+e^-$ decays.
Neutrinos are heated only through weak interactions and remain close to a
thermal spectrum. &
Benchmark neutrino thermalization, energy transfer, and weak freeze-out when
reheating overlaps neutrino decoupling; provide an independent \qkemethod
cross-check of the \nudsmc oscillation treatment.
\\
\hline
\textbf{\setupname{II}}\newline
{\footnotesize\itshape Neutrinophilic decays} &
An at-rest scalar injects neutrino pairs with equal branching fractions into
the three flavors.  The main scan varies its mass for the two lifetime choices
$\tau_X=0.1\,\s$ and $1\,\s$; a separate fixed-mass scan varies the abundance, and inverse decays
are omitted. &
Test whether the \thermalmethod two-moment Fermi-Dirac description reproduces
momentum-resolved transport; \qcbemethod independently checks the
mass scan with neutrino oscillations disabled.
\\
\hline
\textbf{\setupname{III}}\newline
{\footnotesize\itshape Through neutrino decoupling} &
An at-rest neutrinophilic scalar decays into neutrinos over the broad ranges
$20\leq m_X/\mev\leq500$ and $0.01\leq\tau_X/\s\leq20$, with inverse decays and
quantum-statistical factors included; pion production is absent. &
Test the consistency of \thermalmethod with momentum-resolved \nudsmc across
lifetimes and determine its dependence on $m_X$.
\\
\hline
\textbf{\setupname{IV}}\newline
{\footnotesize\itshape Inverse decays and pions} &
Across $6\,\mev\leq m_X\leq10^4\,\mev$, matched \nudsmc calculations including neutrino oscillations isolate inverse decays in the low-mass region
and the passive charged-pion source in the high-mass region. &
Show the importance of these additions relative to the direct-decay neutrino
prediction and identify the low- and high-mass domains in which they matter.
\\
\hline
\end{tabular}
\end{table*}

\begin{itemize}
\item[--] \textbf{\setupname{I}}: \resp{an established calculation of electromagnetic reheating} following Ref.~\cite{Barbieri:2025moq}.
A scalar with $m_X=100\,\mev$ decays into $e^+e^-$, and the reheating temperature specifies the decay time through $\Gamma_X=3H(T_{\rm reh})$.
The scalar injects energy into the electromagnetic plasma.
Neutrinos receive energy only through electron-positron annihilation and neutrino-electron scattering with the plasma, so their spectra are expected to remain close to a thermal shape.
Comparing \nudsmc with the \thermalmethod, \qcbemethod, and \qkemethod calculations tests this description of neutrino thermalization and its consequences for flavor conversion, energy transfer, and weak freeze-out when the decay overlaps neutrino decoupling.

\item[--] \textbf{\setupname{II}}: decays of a neutrinophilic scalar $X\to\nu\bar\nu$ at rest, with equal branching fractions into the three flavors, $3\,\mev\leq m_X\leq200\,\mev$, and $\tau_X=0.1\,\s$ or $1\,\s$.
Since $m_X$ fixes \resp{the energy of the daughter neutrinos} and $\tau_X$ fixes the injection epoch, we fix the initial abundance at $T=\Tini=20\,\mev$ so that the resulting $|\Delta N_{\rm eff}|\lesssim0.2$ throughout the scan.
This is an observationally relevant range probed by recent CMB analyses~\cite{AtacamaCosmologyTelescope:2025nti,SPT-3G:2025bzu}.
Specifically, we choose
    \begin{equation}
    \begin{gathered}
    m_XY_X(20\,\mev)=A_\tau
    \dfrac{x}{\left(1+x^4\right)^{1/4}},\\
    A_{0.1\,\mathrm{s}}=0.075\,\mev,
    \qquad A_{1\,\mathrm{s}}=0.025\,\mev .
    \end{gathered}
    \label{eq:setup-II-abundance}
    \end{equation}
where $x\equiv m_X/(10\,\mev)$.
Here and below, we do not specify \resp{the underlying mechanism of relic production}, as it is irrelevant for this study.
For presentation, the factor $x/(1+x^4)^{1/4}$ smoothly connects constant $Y_X$ at low mass to constant injected rest energy per entropy, $m_XY_X$, at high mass, so the calculated observables vary smoothly with mass without a piecewise change in normalization.
Inverse decays are omitted, so the daughter spectrum is processed only by the weak $2\leftrightarrow2$ reactions.
A complementary fixed-mass abundance comparison takes $\tau_X=0.1\,\s$ and $m_X=100$ or $200\,\mev$ while varying the initial abundance $Y_X(20\,\mev)$.
The comparison with the \thermalmethod and \qcbemethod calculations asks whether two moments reproduce the momentum-dependent thermalization of an increasingly hard injected neutrino population.

\item[--] \textbf{\setupname{III}}: a systematic study of an at-rest neutrinophilic particle with $20\leq m_X/\mev\leq500$ and $0.01\leq\tau_X/\s\leq20$, with equal branching fractions into the three neutrino flavors.
Decays, inverse decays, and their quantum-statistical factors are included.
The abundance is
    \begin{equation}
      \begin{aligned}
      Y_X(20\,\mev)&=Y_{\rm ref}\frac{20\,\mev}{m_X}
      \sqrt{\frac{t_*+0.1\,\s}{t_*+\tau_X}},\\
      Y_{\rm ref}&=0.0035,\qquad t_*=0.001845\,\s .
      \end{aligned}
      \label{eq:setup-III-abundance}
    \end{equation}
Here, $t_*$ is the approximate cosmic age at the initial temperature $20\,\mev$, obtained from the radiation-dominated relation $t\simeq0.738\,\s\,(\mev/T)^2$ with $g_*=10.75$.
We define $T_d/\mev=\sqrt{0.738\,\s/(t_*+\tau_X)}$ as the corresponding temperature one vacuum lifetime after initialization.
This reference estimate sets the abundance prescription; the plasma temperature and relic population evolve in the transport calculation.
When $m_X<3T_d$, the right-hand side is capped at $Y_{\rm ref}$ to avoid extrapolation into the relativistic regime.
Pion production is absent.
The mass sets the daughter energy, and the lifetime fixes the vacuum decay rate.
Efficient inverse decays can replenish the $X$ population and thereby change \resp{the rate and duration of neutrino injection}.

The purpose of this scenario is to test the consistency between \thermalmethod and the momentum-resolved \nudsmc calculation across a broad range of lifetimes, and to establish how that consistency depends on $m_X$.

\item[--] \textbf{\setupname{IV}}: neutrinophilic decays of a particle spanning the broad mass range $6\,\mev\leq m_X\leq10^4\,\mev$.
At low mass, otherwise identical \nudsmc transports with neutrino oscillations enabled are compared with inverse decays omitted or included.
At high mass, where inverse decays are negligible, the same time-dependent neutrino distributions and expansion rate enter two BBN calculations, with the charged-pion source omitted or included.

\end{itemize}

We define the shifts used throughout the comparison figures by
\begin{subequations}\label{eq:comparison-deltas}
\begin{align}
 \dneff&\equiv \neff-\neffsbbn,\label{eq:delta-neff}\\
 \dyp&\equiv \yp-\ypsbbn,\label{eq:delta-yp}\\
 \ddhfive&\equiv10^5\!\left[({\rm D/H})-\dhsbbn\right].
 \label{eq:delta-dh}
\end{align}
\end{subequations}
The comparison figures use $\ypsbbn=0.2468$, $\dhsbbn=2.46\times10^{-5}$, and $\neffsbbn=3.044$, except for the helium reference used in the \setupname{I} \qkemethod comparison, defined below.
The \nudsmc and \thermalmethod abundance calculations use $\eta_B=6.109\times10^{-10}$ and the neutron lifetime $\tau_n=878.4\,\s$.
The baryon-to-photon ratio corresponds approximately to $\Omega_bh^2=0.02230$.
Both calculations include all six proton-neutron conversion channels.

For comparison with light-element observations, we use $Y_P^{\rm obs}=0.2458\pm0.0013$~\cite{Yeh:2026lbt} and $10^5({\rm D/H})^{\rm obs}=2.55\pm0.03$~\cite{ParticleDataGroup:2024cfk}.
The quoted observational uncertainties are at $1\sigma$.
For D/H, we also include \resp{a $1\sigma$ uncertainty from nuclear reaction rates} of $0.07$ and combine it in quadrature with the observational uncertainty~\cite{Escudero:2026mgw}.
Relative to the SBBN baselines above, the resulting $2\sigma$ intervals are
\begin{align}
 -0.0036&\leq \dyp \leq 0.0016,
 \label{eq:observational-yp}\\
 -0.0623&\leq \ddhfive \leq 0.2423.
 \label{eq:observational-dh}
\end{align}

\subsection{\setupname{I}: EM late reheating}
\label{sec:setup-I-results}

\begin{figure*}[t!]
\centering
\includegraphics[width=0.495\textwidth]{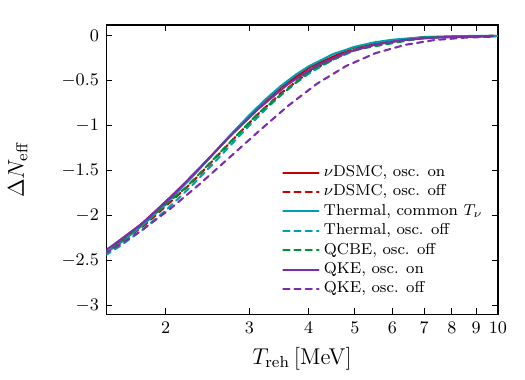}\hfill
\includegraphics[width=0.495\textwidth]{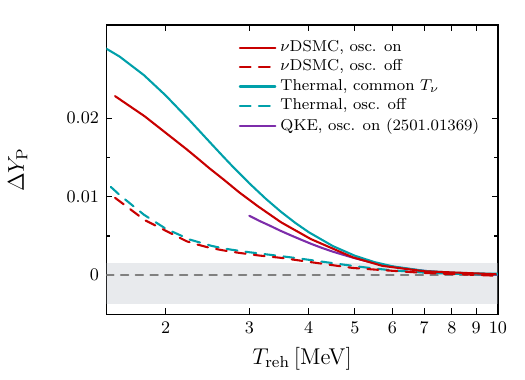}

\vspace{0.25em}
\begin{minipage}[t]{0.495\textwidth}
\vspace{0pt}
\centering
\includegraphics[width=\linewidth]{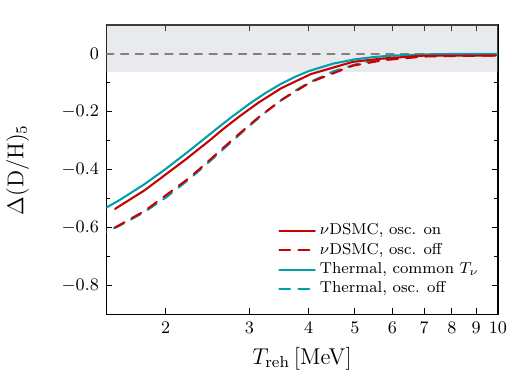}
\end{minipage}\hfill%
\begin{minipage}[t]{0.495\textwidth}
\vspace{0pt}
\centering
{\small $T_\gamma\simeq1\,\mev,\quad T_{\rm reh}=2.65\,\mev$}\par\vspace{0.15em}
\includegraphics[width=0.5\linewidth]{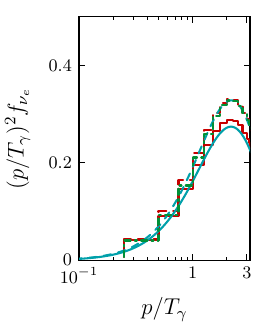}%
\includegraphics[width=0.5\linewidth]{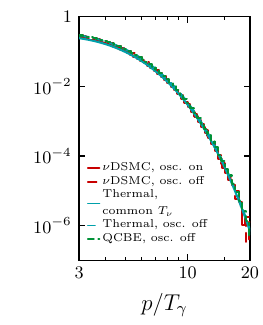}
\end{minipage}
\caption{Late electromagnetic reheating in \setupname{I} (Tab.~\ref{tab:physics-setups}).
The upper panels show \dneff and \dyp; the lower-left panel shows \ddhfive, and the lower-right panel shows the electron-neutrino spectrum around its peak and through its high-energy tail.
Here and below, gray shading marks the $2\sigma$ observational intervals for primordial helium and deuterium defined in Eqs.~\eqref{eq:observational-yp} and \eqref{eq:observational-dh}; the D/H interval includes the nuclear-rate uncertainty. Logarithmic panels show only the positive part of these intervals.
Red, dark-cyan, green, and violet denote, respectively, \nudsmc, \thermalmethod, \qcbemethod, and \qkemethod from Tab.~\ref{tab:approaches}.
Both \nudsmc and \thermalmethod calculate the abundances with the common finite-mass weak-rate prescription of Sec.~\ref{sec:bbn-framework}.
The \qkemethod results use the BBN prescription of Ref.~\cite{Barbieri:2025moq}, with the helium reference subtraction defined in the text.
Solid red and violet curves show calculations with neutrino oscillations enabled; dashed curves show the corresponding calculations with them disabled.
Solid dark-cyan curves show common-$T_\nu$ \thermalmethod, while dashed dark-cyan curves show \thermalmethod with neutrino oscillations disabled.
\qcbemethod is also shown with neutrino oscillations disabled.
For FortEPiaNO, differences in the collision prescription may also contribute to the separation of the two curves (Appendix~\ref{app:setup-I-qke-comparison}).}
\label{fig:Setup-I}
\end{figure*}

When electromagnetic reheating overlaps neutrino decoupling, incomplete energy transfer to neutrinos lowers the final ratio $\rho_\nu/\rho_\gamma$ and hence $N_{\rm eff}$.
In \setupname{I}, we compare the resulting radiation density, abundances, and electron-neutrino spectrum across the four methods (Fig.~\ref{fig:Setup-I}).

Calculations without neutrino oscillations isolate collision transport from flavor conversion: the separate-flavor \thermalmethod calculation, the momentum-resolved \qcbemethod calculation, and \nudsmc.
At high reheating temperatures, the decay completes before neutrino decoupling, and these calculations approach the standard result.
When reheating overlaps decoupling, weak interactions can no longer transfer all of the injected electromagnetic energy to neutrinos.
At each reheating temperature evaluated by both \nudsmc and \qcbemethod, the two $N_{\rm eff}$ predictions differ by at most $5.5\times10^{-3}$.
The separate-flavor, two-moment \thermalmethod calculation differs from the corresponding \nudsmc result by at most $0.038$.

The agreement of \nudsmc with \thermalmethod and \qcbemethod extends beyond the integrated quantity $N_{\rm eff}$ to the electron-neutrino energy distribution shown in the paired lower-right views of Fig.~\ref{fig:Setup-I}.
At $T_\gamma\simeq1\,\mev$ and $T_{\rm RH}=2.65\,\mev$, the dashed red \nudsmc curve with neutrino oscillations disabled and the green \qcbemethod curve reproduce the same nonthermal electron-neutrino spectrum at approximately the percent level over the momentum range that dominates the neutrino energy density.
Their energy-weighted moments differ by only about $0.15\%$.
The remaining visible difference is concentrated in the four coarse bins at $p/T_\gamma<1$, which together contain about $1.4\%$ of the plotted electron-neutrino energy and $6.3\%$ of its particle number.

The \nudsmc result with neutrino oscillations enabled can be compared with \resp{the common-$T_\nu$ \thermalmethod calculation, which assumes equilibrated flavors and evolves their number and energy densities}.
The two results differ by at most $0.034$ in $N_{\rm eff}$.
Electron neutrinos interact with the electron-positron component of the electromagnetic plasma through both charged- and neutral-current reactions.
Muon and tau neutrinos interact with it through neutral-current reactions.
Oscillations redistribute the more efficiently produced electron neutrinos among all three flavors, leaving additional phase space for electron-neutrino production and thereby increasing the total energy transferred to the neutrino sector.
In \nudsmc, the resulting increase in $N_{\rm eff}$ reaches $0.077$ near $T_{\rm RH}=2.86\,\mev$.
Together with the comparison above with neutrino oscillations disabled, this agreement shows that a two-moment, thermal-like description is sufficiently accurate for $N_{\rm eff}$ in \resp{this scenario with electromagnetic reheating} despite the spectral distortions produced during reheating.

The momentum-dependent \qkemethod provides a further comparison.
With neutrino oscillations enabled, \nudsmc agrees well with \qkemethod, although their difference reaches $3.3\times10^{-2}$ at the most discrepant point, $T_{\rm RH}=4.95\,\mev$.
This separation cannot be explained by the different numerical constants used in the two calculations: replacing the \nudsmc oscillation parameters by the public FortEPiaNO values or varying the Weinberg angle $\sin^2\theta_W$ changes $N_{\rm eff}$ only at the $10^{-3}$ level.
Other tested convention changes are likewise too small.
In the public low-reheating FortEPiaNO configuration, the collision terms governing the off-diagonal entries of the neutrino density matrix in the flavor basis are approximated by fitted damping rates.
This configuration omits neutrino-neutrino interactions both from these damping rates and from the collision integrals for the diagonal neutrino occupation numbers.
Restoring the fitted off-diagonal $\nu\nu$ damping with all other inputs fixed shifts $N_{\rm eff}$ by $0.047$, making the collision prescription a possible source of the observed separation.
Appendix~\ref{app:setup-I-qke-comparison} gives the detailed comparison.

The FortEPiaNO result without oscillations lies below both \nudsmc and \qcbemethod, differing from \nudsmc by as much as $0.21$ near $T_{\rm RH}\simeq3.5\,\mev$.
This is unexpected because, for the same collision operator and inputs, removing flavor coherences from the \qkemethod should leave the Boltzmann-like evolution of the diagonal momentum distributions.
By contrast, an independent momentum-dependent QKE calculation including neutrino self-interactions finds a much smaller effect from switching off oscillations: the resulting $N_{\rm eff}(T_{\rm RH})$ curve remains close to the curve with neutrino oscillations enabled, as in \nudsmc~\cite{Hasegawa:2019jsa}.
Differences in the reheating convention and flavor treatment nevertheless preclude a point-by-point comparison.

At large $T_{\rm RH}$, reheating completes before neutrino decoupling and the neutrino spectra remain close to thermal.
The helium abundances then approach their respective SBBN limits.
As $T_{\rm RH}$ is lowered through the decoupling epoch, the energy dependence of the neutrino-electromagnetic interaction rates causes different momentum modes to depart from equilibrium by different amounts.
The resulting nonthermal shape is visible in the paired lower-right views at $T_\gamma\simeq1\,\mev$ and $\Treh=2.65\,\mev$.
The \thermalmethod approach reconstructs the distributions from their number and energy densities and cannot describe these spectral distortions.

Oscillations have a much larger relative effect on $Y_P$ than on $N_{\rm eff}$.
The latter depends on the final energy density summed over all flavors, whereas the weak proton-neutron conversion rates probe the time-dependent $\nu_e$ and $\bar\nu_e$ spectra directly.
The energy density is weighted by $\int dE_\nu\,E_\nu^3f_\nu$.
For $m_e,\Delta m_{np}\ll E_\nu\ll m_N$, where $\Delta m_{np}=m_n-m_p$ and $m_N=(m_n+m_p)/2$, the incoming-neutrino contribution to charged-current capture scales as $\int dE_\nu\,E_\nu^4f_\nu$ at leading Born order.
Recoil and the nucleon form factors modify this weighting at larger energies.
The high-energy tail can therefore change $Y_P$ substantially while producing a more modest change in $N_{\rm eff}$.
In electromagnetic reheating, oscillations reshape the electron-flavor spectra over the momentum range relevant to the weak rates, as the solid and dashed red curves in the lower-right panel illustrate.

With neutrino oscillations enabled, the two momentum-resolved calculations are \nudsmc and the \qkemethod implementation in FortEPiaNO.
The \qkemethod abundance calculation~\cite{Barbieri:2025moq} uses different weak and nuclear rates and cosmological inputs.
We subtract each calculation's SBBN reference: $0.2468$ for \nudsmc and $0.2481$ for FortEPiaNO, whose reference is evaluated at $T_{\rm RH}=20\,\mev$.
The resulting helium shifts differ by up to about $2\times10^{-3}$; this comparison includes both transport and BBN differences.

Finally, the lower-left panel shows the deuterium shift \ddhfive.
For a nearly thermal perturbation, the textbook correlation, expressed both as a fractional abundance shift and using the definition in Eq.~\eqref{eq:delta-dh}, reads
\begin{equation}
 \begin{aligned}
 \frac{\Delta({\rm D/H})}{({\rm D/H})_{\rm SBBN}}
 &\simeq 0.11\,\dneff\,,\\
 \ddhfive&\simeq0.27\,\dneff\,.
 \end{aligned}
 \label{eq:dh-neff-correlation}
\end{equation}
Strongly depleted spectra require evolving the weak rates and nuclear abundances directly.

\subsection{\setupname{II}: neutrinophilic decays}
\label{sec:setup-II-results}
\setcounter{dbltopnumber}{1}
\renewcommand{\dblfloatpagefraction}{0.99}

The mass scan defined by Eq.~\eqref{eq:setup-II-abundance} is organized by lifetime in Figs.~\ref{fig:setup-II-tau0p1} and \ref{fig:setup-II-tau1}.
Each figure shows \dneff, \dyp, \ddhfive, and a representative electron-neutrino spectrum.
For both lifetimes, we compare \nudsmc calculations with neutrino oscillations enabled or disabled against common-$T_\nu$ \thermalmethod and \thermalmethod with neutrino oscillations disabled.
For \dneff, we also show \qcbemethod with neutrino oscillations disabled\resp{, together with the \qkemethod results}.

At low mass, the initial number abundance approaches a constant; at high mass, the injected rest energy per entropy approaches a constant.
Dividing Eq.~\eqref{eq:setup-II-abundance} by $m_X$ gives $Y_X(20\,\mev)=A_\tau(10\,\mev)^{-1}(1+x^4)^{-1/4}$.
For $x\ll1$, $Y_X$ approaches a constant and $m_XY_X\simeq A_\tau x$ decreases linearly with $m_X$.
Since $X$ is initially at rest, $\rho_X(20\,\mev)=s(20\,\mev)m_XY_X(20\,\mev)$, so the energy released by the decays decreases in the same way.
Thus, $m_XY_X$ fixes the total injected energy per entropy, while the daughter energy $E_\nu=m_X/2$ fixes its spectral hardness.
For $x\gg1$, the injected rest energy per entropy approaches the constant $m_XY_X\simeq A_\tau$.
The subsequent decrease of \dneff is dynamical: increasing $m_X$ raises the daughter energy, strengthens \resp{the transfer of energy from neutrinos to the electromagnetic plasma}, and leaves a smaller fraction of the injected energy in the neutrino sector.

\subsubsection{Fixed abundance case: $\tau_X=0.1\,\s$}

\begin{figure*}[t!]
\centering
  \includegraphics[width=0.495\textwidth]{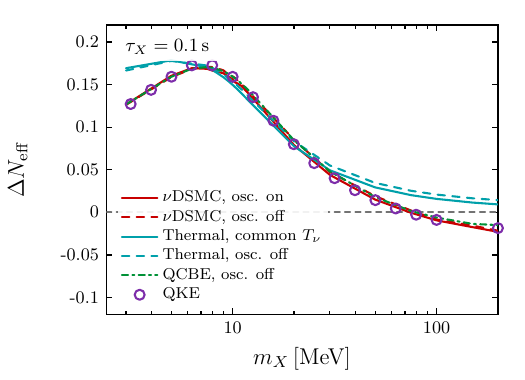}\hfill%
\includegraphics[width=0.495\textwidth]{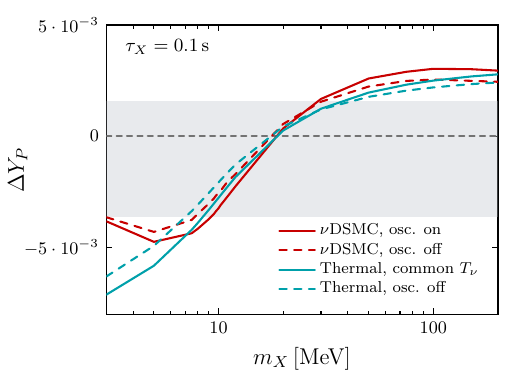}

\vspace{0.25em}
\begin{minipage}[t]{0.495\textwidth}
\vspace{0pt}
\centering
\includegraphics[width=\linewidth]{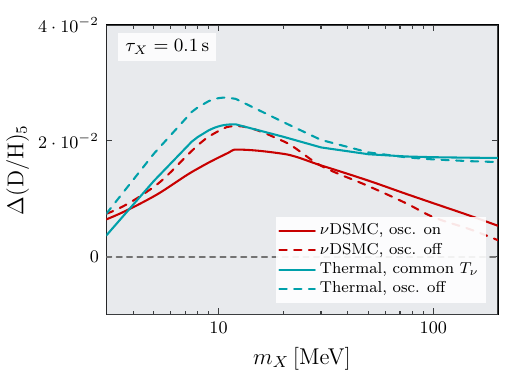}
  \end{minipage}\hfill%
\begin{minipage}[t]{0.495\textwidth}
\vspace{0pt}
\centering
{\small $m_X=100\,\mev,\quad \tau_X=0.1\,\s,\quad T_\gamma=1\,\mev$}\par\vspace{0.15em}
    \includegraphics[width=0.5\linewidth]{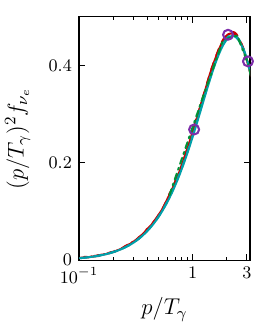}%
\includegraphics[width=0.5\linewidth]{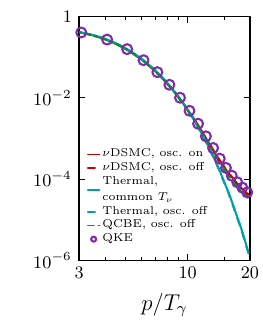}
\end{minipage}
\caption{\setupname{II} mass scan at $\tau_X=0.1\,\s$ defined by Eq.~\eqref{eq:setup-II-abundance}.
The upper panels show \dneff (left) and \dyp (right), while the lower-left panel shows \ddhfive.
The lower-right panel shows the electron-neutrino spectrum around its peak and through its high-energy tail; the vertical coordinate is $(p/T_\gamma)^2f_{\nu_e}$.
Solid and dashed red curves denote \nudsmc with neutrino oscillations enabled and disabled, respectively; the corresponding dark-cyan curves denote common-$T_\nu$ \thermalmethod and \thermalmethod with neutrino oscillations disabled.
The green curve is \qcbemethod with neutrino oscillations disabled.
\resp{Violet open circles in the upper-left and lower-right panels show the \qkemethod results.}}
\label{fig:setup-II-tau0p1}
\end{figure*}

In the upper-left panel of Fig.~\ref{fig:setup-II-tau0p1}, the \nudsmc calculation with neutrino oscillations disabled and the \qcbemethod curve agree closely throughout the scan and cross $\Delta N_{\rm eff}=0$ together on the high-mass branch.
\resp{The \qkemethod results agree very well with \nudsmc and \qcbemethod, reproducing the turnover and the negative \dneff at high mass.}

At the lowest masses, \thermalmethod predicts a larger \dneff than \nudsmc and \qcbemethod.
This ordering can be understood from momentum-dependent Pauli blocking.
The injected line $E_\nu=m_X/2$ lies in the highly occupied low-momentum part of the neutrino bath, where the decays create a localized excess.
In \nudsmc and \qcbemethod, this excess suppresses $e^+e^-\to\nu\bar\nu$ through the final-state factors $(1-f_\nu)(1-f_{\bar\nu})$ and enhances the reverse $\nu\bar\nu\to e^+e^-$ reaction through the larger initial-state occupations.
Relative to the smooth thermal closure, this localized distortion shifts the pair energy exchange toward the electromagnetic plasma.
The \thermalmethod closure evolves the number and energy moments and reconstructs a smooth Fermi-Dirac spectrum, so it cannot represent this localized blocking and yields less photon heating.
This Pauli-blocking contribution weakens as the injected line moves beyond the highly occupied part of the spectrum.

Enabling neutrino oscillations in \nudsmc moves the zero crossing to a slightly smaller mass.
Neutrino flavor conversion continually repopulates the more strongly coupled electron flavor while neutrino-neutrino scattering redistributes the injected population within the neutrino sector.
The resulting increase in energy exchange with the electromagnetic plasma lowers $N_{\rm eff}$.
At this short lifetime, electromagnetic interactions keep electrons, positrons, and photons in thermal equilibrium.
Hard injected neutrinos then transfer enough energy to this bath for photon heating to outweigh the surviving excess neutrino energy and drive \dneff below zero~\cite{Boyarsky:2021yoh,Ovchynnikov:2024rfu,Ovchynnikov:2024xyd, Akita:2024nam,Akita:2024ork}.
\thermalmethod follows the same downward trend and remains at $\Delta N_{\rm eff}>0$ across the displayed range.

The \qcbemethod agreement also holds for the spectrum in the lower-right panel.
At $m_X=100\,\mev$ and $T_\gamma=1\,\mev$, \nudsmc and \qcbemethod, both with neutrino oscillations disabled, give nearly identical electron-neutrino distributions.
Their agreement therefore extends from the integrated $N_{\rm eff}$ to the momentum distribution entering the collision integrals.

\subsubsection{Fixed abundance case: $\tau_X=1\,\s$}

\begin{figure*}[t!]
\centering
  \includegraphics[width=0.495\textwidth]{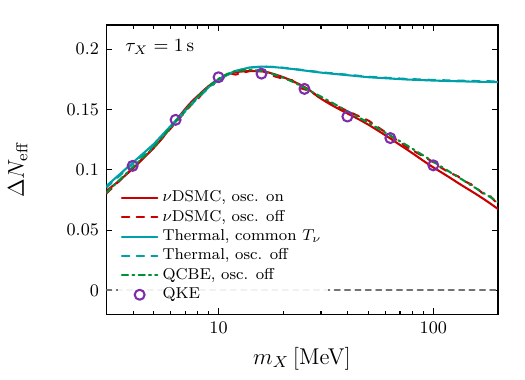}\hfill%
\includegraphics[width=0.495\textwidth]{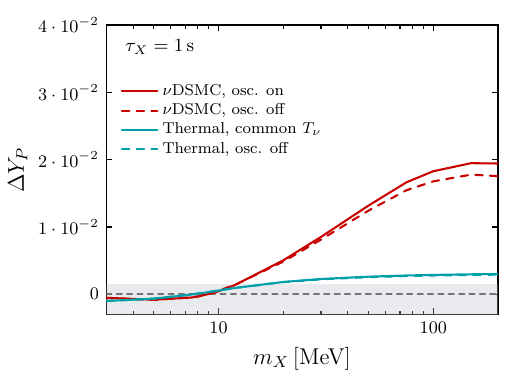}

\vspace{0.25em}
\begin{minipage}[t]{0.495\textwidth}
\vspace{0pt}
\centering
\includegraphics[width=\linewidth]{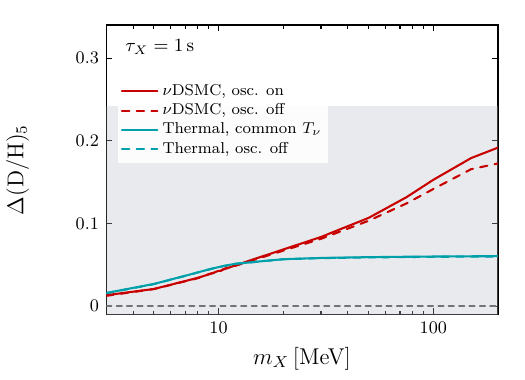}
  \end{minipage}\hfill%
\begin{minipage}[t]{0.495\textwidth}
\vspace{0pt}
\centering
{\small $m_X=100\,\mev,\quad \tau_X=1\,\s,\quad T_\gamma=0.5\,\mev$}\par\vspace{0.15em}
    \includegraphics[width=0.5\linewidth]{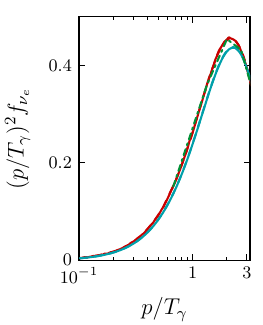}%
\includegraphics[width=0.5\linewidth]{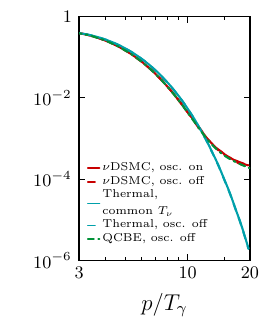}
\end{minipage}
\caption{The same \setupname{II} observables as in Fig.~\ref{fig:setup-II-tau0p1}, now for $\tau_X=1\,\s$.
The upper panels show \dneff and \dyp, the lower-left panel shows \ddhfive, and the lower-right panel shows the $m_X=100\,\mev$ electron-neutrino spectrum at $T_\gamma=0.5\,\mev$.
Solid and dashed red curves denote \nudsmc with neutrino oscillations enabled and disabled, respectively; solid and dashed dark-cyan curves denote common-$T_\nu$ \thermalmethod and \thermalmethod with neutrino oscillations disabled.
The green curves in the upper-left and lower-right panels denote \qcbemethod with neutrino oscillations disabled.
\resp{Violet open circles in the upper-left panel show the \qkemethod results.}}
\label{fig:setup-II-tau1}
\end{figure*}

At $\tau_X=1\,\s$, energy injection extends through the late stages of neutrino decoupling and beyond, when the thermal neutrino bath exchanges energy inefficiently with the electromagnetic plasma.
\qcbemethod and \nudsmc with neutrino oscillations disabled differ by at most $3\times10^{-3}$ in \dneff throughout the mass scan (Fig.~\ref{fig:setup-II-tau1}).
Both reach a maximum near $m_X=14\,\mev$ and then continue to decrease with mass.
At high mass, the injected energy in Eq.~\eqref{eq:setup-II-abundance} approaches a constant, while the daughter energy continues to grow.
The agreement of the two momentum-resolved calculations therefore shows that the decreasing \dneff follows from the hard neutrino spectrum: weak energy-transfer rates grow with neutrino energy and deposit an increasing fraction of the injected energy in the electromagnetic plasma.
The \thermalmethod closure reconstructs a smooth spectrum from integrated neutrino moments and loses this continuing sensitivity to the daughter energy, producing the high-mass plateau.
Neutrino oscillations shift the \nudsmc prediction but preserve the decreasing high-mass trend.
All displayed values remain positive.
\resp{The \qkemethod results closely follow \nudsmc and \qcbemethod over $4\lesssim m_X/\mev\leq100$, reproducing the maximum and the decrease at higher masses.
The agreement in $N_{\rm eff}$ therefore extends to decays during the later stages of neutrino decoupling.}

At $m_X=100\,\mev$, \nudsmc and \qcbemethod, both with neutrino oscillations disabled, give nearly identical electron-neutrino spectra around the thermal peak and through the high-energy tail (lower-right panel of Fig.~\ref{fig:setup-II-tau1}).

In \thermalmethod, both \dyp and \dneff approach plateaus as the neutrino number and energy densities become less sensitive to the relic mass (Fig.~\ref{fig:setup-II-tau1}).
Once those moments change little with mass, the reconstructed electron-flavor distribution and the resulting proton-neutron conversion rates also change little.

The \nudsmc result behaves differently because the later decays produce a high-energy tail of electron neutrinos and antineutrinos that survives during neutron-proton freeze-out and is absent from the \thermalmethod reconstruction.
The lower-right panel of Fig.~\ref{fig:setup-II-tau1} shows this hard tail directly.
The stronger energy weighting of the proton-neutron rates, discussed in Sec.~\ref{sec:setup-I-results}, makes this tail increasingly important as the mass grows.
The helium shift predicted by \nudsmc rises toward a broad maximum near $150\,\mev$ and decreases slightly at $200\,\mev$.
Its height is about $0.02$ with either oscillation treatment, several times the value predicted by the \thermalmethod approach.

The deuterium shift predicted by the \thermalmethod approach levels off and approximately follows Eq.~\ref{eq:dh-neff-correlation}.
\nudsmc instead predicts a deuterium shift that grows with mass even as \dneff decreases.
This behavior arises from neutron production by nonthermal neutrinos after ordinary thermal weak conversion has become inefficient.
A hard electron-antineutrino population survives, and $\bar\nu_e+p\to n+e^+$ remains efficient after the corresponding thermal rate has become small.
The reverse channel $\nu_e+n\to p+e^-$ is included, but protons are much more abundant, so even a charge-symmetric injected population can provide a net late neutron source.
These neutrons enter $p(n,\gamma)d$ after deuterium destruction becomes inefficient.
Thus, \ddhfive cannot, in general, be inferred from the final \dneff alone.

\subsubsection{Fixed mass case}

\begin{figure*}[t!]
\centering
  \includegraphics[width=0.5\textwidth]{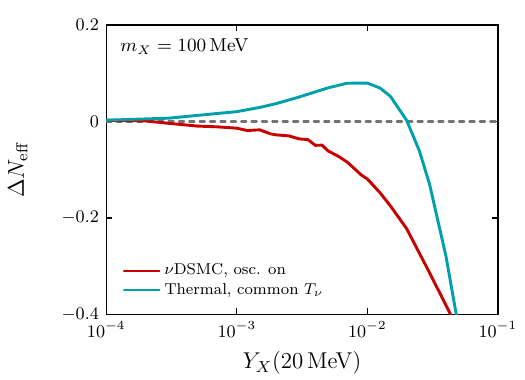}%
\includegraphics[width=0.5\textwidth]{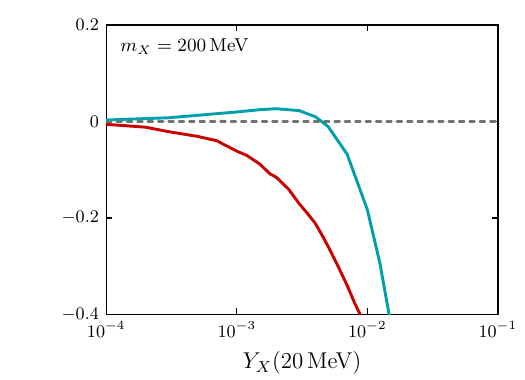}
\caption{\setupname{II} fixed-mass abundance scan at $\tau_X=0.1\,\s$.
The panels show \dneff as a function of the initial abundance for $m_X=100$ and $200\,\mev$.
Red curves denote \nudsmc with neutrino oscillations enabled, and dark-cyan curves denote common-$T_\nu$ \thermalmethod.
}
\label{fig:setup-II-fixed-y-hard-tail}
\end{figure*}

The complementary fixed-mass abundance scan tests whether the difference between \nudsmc and \thermalmethod in \dneff at $\tau_X=0.1\,\s$ depends on the normalization chosen for the mass scan.
The two panels of Fig.~\ref{fig:setup-II-fixed-y-hard-tail} show \dneff for $m_X=100$ and $200\,\mev$.

Once the injected abundance becomes appreciable, \nudsmc predicts an increasingly negative \dneff at both masses.
\thermalmethod first predicts a positive shift and turns downward only at larger abundance.
The agreement in the weak-injection limit therefore deteriorates systematically with source strength, and it deteriorates earlier for the harder $m_X=200\,\mev$ injection.
The fixed-mass scan removes the mass-dependent normalization in Eq.~\eqref{eq:setup-II-abundance} and confirms that the disagreement grows with both spectral hardness and injected abundance.

\resp{Energy transferred from neutrinos to the electromagnetic plasma thermalizes almost immediately.
The injected neutrinos take much longer to reach a thermal distribution.
Their weak cross sections grow with the invariant mass of the colliding pair [Eq.~\eqref{eq:weak-cross-section-scaling}], so sufficiently energetic neutrinos continue to heat the plasma as their distribution relaxes.
This photon heating can reduce $\rho_\nu/\rho_\gamma$ below its value in the cosmology without $X$, giving negative \dneff.}

\resp{We test the role of the interaction strength by setting the couplings of electron neutrinos to electrons equal to those of muon and tau neutrinos.
Removing the charged-current enhancement weakens the transfer of energy to the plasma.
It preserves both rapid electromagnetic thermalization and the growth of weak cross sections with energy.
With these reduced couplings, \dneff still becomes negative as $m_X$ increases, although the sign change occurs at a larger mass for the same lifetime and initial abundance (Appendix~\ref{app:modified-coupling-check}).
The mechanism therefore also operates when all neutrino flavors have the weaker coupling to electrons.}

\subsection{\setupname{III}: through neutrino decoupling}
\label{sec:setup-III-results}

\begin{figure*}[t!]
\centering
  \includegraphics[width=0.333333\textwidth]{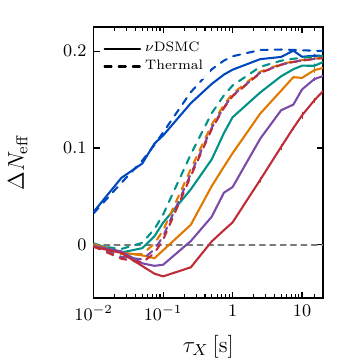}%
  \includegraphics[width=0.333333\textwidth]{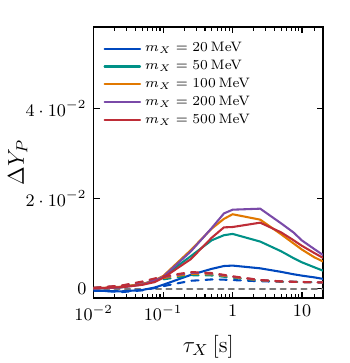}%
\includegraphics[width=0.333334\textwidth]{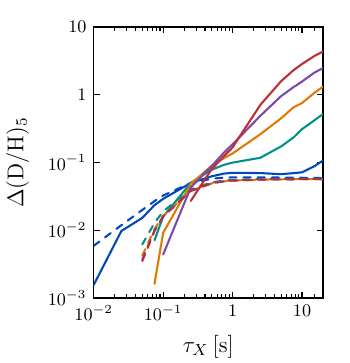}
\caption{\resp{\setupname{III} comparison across lifetimes} (Tab.~\ref{tab:physics-setups}) for five masses, using Eq.~\eqref{eq:setup-III-abundance}.
From left to right, the panels show \dneff, \dyp, and \ddhfive as functions of $\tau_X$.
Colors distinguish the masses listed inside the middle panel; solid and dashed curves denote momentum-resolved \nudsmc and common-$T_\nu$ \thermalmethod, respectively.}
\label{fig:setup-III-controlled}
\end{figure*}

The agreement between a thermal description and momentum-resolved transport depends on when the relic decays and on the energy of its daughters.
\setupname{III} tests this dependence for five masses and $0.01\,\s\leq\tau_X\leq20\,\s$, spanning decays before, during, and after neutrino decoupling (Fig.~\ref{fig:setup-III-controlled}).
Decays, inverse decays, quantum-statistical factors, and flavor conversion are included throughout.
Inverse decays are included in both methods here; their separate effect is studied in \setupname{IV}.
Secondary charged-pion production during neutrino thermalization is negligible over $20\leq m_X/\mev\leq 500$ and is therefore switched off in this scan.

The mass controls the spectral hardness through $E_\nu\simeq m_X/2$, and the lifetime fixes the vacuum decay rate.
Together with inverse decays, these parameters determine \resp{the rate and duration of neutrino injection}.
The abundance prescription limits the variation of the total source energy.

For \dneff, \thermalmethod and \nudsmc are nearly coincident at $\tau_X=0.01\,\s$ throughout the mass scan.
Two regimes produce this agreement.
At low mass, inverse decays efficiently replenish the $X$ population.
The resulting $1\leftrightarrow2$ cycling provides an additional thermalization mechanism, independent of weak $2\leftrightarrow2$ scattering: repeated absorption and reinjection of neutrino pairs maintain a nearly thermal neutrino distribution.
At high mass, inverse decays are Boltzmann suppressed.
The short vacuum lifetime then depletes the initial $X$ population while weak $2\leftrightarrow2$ reactions can still efficiently redistribute the daughter neutrinos.

As $\tau_X$ increases into \resp{the epoch of neutrino decoupling}, the agreement develops a clear mass dependence.
Low-mass daughters lie closer to the thermal scale, and repeated inverse decays and decays provide an additional number-changing redistribution channel.
At high mass, inverse decays are ineffective and the injected hard population survives while continuing to transfer energy to the electromagnetic plasma.
By replacing this population with a Fermi-Dirac distribution, \thermalmethod overestimates the neutrino energy density and can predict the opposite sign of \dneff.
At the largest $\tau_X$ shown, both calculations trend toward positive plateaus, but no common limit is reached within the scan.

The $m_X=20\,\mev$ curves make the low-mass agreement explicit.
For $\tau_X\leq0.1\,\s$, the methods predict similar abundances at $m_X=20\,\mev$.
At later lifetimes the \dneff agreement remains comparatively good, while the light-element predictions develop \resp{differences that depend on the observable} as late nonthermal weak conversion becomes important.

The helium excess predicted by \nudsmc peaks around $\tau_X=1\,\s$ for the lighter parents and shifts toward a few seconds for the heavier ones.
The peak height turns over between $200$ and $500\,\mev$.

In the \thermalmethod approach, the helium shift reaches smaller maxima at earlier lifetimes, before $1\,\s$.
At long lifetimes, the predicted helium shift becomes nearly independent of the parent mass.
The helium shift is largest when the injection of non-thermal electron-flavor neutrinos overlaps the neutron-proton freeze-out.
\thermalmethod instead smooths the injected population into a Fermi-Dirac spectrum.
The stronger sensitivity of the weak conversion rates to the high-energy tail, explained in Sec.~\ref{sec:setup-I-results}, accounts for the large helium shift despite the smaller change in \dneff.

The deuterium correction as predicted by \nudsmc continues to grow at long lifetimes and increases strongly with parent mass, whereas \thermalmethod predicts a plateau.
This is the late nonthermal proton-to-neutron mechanism identified in Sec.~\ref{sec:setup-II-results}.
Deuterium remains sensitive to neutrons injected after efficient deuterium destruction ends, while helium depends mainly on the neutron population established earlier.

\begin{figure}[t!]
\centering
\includegraphics[width=\columnwidth]{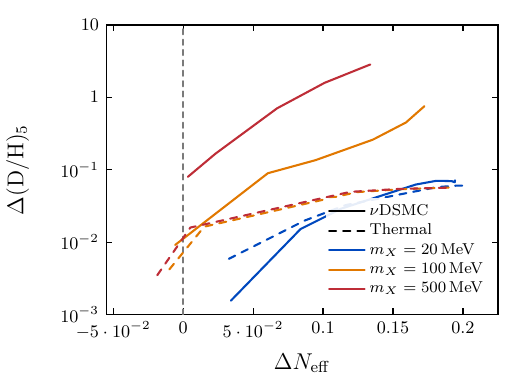}
\caption{\setupname{III} \ddhfive as a function of the final \dneff.
Colors distinguish $m_X=20$, $100$, and $500\,\mev$, while solid and dashed curves denote \nudsmc and \thermalmethod, respectively, for $0.01\,\mathrm{s}\leq\tau_X\leq 10\,\mathrm{s}$; the $20\,\mathrm{s}$ results are shown in Fig.~\ref{fig:setup-III-controlled}.
The mass dependence shows that neutrinophilic decays break the nearly thermal correlation in Eq.~\eqref{eq:dh-neff-correlation}, making the relation between \ddhfive and \dneff model-dependent.}
\label{fig:setup-III-dh-neff}
\end{figure}

Different relic masses and lifetimes can give different helium and deuterium abundances at the same final \dneff (Fig.~\ref{fig:setup-III-dh-neff}).
The separation reflects the dependence of late neutron production on the timing of injection and the electron-antineutrino spectrum.
Common-$T_\nu$ \thermalmethod approximately follows Eq.~\eqref{eq:dh-neff-correlation} because it replaces those spectra with thermal distributions.

At long lifetimes, the thermal approximation allows relic parameters for which momentum-resolved transport predicts deuterium overproduction.
This difference depends on the initial relic abundance: fewer relics inject less total energy while each daughter remains equally energetic.

\subsection{\setupname{IV}: effects of inverse decays and pions}
\label{sec:setup-IV-results}

\begin{figure*}[t!]
\centering
  \includegraphics[width=0.333333\textwidth]{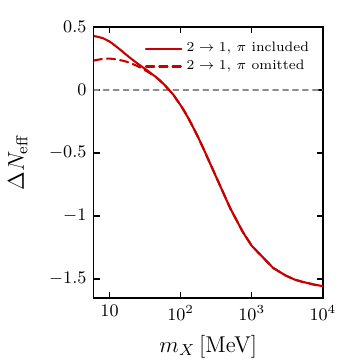}%
  \includegraphics[width=0.333333\textwidth]{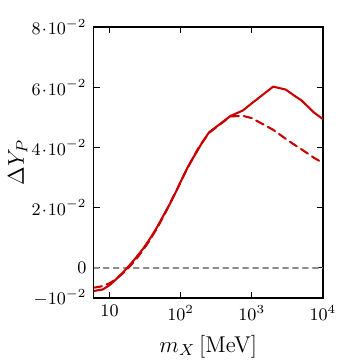}%
\includegraphics[width=0.333334\textwidth]{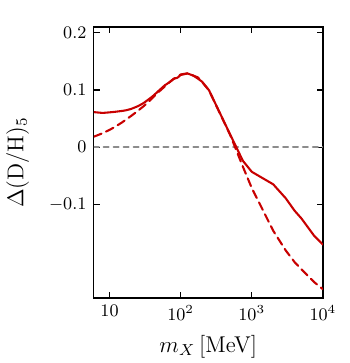}
\caption{\resp{\setupname{IV} comparison isolating inverse decays and pion production} (Tab.~\ref{tab:physics-setups}).
From left to right, the panels show \dneff, \dyp, and \ddhfive as functions of $m_X$.
Solid and dashed red curves denote the relevant additional process included and omitted, respectively: inverse decays for $m_X\leq50\,\mev$, and the passive charged-pion source for $m_X\geq500\,\mev$.
At intermediate masses, the solid curve includes inverse decays and pions, while the dashed curve includes direct decays alone.
Neutrino oscillations are included throughout.
At high mass, \resp{the results with and without pion production} use the same neutrino and photon energy densities, so their $N_{\rm eff}$ values are identical.}
\label{fig:setup-IV-full-mass}
\end{figure*}

Fig.~\ref{fig:setup-IV-full-mass} combines \resp{two comparisons that separately isolate inverse decays and pion production}.
For $m_X\leq50\,\mev$, otherwise identical \nudsmc calculations with neutrino oscillations enabled are compared with inverse decays omitted or included.
For $m_X\geq500\,\mev$, where inverse decays are negligible, the same time-dependent neutrino distributions and expansion rate enter two BBN calculations, with the charged-pion source omitted or included.
For $50\,\mev<m_X<500\,\mev$, the curves compare direct decays alone with inverse decays and pion production also included.
All calculations use $\tau_X=0.1\,\s$, $Y_X(20\,\mev)=0.01$, particles at rest, and equal decay branching fractions into the three neutrino flavors.

At intermediate masses, the two abundance predictions nearly coincide.

Below about $50\,\mev$, inverse decays regenerate $X$ from thermal neutrinos, while charged-pion production is negligible.
Inverse decays lower helium and raise D/H; their deuterium and radiation-density changes approximately follow Eq.~\ref{eq:dh-neff-correlation}.
The effect decreases as the relic mass rises above the thermal scale.
Because inverse decays regenerate $X$ from the neutrino bath, they can dominate the shifts in all three observables when $Y_X(T_{\rm ini})$ is very small.
Such a suppressed initial abundance may arise, for example, in a late-reheating cosmology.

Above about $500\,\mev$, inverse decays are suppressed, while the hard electron-flavor neutrinos produce charged pions through Eq.~\eqref{eq:pion-producting-process}.
Pion absorption then changes the BBN abundances relative to the neutrino-only prediction.
The pion yield stays below $3\times10^{-4}$ per relic decay, consistent with neglecting its effect on neutrino transport.
At GeV masses, pion absorption reduces the deuterium deficit without reversing its sign and contributes about a third of the neutrino-induced helium shift.
Neutrino-induced conversion remains dominant for both abundances over the displayed masses.

\subsection{Effect of finite nucleon masses}
\label{sec:finite-mass-comparison}

\begin{figure}[t!]
\centering
\includegraphics[width=\columnwidth]{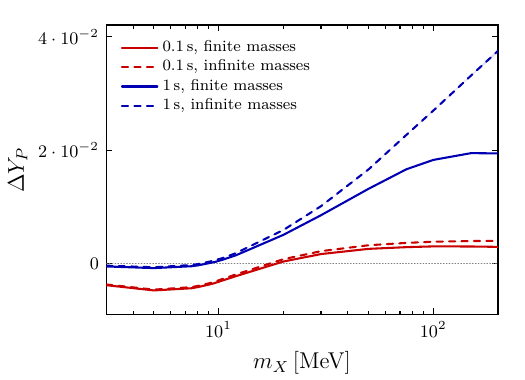}
\caption{Helium correction $\Delta Y_P=Y_P-Y_{P,{\rm SBBN}}$ in \setupname{II}, defined by Eq.~\ref{eq:delta-yp}, with oscillations included.
Solid and dashed curves use finite and infinite nucleon masses, respectively; colors distinguish the lifetimes.
The finite-mass rates are given by Eq.~\ref{eq:weak-conversion-kernels} and include the radiative prescription of Sec.~\ref{sec:bbn-framework}.
Bands show standard errors of the means.}
\label{fig:finite-mass-rates}
\end{figure}

Finite nucleon masses modify neutron-proton conversion by energetic neutrinos.
Fig.~\ref{fig:finite-mass-rates} compares the finite- and infinite-nucleon-mass prescriptions of Sec.~\ref{sec:bbn-framework} in \setupname{II}, with the two lifetimes and mass-dependent initial abundance defined in Eq.~\ref{eq:setup-II-abundance}.
This scenario excludes inverse decays and pion production.
Both BBN calculations use the same evolving neutrino spectra, expansion rate, plasma temperature, and nuclear reaction rates.

The finite-mass calculation gives a smaller helium shift at large relic masses.
Recoil and the momentum dependence of the nucleon charged-current form factors reduce the conversion rates for energetic neutrinos.
The helium shift consequently levels off, while the infinite-mass approximation predicts continued growth.
The reduction is largest for $\tau_X=1\,\s$, when energetic neutrinos survive during neutron conversion.
Earlier injection thermalizes more efficiently and produces a smaller difference between the two prescriptions.

\section{Code release, performance, and limitations}
\label{sec:performance-limitations}

\subsection{Code release}
\label{sec:code-release}

The publicly released \textsc{Mathematica} implementation of \nudsmc contains the framework developed in this work and allows its physics results to be reproduced~\cite{nuDSMC:repository}.
The package also includes the BBN code, which takes the \nudsmc results as input and computes the primordial nuclear abundances.
We have also developed a complementary multithreaded \textsc{C++} implementation, which gives consistent physics results and is used for the performance benchmarks below.
The \textsc{C++} version will be publicly released in time for publication of this work.

\subsection{Performance}
\label{sec:performance}

We use the \textsc{C++} implementation for the timing comparison because it is the optimized version that will be publicly available at publication of this work.
Its transport loop and parallel collision calculations run within a single compiled program.
Agreement with the \textsc{Mathematica} implementation provides an independent check of the predicted observables.
For matched benchmarks, the \textsc{Mathematica} implementation is typically slower than the \textsc{C++} implementation by a factor of a few.
The three benchmarks exercise electromagnetic reheating, hard neutrinophilic injection, and active decays and inverse decays.

\begin{table*}[t!]
\centering
\begin{ruledtabular}
\begin{tabular}{lccc}
Method & EM reheating & Hard neutrino injection &
Active $1\leftrightarrow2$ processes \\
\hline
\nudsmc (Air) & $23.5$ & $8.3$ & $13.6$ \\
\thermalmethod (Air) & $0.54$ & $0.21$ & $0.29$ \\
\qcbemethod (Air) & $59.7$ & $1687.2$ & $52.1$ \\
\end{tabular}
\end{ruledtabular}
\caption{Transport wall time per calculation in seconds, for the benchmarks defined in Sec.~\ref{sec:performance-limitations}.
The three approaches are defined in Tab.~\ref{tab:approaches}.
The \nudsmc Air entries are mean times per realization from five sequential runs, the \thermalmethod entries are medians of seven runs, and the \qcbemethod entries are single runs.
Air denotes the MacBook Air M4, using ten threads for \nudsmc and one for \qcbemethod.
These two methods include initialization and thermal pre-evolution; spectrum exports and BBN are excluded.
The \thermalmethod entries cover its transport integration.}
\label{tab:performance}
\end{table*}

For the timing comparison in Tab.~\ref{tab:performance}, we define three fixed benchmarks:
\begin{enumerate}
\item \emph{EM reheating}: $m_X=100\,\mev$, $T_{\rm RH}=3.5\,\mev$ ($\tau_X=0.0402\,\s$), ${\rm Br}_{X\to\gamma\gamma}=1$, and the initial condition $\rho_X/\rho_{\rm rad}=50$ at $T_0=10.14\,\mev$.
\item \emph{Hard neutrino injection}: $m_X=200\,\mev$, $\tau_X=1\,\s$, and $Y_X(20\,\mev)=1.25\times10^{-4}$, with inverse decays disabled.
\item \emph{Active $1\leftrightarrow2$ processes}: $m_X=10\,\mev$, $\tau_X=0.1\,\s$, and $Y_X(20\,\mev)=0.01$, with inverse decays enabled.
\end{enumerate}
For the first benchmark, this energy-density ratio specifies the initial $X$ abundance without an independent $Y_X$.
The last two benchmarks use flavor-democratic $X\to\nu\bar\nu$.
These physical inputs define the corresponding calculations across the three methods.
The \nudsmc and \qcbemethod benchmarks both disable neutrino oscillations and secondary pion production.
Their initial $X$ number densities and cosmic times are matched.
Both use thermal pre-evolution down to $5\,\mev$, followed by momentum-resolved transport.

The \nudsmc and \qcbemethod timings cover the complete transport calculation through $T_\gamma=0.024\,\mev$, including initialization and thermal pre-evolution.
Neither calculation exports neutrino spectra during the timing measurement.
The BBN calculation is performed separately and is excluded from these timings.

Each \nudsmc timing is a five-seed mean for $1.5\times10^6$ unit-weight Monte Carlo neutrinos, or approximately $5\times10^5$ particles per CP-symmetric flavor sector, \resp{distributed over 100 collision cells}.
The \nudsmc and \qcbemethod timings are measured on the same MacBook Air M4, with ten threads for \nudsmc and one for \qcbemethod.
The latter matches the serial execution mode of the public source.
Parallelizing its collision routine can reduce the \qcbemethod runtime.

For the timing comparison, we test convergence of both $N_{\rm eff}$ and $Y_P$ and require the \qcbemethod numerical error in $N_{\rm eff}$ to be no larger than the statistical error of \nudsmc.
The empirical standard deviation of $N_{\rm eff}$ across the five \nudsmc seeds is below $0.006$ in each benchmark.
For hard injection, the finite-mass calculation gives a standard error of about $3\times10^{-4}$ in $Y_P$.

The grids, integration tolerances, and decay cutoff used for Tab.~\ref{tab:performance} are specified in Appendix~\ref{app:performance-settings}.
The \qcbemethod normalization uses the thermal Fermi-Dirac moment evaluated on the same grid as the evolving distributions.
Both methods use their time-dependent spectra and expansion rates in the BBN calculation of Sec.~\ref{sec:bbn-framework}.

For hard neutrino injection, convergence of $N_{\rm eff}$ alone does not ensure convergence of the primordial abundances.
The total neutrino energy density can already be stable while small changes in the high-energy electron-neutrino and antineutrino populations still alter the neutron-proton conversion rates and hence $Y_P$ and D/H.
Resolving this tail requires tighter control of the small neutrino occupations than is needed for $N_{\rm eff}$.
The momentum range must contain the populated decay products until the residual relic energy density is negligible.
Ending the calculation while relic decays are still appreciable can change the abundance prediction even when the neutrino energy moments appear stable.

Rapid inverse decays can also increase the cost at fixed $N_p$.
The \thermalmethod values describe transport integration with a common neutrino temperature, excluding initialization and BBN.
These momentum-averaged calculations finish in under one second, with the accuracy restrictions discussed in Secs.~\ref{sec:setup-II-results} and \ref{sec:setup-III-results}.

\subsection{Limitations}
\label{sec:limitations}

The flavor treatment assumes CP-symmetric active-neutrino distributions, $f_{\nu_\alpha}(p)=f_{\bar\nu_\alpha}(p)$, negligible charged-lepton asymmetries, and the propagation and collision approximations specified in Sec.~\ref{sec:oscillations}.
For sizable lepton asymmetries, the CP-odd neutrino self-potential depends on the evolving flavor density matrices and requires momentum-dependent QKE evolution.
A hybrid calculation combines QKE flavor evolution with energy transfer calculated by \nudsmc for heavy neutral leptons in the presence of lepton flavor asymmetries~\cite{Akita:2026gee}.

The collision kernel includes the Standard Model $2\leftrightarrow2$ reactions and the specified $X$ decays and inverse decays.
We neglect additional $2\leftrightarrow2$ neutrino scattering mediated by $X$.
This approximation is typically valid for the lifetimes studied here, $\tau_X\gtrsim10^{-2}\,\s$.
For a renormalizable two-body $X\to\nu\bar\nu$ decay away from threshold, a dimensionless interaction coupling $y_X$ gives $\tau_X^{-1}\sim y_X^2m_X$ and an off-shell effective coupling $G_X\sim y_X^2/m_X^2$.
This is the $\alpha=1$ case of the model-dependent scaling $\tau_X\propto y_X^{-2}m_X^{-\alpha}$ and gives, up to spin-dependent factors,
\begin{equation*}
 \frac{G_X}{G_F}\lesssim\mathcal O(10^{-7})
 \left(\frac{10^{-3}\,\s}{\tau_X}\right)
 \left(\frac{3\,\mev}{m_X}\right)^3 .
\end{equation*}
Thus, for $m_X\geq3\,\mev$ and these lifetimes, the omitted off-shell neutrino self-scattering is negligible unless it is parametrically enhanced.
The on-shell channel is already included through inverse decays.
This estimate does not cover threshold or resonant enhancement, higher-dimensional interactions, a lighter mediator, or neutrino-electron scattering controlled by an independent electron coupling.
Such models and the regime $\tau_X\ll10^{-3}\,\s$ require a separate treatment.

\section{Conclusions}
\label{sec:conclusions}

We have extended Neutrino Direct Simulation Monte Carlo (\nudsmc) into a framework for predicting the cosmological consequences of relic decays at MeV temperatures.
The calculation jointly follows the relic population, neutrino momentum distributions, electromagnetic plasma, and cosmic expansion, and couples this evolution to Big Bang nucleosynthesis (Secs.~\ref{sec:new-algorithm} and \ref{sec:bbn-framework}).
Flavor conversion uses adiabatic propagation with averaged phases and includes interference between flavor amplitudes in collisions, in the regime of negligible lepton asymmetries.
The secondary charged-pion source enters the BBN reactions while its small energy yield permits neglecting its backreaction on neutrino transport.

Equilibrium and conservation tests check the transport algorithm (Sec.~\ref{sec:cross-checks}).
To investigate the physics of neutrino thermalization and cross-check \nudsmc, we consider the four physical scenarios in Tab.~\ref{tab:physics-setups} and compare independent calculations with the same physical inputs (Sec.~\ref{sec:case-studies}).
The comparison uses the complementary approaches defined in Tab.~\ref{tab:approaches}: momentum-averaged \thermalmethod calculations, quasi-classical Boltzmann equations (\qcbemethod), and quantum kinetic equations (\qkemethod).
These scenarios represent common possibilities in extensions of the Standard Model and can be generalized to include model-specific processes, as demonstrated for heavy neutral leptons and lepton flavor asymmetries in Ref.~\cite{Akita:2026gee}.
Without oscillations, \qcbemethod and \nudsmc agree on the radiation density and neutrino momentum distributions.
Their abundance predictions also agree in the electromagnetic reheating benchmark and the benchmark with decays and inverse decays (Sec.~\ref{sec:performance-limitations}).
The \qkemethod comparisons include flavor coherence; they agree closely for neutrinophilic decays, while a residual difference in electromagnetic reheating is examined in Appendix~\ref{app:osc-checks}.

Particles decaying near neutrino decoupling cannot generally be characterized by their final contribution to the radiation density alone.
Persistent neutrino spectral distortions can reverse the predicted sign of \dneff and substantially change the primordial abundances (Figs.~\ref{fig:setup-II-tau0p1}, \ref{fig:setup-II-tau1}, and \ref{fig:setup-III-controlled}).
The energetic electron-antineutrino population continues to produce neutrons after thermal weak conversion becomes inefficient, driving the enhancement of deuterium at long lifetimes in the right panel of Fig.~\ref{fig:setup-III-controlled}.
Relic decays with similar final $N_{\rm eff}$ can therefore yield different $Y_P$ and D/H (Fig.~\ref{fig:setup-III-dh-neff}).
In particular, a small shift in $N_{\rm eff}$ can accompany a helium or deuterium abundance outside the observationally allowed range.
High-energy neutrinos can also produce charged pions as they thermalize.
Even for relics as heavy as $10\,\gev$, neutrino-induced neutron-proton conversion produces a larger helium shift than secondary mesons.
At GeV relic masses, secondary pions reduce the deuterium deficit while preserving its sign (Fig.~\ref{fig:setup-IV-full-mass}).
Including nucleon recoil and structure reduces the helium shift caused by energetic neutrinos and weakens its growth with relic mass (Fig.~\ref{fig:finite-mass-rates}).
A dedicated study of the transition region $10\,\gev\lesssim m_X\lesssim10\,\mathrm{TeV}$ and $0.1\,\s\lesssim\tau_X\lesssim100\,\s$ would clarify when secondary hadrons dominate over the effects of neutrinos on primordial abundances.
We leave this to future work, complementing Ref.~\cite{Bianco:2025boy}, which studies neutrinophilic relics up to $10\,\mathrm{TeV}$ at much longer lifetimes, $\tau_X\gtrsim10^4\,\s$.

In the timing benchmarks, \nudsmc is faster than the calculation on a discretized momentum grid, with the largest gain for hard neutrino injection (Tab.~\ref{tab:performance}).
The speed of \nudsmc makes broad parameter scans practical while resolving the nonthermal spectra that affect $N_{\rm eff}$ and the primordial abundances.

Comparisons of neutrino transport require matched physical inputs and agreement within numerical uncertainties on the final $N_{\rm eff}$, the neutrino momentum distributions and their moments throughout decoupling, and the resulting primordial nuclear abundances.
The precision of measurements of $Y_P$ and D/H makes this broader validation essential for using BBN to constrain nonstandard cosmologies.

\par\vspace{0.5\baselineskip}
\section*{Acknowledgements}

The authors thank Julien Froustey for reading the manuscript and for useful discussions.
KA was supported by JSPS KAKENHI Grant Number 24KJ0060.
MO received support from the European Union's Horizon Europe research and innovation programme under the Marie Sklodowska-Curie grant agreement No~101204216.

The authors used Astra 6 for assistance with refinements to the \textsc{C++} code and other technical tasks.
The authors take full responsibility for the results.

\bibliography{main.bib}

\clearpage

\onecolumngrid

\appendix

\resp{The appendices give the reaction rates and tests underlying the transport calculation.
Appendix~\ref{app:thermal-1-2} derives the inverse-decay rates and checks that efficient $1\leftrightarrow2$ reactions drive the simulated particles toward the expected thermal equilibrium.
It includes thermal-mass and quantum-statistical corrections for bosonic and fermionic relics and their different decay channels.
Appendix~\ref{app:modified-coupling-check} tests how reducing the electron-neutrino couplings to the electromagnetic plasma changes the negative-\dneff branch.
Appendix~\ref{app:osc-checks} examines the averaged-oscillation and adiabatic approximations, flavor interference in collision rates, and the comparison with a quantum kinetic calculation of electromagnetic reheating.
Appendix~\ref{app:performance-settings} gives the numerical settings for the \qcbemethod performance benchmarks.}

\section{The $1\leftrightarrow 2$ processes}
\label{app:thermal-1-2}

\subsection{Inverse rate}

Consider the inverse decay process
\begin{equation}
1(p_1)+2(p_2)\to X(p_X)\,,
\qquad
E_a^2 = |\mathbf p_a|^2 + m_a^2\,.
\end{equation}
We denote by $g_a$ the number of polarization states of particle $a$, and by $g_X$ the number of polarization states of $X$.
As in Sec.~\ref{sec:general}, $f_X(E)$ denotes the occupation number per polarization state of $X$.
Thus, for bosonic $X$, the final-state stimulation factor is
\begin{equation}
1+f_X(E_X)\,,
\qquad
E_X=E_1+E_2\,.
\end{equation}

We start from the invariant expression for the Møller-weighted cross section,
\begin{equation}
\sigma_{12\to X} v_{\rm M}
=
\frac{1}{2E_1\,2E_2}
\int
\frac{d^3 p_X}{(2\pi)^3 2E_X}
(2\pi)^4
\delta^{(4)}(p_1+p_2-p_X)\,
\overline{|\mathcal M_{12\to X}|^2}\,
\left[1+f_X(E_X)\right]\,.
\end{equation}
Here,
\begin{equation}
\overline{|\mathcal M_{12\to X}|^2}
\equiv
\frac{1}{g_1 g_2}
\sum_{\lambda_1,\lambda_2,\lambda_X}
|\mathcal M_{\lambda_1\lambda_2\to\lambda_X}|^2
\end{equation}
is averaged over the initial polarizations and summed over the final polarizations.
The one-particle phase-space integral gives
\begin{equation}
\int
\frac{d^3 p_X}{(2\pi)^3 2E_X}
(2\pi)^4
\delta^{(4)}(p_1+p_2-p_X)
=
2\pi\,\delta\!\left((p_1+p_2)^2-m_X^2\right)\,,
\end{equation}
so that
\begin{equation}
\sigma_{12\to X} v_{\rm M}
=
\frac{\pi}{2E_1E_2}\,
\overline{|\mathcal M_{12\to X}|^2}\,
\delta\!\left(s-m_X^2\right)
\left[1+f_X(E_1+E_2)\right]\,,
\end{equation}
where
\begin{equation}
s=(p_1+p_2)^2
=
m_1^2+m_2^2
+
2\left(E_1E_2-|\mathbf p_1||\mathbf p_2|\cos\theta\right)\,.
\end{equation}

For an isotropic ensemble, the relative angle is averaged as
\begin{equation}
\left\langle \sigma_{12\to X}v_{\rm M}\right\rangle_{\theta}
=
\frac{1}{2}
\int_{-1}^{1} d\cos\theta\,
\sigma_{12\to X}v_{\rm M}\,.
\end{equation}
Defining
\begin{equation}
c_\star
\equiv
\frac{
m_1^2+m_2^2+2E_1E_2-m_X^2
}{
2|\mathbf p_1||\mathbf p_2|
}\,,
\end{equation}
one obtains
\begin{equation}
\frac{1}{2}
\int_{-1}^{1}d\cos\theta\,
\delta(s-m_X^2)
=
\frac{1}{4|\mathbf p_1||\mathbf p_2|}
\,
\Theta(1-|c_\star|)\,.
\end{equation}
Therefore,
\begin{equation}
\left\langle \sigma_{12\to X}v_{\rm M}\right\rangle_{\theta}
=
\frac{\pi}{8E_1E_2|\mathbf p_1||\mathbf p_2|}
\,
\overline{|\mathcal M_{12\to X}|^2}
\,
\Theta(1-|c_\star|)
\left[1+f_X(E_1+E_2)\right]\,.
\end{equation}
For massless initial particles, this reduces to
\begin{equation}
\left\langle \sigma_{12\to X}v_{\rm M}\right\rangle_{\theta}
=
\frac{\pi}{8E_1^2E_2^2}
\,
\overline{|\mathcal M_{12\to X}|^2}
\,
\Theta(4E_1E_2-m_X^2)
\left[1+f_X(E_1+E_2)\right]\,.
\end{equation}
This is the origin of the characteristic $1/(E_1^2E_2^2)$ factor in the inverse-decay rate.

The squared matrix element can be expressed through the partial decay width of $X$.
Let
\begin{equation}
\lambda_X
\equiv
\lambda(m_X^2,m_1^2,m_2^2)
=
\left[m_X^2-(m_1+m_2)^2\right]
\left[m_X^2-(m_1-m_2)^2\right]\,.
\end{equation}
The partial decay width is
\begin{equation}
\Gamma_{X\to 12}
=
\frac{1}{2m_X}\,
\frac{1}{g_X}\,
\frac{1}{S_{12}}
\sum_{\lambda_X,\lambda_1,\lambda_2}
\int d\Pi_1\,d\Pi_2\,
(2\pi)^4
\delta^{(4)}(p_X-p_1-p_2)
|\mathcal M_{\lambda_X\to\lambda_1\lambda_2}|^2\,,
\label{eq:rest-frame-partial-width}
\end{equation}
where
\begin{equation}
d\Pi_a\equiv \frac{d^3p_a}{(2\pi)^3 2E_a}\,,
\end{equation}
and $S_{12}=1$ for distinguishable final particles, while $S_{12}=2!$ for identical final particles.
Eq.~\eqref{eq:rest-frame-partial-width} is evaluated in the $X$ rest frame, $p_X=(m_X,\mathbf 0)$, and excludes occupation factors.
Using the thermal daughter masses $m_1(T)$ and $m_2(T)$ and the chosen decay matrix element defines $\Gamma^{\rm kin}_{X\to12}(T)$ in Eq.~\eqref{eq:Xs-thermal-width}.
It reduces to ${\rm Br}_{12}/\tau_X$ in vacuum and vanishes when $m_X<m_1(T)+m_2(T)$.
Its thermal-mass dependence is channel dependent, separate from time dilation and the occupation factors.
The electron thermal mass reduces the phase space for scalar decay and can close the $e^+e^-$ channel.
For the CP-even interaction $-y_eX\bar e e$, we fix $y_e$ through the vacuum width and use the mass in Eq.~\ref{eq:thermal-electron-mass}.
For a channel open in vacuum, the quasiparticle prescription gives
\begin{equation}
\frac{\Gamma^{\rm kin}_{X\to e^+e^-}(T)}{\Gamma_{X\to e^+e^-}}
=
\left[\frac{1-4m_{e,{\rm th}}^2(T)/m_X^2}
{1-4m_e^2/m_X^2}\right]^{3/2}
\Theta\!\left(m_X-2m_{e,{\rm th}}(T)\right).
\label{eq:scalar-ee-thermal-width}
\end{equation}
The bracket is evaluated only above threshold.
Pauli blocking and the moving-parent factor $m_X/E_X$ enter separately through Eq.~\ref{eq:Xs-thermal-width}.
The inverse reaction uses the same in-medium matrix element and kinematics.
For the isotropic two-body decays sampled here, the normalized average in Eq.~\eqref{eq:Xs-thermal-width} is
\[
\left\langle\Pi_Q\right\rangle_{\rm dec}
=\frac12\int_{-1}^{1}d\cos\theta_*\,
\Pi_Q\bigl(E_1(\theta_*),E_2(\theta_*)\bigr)\,.
\]
Here, $\theta_*$ is the daughter emission angle relative to the boost direction in the $X$ rest frame; $E_1$ and $E_2$ are the daughter energies after the boost to the plasma frame.
Sampling the decay direction and applying the statistical acceptance test evaluates this average stochastically.
When thermal masses are included in the inverse-rate formulas below, the partial width is likewise replaced by $\Gamma^{\rm kin}_{X\to12}(T)$.

For a two-body decay,
\begin{equation}
\frac{1}{g_X}
\frac{1}{S_{12}}
\sum_{\lambda_X,\lambda_1,\lambda_2}
|\mathcal M_{\lambda_X\to\lambda_1\lambda_2}|^2
=
\frac{16\pi m_X^3}{\sqrt{\lambda_X}}\,
\Gamma_{X\to 12}\,.
\end{equation}
Equivalently,
\begin{equation}
\overline{|\mathcal M_{12\to X}|^2}
=
\frac{g_X S_{12}}{g_1g_2}
\frac{16\pi m_X^3}{\sqrt{\lambda_X}}\,
\Gamma_{X\to 12}\,.
\end{equation}

In the discrete-particle description, we define the sums over particles including their polarization degeneracies as
\begin{equation}
\sum_{i\in 1}
\equiv
g_1
\sum_{i\in 1,{\rm pol}}\,,
\qquad
\sum_{j\in 2}
\equiv
g_2
\sum_{j\in 2,{\rm pol}}\,.
\end{equation}
With this convention, the factors $1/g_1$ and $1/g_2$ from the polarization average in $\overline{|\mathcal M|^2}$ are cancelled by the polarization sums.

The inverse-decay rate is then
\begin{equation}
\Gamma_{\rm inv}
=
\frac{1}{V}
\sum_{i\in 1}
\sum_{j\in 2}^{\rm pairs}
\left\langle \sigma_{ij\to X}v_{\rm M}\right\rangle_{\theta}\,,
\end{equation}
where the pair sum is understood as
\begin{equation}
\sum_{i\in 1}
\sum_{j\in 2}^{\rm pairs}
=
\begin{cases}
\displaystyle
\sum_{i\in 1}\sum_{j\in 2}\,,
& 1\not\equiv 2\,,
\\[1.2em]
\displaystyle
\frac{1}{2}
\sum_{i,j\in 1}^{i\neq j}\,,
& 1\equiv 2\,.
\end{cases}
\end{equation}
Thus, after replacing the matrix element by the lifetime, one obtains
\begin{equation}
\Gamma_{\rm inv}
=
\frac{2\pi^2 g_X S_{12} m_X^3}{V\sqrt{\lambda_X}}\,
\Gamma_{X\to 12}
\sum_{i\in 1}
\sum_{j\in 2}^{\rm pairs}
\frac{
\Theta(1-|c_{\star,ij}|)
}{
E_iE_j|\mathbf p_i||\mathbf p_j|
}
\left[
1+f_X(E_i+E_j)
\right]\,,
\end{equation}
with
\begin{equation}
c_{\star,ij}
=
\frac{
m_1^2+m_2^2+2E_iE_j-m_X^2
}{
2|\mathbf p_i||\mathbf p_j|
}\,.
\end{equation}
In vacuum, the partial width is fixed by the branching fraction ${\rm Br}_{12}$ and lifetime $\tau_X$:
\begin{equation}
\Gamma_{X\to 12}
=
\frac{{\rm Br}_{12}}{\tau_X}\,.
\end{equation}

For massless initial particles, $m_1=m_2=0$, this simplifies to
\begin{equation}
\Gamma_{\rm inv}
=
\frac{2\pi^2 g_X S_{12} m_X}{V\tau_X}\,
{\rm Br}_{12}
\sum_{i\in 1}
\sum_{j\in 2}^{\rm pairs}
\frac{
\Theta(4E_iE_j-m_X^2)
}{
E_i^2E_j^2
}
\left[
1+f_X(E_i+E_j)
\right]
\,.
\end{equation}

\subsection{Counting pairs in the decay line}
\label{app:line-pair-counting}

The compact decay line of Sec.~\ref{sec:zero-mode} represents neutrinos and antineutrinos by a combined CP-symmetric count for each propagation label $i$.
To derive its occupation and pair weights, let $L_i\equiv N_{{\rm line},i}$ and let $G_\nu$ be the number of states in the line bin for one species with one helicity.
The phase-space volume gives
\begin{equation}
G_\nu=\frac{V_{\rm cell}}{6\pi^2}
\left(E_{{\rm line},+}^3-E_{{\rm line},-}^3\right),
\qquad
G_{\rm line}=2G_\nu,
\qquad
f_{{\rm line},i}=\frac{L_i/2}{G_\nu}=\frac{L_i}{G_{\rm line}}.
\label{eq:line-combined-count}
\end{equation}
Thus, one unit in the combined Monte Carlo count changes the common occupation by $\Delta f_{\rm line}=1/G_{\rm line}$, while a complete decay adds two units.
The compact sampler draws two distinct entries from this combined population.
For inverse decay within one bin, there are $L_i$ choices for the first entry and $L_i-1$ for the second.
For direct decay, the corresponding available capacity is $H_i=G_{\rm line}(1-f_{{\rm bin},i})$, where $f_{{\rm bin},i}$ includes the individually represented neutrinos in that bin.
Dividing the ordered pair weights by $G_{\rm line}^2$ yields
\begin{align}
\frac{L_i\max(L_i-1,0)}{G_{\rm line}^2}
&=f_{{\rm line},i}\max\!\left(f_{{\rm line},i}-\Delta f_{\rm line},0\right),
\label{eq:line-inverse-pair-weight}\\
\frac{H_i\max(H_i-1,0)}{G_{\rm line}^2}
&=(1-f_{{\rm bin},i})\max\!\left(1-f_{{\rm bin},i}-\Delta f_{\rm line},0\right).
\label{eq:line-forward-pair-weight}
\end{align}
For particles in different propagation bins, the two counts are independent and their weights multiply directly.
The flavor probabilities multiply these pair weights through $\Pi_{\alpha i}\Pi_{\alpha j}$, as in Sec.~\ref{sec:zero-mode}.

The terms proportional to $\Delta f_{\rm line}$ arise from sampling distinct entries in the combined Monte Carlo population.
For separately enumerated physical neutrinos and antineutrinos, the inverse-decay pair count is $N_{\nu_i}N_{\bar\nu_i}$, and each species has its own Pauli factor.
The combined representation uses the spin and charge normalization of the continuum rates in Eqs.~\eqref{eq:zero-mode-direct} and \eqref{eq:zero-mode-inverse}.
Its finite sampling corrections vanish as $G_{\rm line}\to\infty$ at fixed occupations, recovering $(1-f_{{\rm bin},i})^2$ and $f_{{\rm line},i}^2$.

\subsection{Thermal equilibration through $1\leftrightarrow2$ reactions}
\label{app:thermal-1-2-checks}

With inverse decays disabled and $m_X\gg3T$, quantum-statistical factors are negligible and every energy shell obeys $dN_X(E_X)/dt=-N_X(E_X)/(\tau_X\gamma_X)$.
The simulated shell populations reproduce this exponential law for each decay mode.

A stronger test follows the system to equilibrium in a fixed volume.
\resp{We set $m_X=3\mev$ and $\tau_X=0.005\s$, take one internal degree of freedom for $X$, initialize the neutrinos and antineutrinos with Fermi-Dirac distributions at $T_0=5\mev$ and $\mu_\nu=0$, set the $X$ abundance to zero, and disable cosmic expansion.}
The equilibrium distributions are
\begin{align}
 f_\nu(p)&=\frac{1}{\exp[(p-\mu_\nu)/T]+1}\,,
 &
 f_X(p)&=\frac{1}{\exp[(\sqrt{p^2+m_X^2}-\mu_X)/T]-1}\,.
 \label{eq:static-equilibrium-distributions}
\end{align}
Their temperature and chemical potentials follow from the conserved energy, the conserved particle-number combinations, and chemical equilibrium.

\begin{figure}[t!]
\centering
  \includegraphics[width=0.5\textwidth]{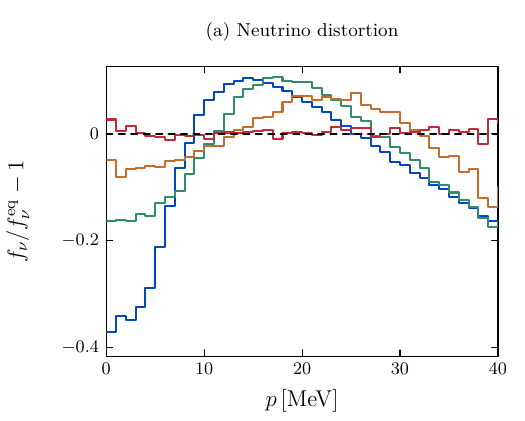}%
\includegraphics[width=0.5\textwidth]{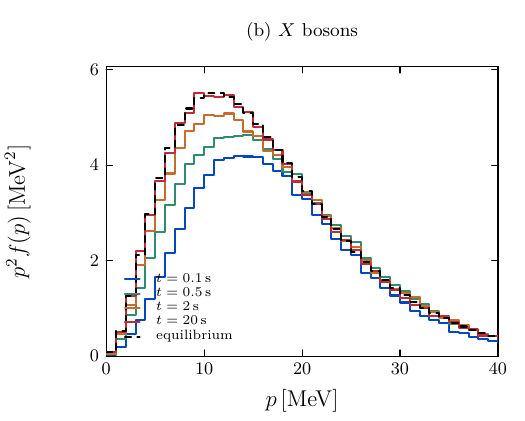}
\caption{Fixed-volume equilibration through $X\leftrightarrow\nu_\alpha\bar\nu_\alpha$.
The left panel shows the fractional neutrino distortion $f_\nu/f_\nu^{\rm eq}-1$.
At each displayed time, $f_\nu^{\rm eq}$ is defined as the Fermi-Dirac distribution with the same neutrino number and energy as the instantaneous simulated spectrum.
Its intermediate-energy excess and depleted low- and high-energy tails display the nonthermal redistribution produced by the inverse-decay/direct-decay cycle.
The right panel shows $p^2f_X(p)$ per internal degree of freedom.
Both panels use one $X$ internal degree of freedom, $m_X=3\mev$, and $\tau_X=0.005\s$, with cosmic expansion and weak $2\leftrightarrow2$ reactions disabled; the decoupled electromagnetic bath remains at $5\mev$.
\resp{The neutrinos initially follow a Fermi-Dirac distribution at $T_0=5\mev$ with $\mu_\nu=0$, and the $X$ sector is empty.}
The explicitly labeled, zero-order histogram curves show $t/\s=0.1,0.5,2,$ and $20$.
The black dashed curves are the thermal references; on the right this is the Bose-Einstein prediction $T_{\rm fin}=5.56\mev$ and $\mu_X=-6.01\mev=2\mu_\nu$, with $\mu_\nu=-3.01\mev$.
At $20\s$, the simulated spectra agree with these predictions.}
\label{fig:static-nu-equilibration}
\end{figure}

For $X\leftrightarrow\nu_\alpha\bar\nu_\alpha$ alone, with equal branching fractions into the three flavors, the conserved number is
\begin{equation}
 Q_\nu=N_\nu+2N_X\,,
 \label{eq:static-nu-conserved-number}
\end{equation}
where $N_\nu$ includes neutrinos and antineutrinos.
CP symmetry and chemical equilibrium give $\mu_X=\mu_\nu+\mu_{\bar\nu}=2\mu_\nu$.
Combining this condition with conservation of $Q_\nu$ and energy predicts
\begin{equation}
 T_{\rm fin}=5.56\mev,\qquad
 \mu_\nu=-3.01\mev,\qquad
 \mu_X=-6.01\mev.
 \label{eq:static-nu-equilibrium-target}
\end{equation}
\resp{Fig.~\ref{fig:static-nu-equilibration} shows equilibration through repeated $X\leftrightarrow\nu_\alpha\bar\nu_\alpha$ cycles.
Inverse decays remove kinematically selected neutrino pairs, and subsequent boosted decays populate different momenta.
Detailed balance drives the coupled system of neutrinos and $X$ particles toward the Fermi-Dirac and Bose-Einstein limits, with the intermediate distortion persisting for several seconds.}
To expose its shape independently of the changing bulk temperature and chemical potential, we define $f_\nu^{\rm eq}(p,t)$ at each snapshot as the unique Fermi-Dirac distribution having the same instantaneous neutrino number and energy as the simulated spectrum, and plot the fractional deviation from it.
The early spectrum contains a broad intermediate-energy shoulder bracketed by deficits in the soft and far-tail domains.
This nonthermal pattern is generated as the nonuniform inverse-decay kernel removes selected neutrino pairs and boosted two-body decays redistribute their energy.
At $t=20\s$, the distortion has fallen to the Monte Carlo noise, and the fitted temperature and chemical potentials agree with Eq.~\eqref{eq:static-nu-equilibrium-target}.

The electromagnetic bath remains a decoupled spectator at $5\mev$ throughout this neutrino-only test.

\begin{figure}[t!]
\centering
  \includegraphics[width=0.5\textwidth]{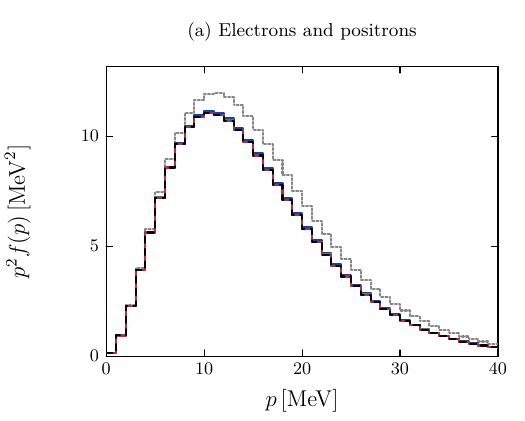}%
\includegraphics[width=0.5\textwidth]{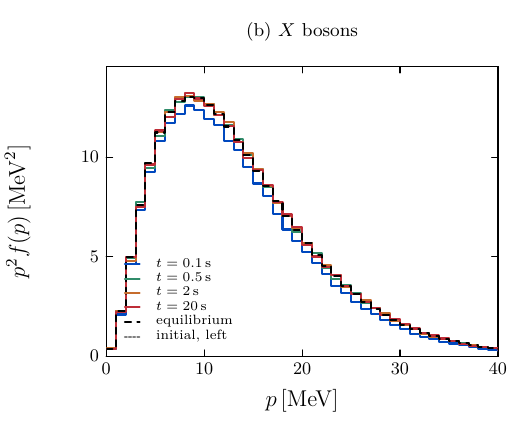}
\caption{Fixed-volume equilibration through $X\leftrightarrow e^+e^-$.
The left panel shows the thermal electron-positron distribution reconstructed from the electromagnetic energy, and the right panel shows the simulated $X$ distribution; the vertical coordinate is $p^2f(p)$ per internal degree of freedom.
Both panels use one $X$ internal degree of freedom, $m_X=3\mev$, and $\tau_X=0.005\s$; the neutrino sector is absent, and cosmic expansion and all other interactions are disabled.
The system starts at $T_0=5\mev$ with no $X$ particles.
The explicitly labeled, zero-order histogram curves show the initial Fermi-Dirac distribution and $t/\s=0.1,0.5,2,$ and $20$.
The black dashed curves are the equilibrium predictions $T_{\rm fin}=4.80\mev$ and $\mu_X=0$.
At $20\s$, the simulated $X$ spectrum and the electromagnetic bath agree with these predictions.
Rapid electromagnetic interactions maintain a thermal plasma as it exchanges energy with $X$, producing the early equilibration visible in both panels.}
\label{fig:static-em-equilibration}
\end{figure}

For $X\leftrightarrow e^+e^-$, the electromagnetic bath remains internally thermal.
Pair symmetry gives $\mu_{e^-}+\mu_{e^+}=0$, so chemical equilibrium fixes $\mu_X=0$.
No independent total-particle-number constraint remains, and energy conservation reduces to
\begin{equation}
 \rho_{\rm EM}(T_{\rm fin})+\rho_X(T_{\rm fin},0)
 =\rho_{\rm EM}(T_0)\,.
 \label{eq:static-em-energy-conservation}
\end{equation}
This gives
\begin{equation}
 T_{\rm fin}=4.80\mev,\qquad \mu_X=0.
 \label{eq:static-em-equilibrium-target}
\end{equation}
The electron-positron spectrum has a Fermi-Dirac form after each energy exchange, allowing the $X$ distribution to reach its Bose-Einstein limit much earlier than in the neutrino-only system.
Fig.~\ref{fig:static-em-equilibration} shows this rapid relaxation.
The late-time bath temperature and fitted $X$ distribution agree with Eq.~\eqref{eq:static-em-equilibrium-target} within Monte Carlo noise.

As a third check, we activate the weak $2\leftrightarrow2$ reactions and start with CP-symmetric neutrino and electromagnetic sectors at $T_0=5\mev$.
Weak pair reactions impose $\mu_\nu+\mu_{\bar\nu}=\mu_{e^-}+\mu_{e^+}=0$, CP symmetry sets the particle and antiparticle chemical potentials equal, scattering equalizes the bath temperatures, and inverse-decay equilibrium gives $\mu_X=0$.
Total-energy conservation predicts $T_{\rm fin}=4.89\mev$.
At $t=20\s$, the neutrino, electromagnetic, and $X$ sectors agree with this common zero-chemical-potential equilibrium within Monte Carlo noise.

Successive timestep refinements leave the evolution of the particle distributions and the equilibrium comparisons unchanged within Monte Carlo noise.
These tests jointly verify detailed balance and the associated number and energy conservation laws.

\section{Modified electron-flavor coupling check}
\label{app:modified-coupling-check}

The mechanism check in Sec.~\ref{sec:setup-II-results} uses $Y_X(20\,\mev)=0.05$, $\tau_X=0.1\,\s$, and neutrino oscillations.
In the neutrino-electron scattering and $\nu\bar\nu\leftrightarrow e^+e^-$ kernels, we set the left-handed electron-neutrino coupling equal to the muon- and tau-neutrino value, $g_{L,e}=\widetilde g_L$; the right-handed coupling $g_R$ is flavor universal.
The crossed reactions use the same replacement.
The calculation contains the direct decays and $2\leftrightarrow2$ transport of \setupname{II}, while inverse decays and \resp{production of secondary particles} are omitted.
The vacuum lifetime, initial abundance, branching fractions, and remaining inputs are unchanged.

Recomputing the $X$-free cosmology with these couplings gives $N_{\rm eff}=3.0279$.
\resp{Defining \dneff with respect to this cosmology without $X$ and with the same modified couplings, we obtain} $\dneff\simeq+0.34,-0.53,-1.32$, and $-1.62$ for $m_X=100,200,500$, and $1000\,\mev$, respectively.
The signs at $100$ and $200\,\mev$ bracket a zero crossing under continuous variation of the mass.
\resp{Reducing the electron-neutrino couplings shifts the crossing upward, while the negative high-mass branch survives.
This result supports the interpretation in Sec.~\ref{sec:setup-II-results}: the energy dependence of weak interactions and rapid electromagnetic thermalization permit negative \dneff even when the transfer of energy from neutrinos to the electromagnetic plasma is weakened.}
The lifetime, abundance, flavor evolution, and included reactions determine the crossing.
All phenomenological results in the main text use the Standard Model couplings.

\section{Checks of the averaged-oscillation and adiabatic approximations}
\label{app:osc-checks}

Adiabatic propagation and phase averaging depend on the relative time scales of oscillations, collisions, and changes in the medium.
Appendix~\ref{app:osc-conditions} expresses these conditions in terms of the in-medium eigenvalues and evaluates them for a thermal background.
Appendix~\ref{app:osc-coherence} isolates the contribution of flavor interference to neutrino-antineutrino annihilation, supplementing the collision prescription in Sec.~\ref{sec:oscillations}.
Appendix~\ref{app:setup-I-qke-comparison} examines the residual difference from the \qkemethod calculation in \setupname{I}.

\subsection{Averaged-oscillation and adiabaticity conditions}
\label{app:osc-conditions}

The flavor-nonuniversal thermal potential enters the treatment in the main text.
Neglecting the flavor-universal part of the refractive Hamiltonian, and working in a CP-symmetric plasma with negligible lepton asymmetries, the charged-lepton contribution is
\begin{equation}
    V_e-V_x
    =
    -\frac{2\sqrt{2}\,G_F\,E_{\nu}}{m_W^2}
    \left(
        \rho_{e^-}+P_{e^-}+\rho_{e^+}+P_{e^+}
    \right).
    \label{eq:Vex-app}
\end{equation}
In the ultra-relativistic limit, $\rho_{e^-}+P_{e^-}+\rho_{e^+}+P_{e^+}\simeq 7\pi^2T^4/45$, giving
\begin{equation}
    V_e-V_x
    \simeq
    -\frac{14\sqrt{2}\pi^2}{45}
    \frac{G_F E_{\nu}T^4}{m_W^2}.
    \label{eq:Vex-ur-app}
\end{equation}
It is useful to introduce the matter potential in mass-squared units,
\begin{equation}
    A_{\rm m}(E_{\nu},T)
    \equiv
    2E_{\nu}(V_e-V_x)
    \simeq
    -\frac{28\sqrt{2}\pi^2}{45}
    \frac{G_F E_{\nu}^2T^4}{m_W^2}.
    \label{eq:Am-app}
\end{equation}

The oscillation Hamiltonian in flavor space is
\begin{equation}
    \Omega_\nu(E_{\nu},T)
    =
    \frac{1}{2E_{\nu}}\,
    U_{\rm PMNS}M^2U_{\rm PMNS}^\dagger
    +
    {\rm diag}(V_e-V_x,0,0),
    \label{eq:Omega-app}
\end{equation}
where the flavor-universal part has been omitted.
At each local value of $(E_\nu,T)$, the matter mixing matrix is defined by
\begin{equation}
    \left(U^{\rm m}\right)^\dagger
    \Omega_\nu
    U^{\rm m}
    =
    {\rm diag}(\lambda_1,\lambda_2,\lambda_3),
    \label{eq:Um-app}
\end{equation}
and the in-medium oscillation frequencies are
\begin{equation}
    \omega_{ij}(E_\nu,T)
    \equiv
    |\lambda_i(E_\nu,T)-\lambda_j(E_\nu,T)|,
    \qquad i\neq j.
    \label{eq:omegaij-app}
\end{equation}

The adiabatic condition is that \resp{the rate of rotation of the basis} is low compared with the relevant eigenvalue splitting,
\begin{equation}
    \eta_{\rm ad}^{ij}
    \equiv
    \frac{
    \left|
    \left[
    \left(U^{\rm m}\right)^\dagger
    D_t U^{\rm m}
    \right]_{ij}
    \right|
    }
    {\omega_{ij}}
    \ll 1,
    \qquad i\neq j,
    \label{eq:etaad-general-app}
\end{equation}
where $D_t$ denotes the derivative along the freely streaming neutrino trajectory.
In the QKE language, if $K$ is the collision term in the flavor basis and $K_{\rm m}\equiv (U^{\rm m})^\dagger K U^{\rm m}$, the averaged-oscillation approximation additionally requires
\begin{equation}
    \begin{aligned}
    \eta_K^{ij}
    &\equiv
    \frac{|(K_{\rm m})_{ij}|}{\omega_{ij}}
    \ll 1,
    \\
    \eta_{\dot K}^{ij}
    &\equiv
    \frac{|D_t\ln (K_{\rm m})_{ij}|}{\omega_{ij}}
    \ll 1.
    \end{aligned}
    \label{eq:etak-general-app}
\end{equation}
These are the matter-basis scale separations underlying the ATAO construction.

The DSMC implementation does not explicitly form the matrix-valued collision term $K_{\rm m}$.
Comparing the eigenvalue splittings with the physical interaction rates tests whether the phases average between collisions.
For a reaction class $r$, we define
\begin{equation}
    R_{\rm osc}^{(r)}(E_\nu,T)
    \equiv
    \min_{(i,j)\in{\cal P}_r}
    \frac{\omega_{ij}(E_\nu,T)}
    {\Gamma_r^{\rm eff}(E_\nu,T)}.
    \label{eq:Rosc-app}
\end{equation}
Here, ${\cal P}_r$ contains the pairs of propagation eigenstates whose relative phases affect reaction $r$.
The averaged propagation treatment requires $R_{\rm osc}^{(r)}\gg1$ for the population relevant to the observable under consideration.
If $R_{\rm osc}^{(r)}\lesssim1$, interactions occur before the relevant relative phase has averaged.
The flavor state then remains correlated between successive interactions.

For orientation, consider first the matter-dominated $e$-$x$ splitting.
Away from accidental level degeneracies,
\begin{equation}
    \omega_{ex}^{\rm mat}(E_\nu,T)
    \simeq
    |V_e-V_x|.
    \label{eq:omega-mat-app}
\end{equation}
For a relativistic neutrino interacting with a thermal background, the weak interaction rate may be estimated as
\begin{equation}
    \Gamma_\alpha(E_\nu,T)
    =
    C_\alpha G_F^2 E_\nu T^4,
    \qquad
    C_e\simeq 1.27,
    \qquad
    C_{\mu,\tau}\simeq 0.92.
    \label{eq:Gamma-alpha-app}
\end{equation}
Combining Eqs.~\eqref{eq:Vex-ur-app} and \eqref{eq:Gamma-alpha-app}, one finds
\begin{equation}
    \frac{\omega_{ex}^{\rm mat}}{\Gamma_\alpha}
    \simeq
    \frac{14\sqrt{2}\pi^2}
    {45\,C_\alpha\,G_Fm_W^2}.
    \label{eq:mat-ratio-app}
\end{equation}
Thus, the leading $E_\nu$ and $T^4$ dependences cancel for this matter-dominated splitting.
Numerically,
\begin{equation}
    \frac{\omega_{ex}^{\rm mat}}{\Gamma_{\nu_e}}
    \simeq
    45,
    \qquad
    \frac{\omega_{ex}^{\rm mat}}{\Gamma_{\nu_{\mu,\tau}}}
    \simeq
    63.
    \label{eq:mat-ratio-numeric-app}
\end{equation}
Within this thermal rate estimate, raising the neutrino energy does not reduce the ratio for the matter-dominated $e$-$x$ splitting.
The estimate does not include inverse decays or the changes in collision rates caused by nonthermal partner distributions.

The transition between vacuum-dominated and matter-dominated behavior for a given mass scale is determined by $|A_{\rm m}(E_\nu,T)|\sim |\Delta m_{ij}^2|$, namely
\begin{equation}
    E_{\rm mat}^{ij}(T)
    \simeq
    \left[
    \frac{
    |\Delta m_{ij}^2|\,m_W^2
    }
    {
    (28\sqrt{2}\pi^2/45)\,G_F\,T^4
    }
    \right]^{1/2}.
    \label{eq:Emat-app}
\end{equation}
The scale in Eq.~\eqref{eq:Emat-app} marks where the corresponding in-medium splitting changes character.
Below this scale the exact eigenvalue splitting smoothly approaches the vacuum form; above it the $e$-$x$ splitting is controlled by the thermal matter potential.
Near this transition, Eq.~\eqref{eq:Rosc-app} must use the exact eigenvalues of $\Omega_\nu$.
The vacuum estimate $\Delta m_{ij}^2/(2E_\nu)$ alone omits the in-medium contribution.

The adiabaticity of the propagation basis can be estimated analytically.
In a two-level reduction with vacuum parameters $(\Delta m^2,\theta)$, the in-medium mixing angle satisfies
\begin{equation}
    \tan 2\widetilde\theta(E_\nu,T)
    =
    \frac{\Delta m^2\sin 2\theta}
    {\Delta m^2\cos 2\theta-A_{\rm m}(E_\nu,T)}.
    \label{eq:theta-tilde-app}
\end{equation}
Differentiating along the trajectory gives
\begin{equation}
    \left|
    D_t\widetilde\theta
    \right|
    =
    \frac{
    |D_t A_{\rm m}|\,
    \Delta m^2|\sin 2\theta|
    }
    {
    2\left[
    \left(\Delta m^2\cos 2\theta-A_{\rm m}\right)^2
    +
    \left(\Delta m^2\sin 2\theta\right)^2
    \right]
    }.
    \label{eq:thetadot-app}
\end{equation}
The corresponding in-medium level splitting is
\begin{equation}
    \omega_{\rm m}(E_\nu,T)
    =
    \frac{1}{2E_\nu}
    \left[
    \left(\Delta m^2\cos 2\theta-A_{\rm m}\right)^2
    +
    \left(\Delta m^2\sin 2\theta\right)^2
    \right]^{1/2}.
    \label{eq:omega-twolevel-app}
\end{equation}
A convenient two-level adiabaticity parameter is therefore
\begin{equation}
    \eta_{\rm ad}(E_\nu,T)
    =
    \frac{|D_t\widetilde\theta|}{\omega_{\rm m}}
    =
    \frac{
    E_\nu |D_t A_{\rm m}|\,
    \Delta m^2|\sin 2\theta|
    }
    {
    \left[
    \left(\Delta m^2\cos 2\theta-A_{\rm m}\right)^2
    +
    \left(\Delta m^2\sin 2\theta\right)^2
    \right]^{3/2}
    }.
    \label{eq:etaad-twolevel-app}
\end{equation}

Since $A_{\rm m}\propto E_\nu^2T^4$, its logarithmic derivative along a freely streaming trajectory is
\begin{equation}
    \frac{D_t A_{\rm m}}{A_{\rm m}}
    =
    2\frac{D_tE_\nu}{E_\nu}
    +
    4\frac{\dot T}{T}
    =
    -2H
    +
    4\frac{\dot T}{T}.
    \label{eq:Am-logder-app}
\end{equation}
Around standard decoupling, $\dot T/T\simeq -H$, and hence
\begin{equation}
    \left|
    \frac{D_t A_{\rm m}}{A_{\rm m}}
    \right|
    \simeq
    6H.
    \label{eq:Am-logder-standard-app}
\end{equation}
Relic decays modify this estimate only through the actual background functions $H(t)$ and $T(t)$, provided no additional flavor-nonuniversal self-potential from a sizeable neutrino asymmetry is included.

A conservative upper estimate of non-adiabaticity is obtained by evaluating Eq.~\eqref{eq:etaad-twolevel-app} near the point where it would be largest, $A_{\rm m}\simeq \Delta m^2\cos2\theta$.
This gives
\begin{equation}
    \eta_{\rm ad}^{\rm max}
    \simeq
    \frac{
    6E_\nu H|\cos2\theta|
    }
    {
    \Delta m^2\sin^2 2\theta
    }.
    \label{eq:etaadmax-app}
\end{equation}
Using $H=1.66\sqrt{g_*}T^2/M_{\rm Pl}$ with $g_*=10.75$, $T=3\,{\rm MeV}$, and $E_\nu\simeq3.15T$, one finds
\begin{equation}
    \eta_{\rm ad}^{(12)}
    \sim
    1.4\times10^{-3},
    \qquad
    \eta_{\rm ad}^{(13)}
    \sim
    1.0\times10^{-3}.
    \label{eq:etaadnums-app}
\end{equation}
The propagation basis therefore evolves adiabatically in the standard thermal regime.

Relic energy injection changes these scale ratios through the time-dependent expansion rate, plasma temperature, and distributions of collision partners.
Its effect on adiabaticity enters Eq.~\eqref{eq:etaad-general-app} through $D_tU^{\rm m}$, while the rates of scattering and inverse decay enter Eq.~\eqref{eq:Rosc-app}.

\subsection{Flavor interference in collision rates}
\label{app:osc-coherence}

Flavor coherences contribute to neutrino-antineutrino annihilation through ${\rm Tr}[G^a\bar\rho_2G^b\rho_1]$, where $\rho_1$ and $\bar\rho_2$ describe the incoming pair and $a,b\in\{L,R\}$ label the weak chiral couplings~\cite{Sigl:1993ctk,deSalas:2016ztq}.
The Standard Model coupling matrices are
\begin{equation}
    G^L={\rm diag}(g_L,\widetilde g_L,\widetilde g_L),
    \qquad G^R=g_R\openone,
    \label{eq:Gmatrices-app}
\end{equation}
where $g_L=\tfrac12+s_W^2$, $\widetilde g_L=-\tfrac12+s_W^2$, and $g_R=s_W^2$.
The off-diagonal elements contribute to the annihilation rate as
\begin{equation}
    \Delta\Gamma_{\rm loss}\propto
    \sum_{a,b=L,R}K_{ab}
    \sum_{\alpha\neq\beta}g_\alpha^ag_\beta^b\,
    {\rm Re}\!\left[(\bar\rho_2)_{\alpha\beta}(\rho_1)_{\beta\alpha}\right],
    \label{eq:bilinear-app}
\end{equation}
where $K_{ab}$ are the kinematic coefficients, with the mixed $LR$ terms proportional to $m_e^2$.
For annihilation summed over flavors, this term is bilinear in the incoming coherences and vanishes if either density matrix is diagonal in flavor, including a flavor-universal equilibrium distribution.
Neutrino-electron scattering couples density matrices at the incoming and outgoing momenta.

Muon-tau coherences contribute through ${\rm Tr}_{\mu\tau}[\bar\rho_2\rho_1]$, which contains the off-diagonal product in Eq.~\ref{eq:bilinear-app} and is invariant under a simultaneous basis rotation.
At $|A_{\rm m}|\gg|\Delta m^2|$, the thermal potential suppresses mixing of the electron flavor with the other two, while muon-tau mixing remains unsuppressed.

\subsection{\setupname{I} comparison with the \qkemethod calculation}
\label{app:setup-I-qke-comparison}

The largest separation between the \nudsmc and \qkemethod curves with neutrino oscillations enabled in Fig.~\ref{fig:Setup-I} occurs at $T_{\rm RH}=4.95\,\mev$, where $N_{\rm eff}^{\nu{\rm DSMC}}-N_{\rm eff}^{\rm QKE}=0.033$.
We tested whether this difference could be explained by conventions that can be varied within \nudsmc.
Replacing the mixing parameters by the public FortEPiaNO values, changing \resp{the relation between $T_{\rm RH}$ and the lifetime}, changing the early matter-dominated initialization, and varying $\sin^2\theta_W$ each changes the central value of $N_{\rm eff}$ by only about $(1$--$2)\times10^{-3}$.
Raising the boundary between the thermal pre-evolution and stochastic transport from $5\,\mev$ to the apparent $7$--$8\,\mev$ plateau changes $N_{\rm eff}$ by about $5\times10^{-3}$.
None of these variations accounts for the observed separation.

A remaining possibility is the treatment of neutrino-neutrino collisions.
These reactions conserve the energy of the neutrino sector while redistributing it over momentum and damping flavor coherences.
They can therefore modify flavor conversion and, indirectly, \resp{the transfer of energy from the electromagnetic plasma to neutrinos}.
In the public low-reheating FortEPiaNO configuration~\cite{Bennett:2020zkv,FortEPiaNO:public-v1.4.0}, the $\nu\nu$ contributions are disabled in both the diagonal collision integral and the fitted off-diagonal damping term.

We tested this dependence at $T_{\rm RH}=4.950\,\mev$ by changing only the $\nu\nu$ damping switch and keeping all other inputs fixed.
With both $\nu\nu$ contributions disabled, the public configuration gives $N_{\rm eff}=2.838$.
Restoring only the fitted off-diagonal $\nu\nu$ damping gives $N_{\rm eff}=2.884$, a shift of $+0.047$.
This is of the same order as, and somewhat larger than, the $0.033$ separation between \nudsmc and \qkemethod.
As a control, the identical switch at $T_{\rm RH}=15\,\mev$ changes $N_{\rm eff}$ by only $-2\times10^{-4}$.
The effect is therefore specific to the low-reheating regime where the two curves separate and vanishes at high reheating temperature, which excludes a common normalization offset.

The test isolates the effect of off-diagonal $\nu\nu$ damping with the diagonal $\nu\nu$ integral omitted in both calculations.
Its magnitude makes neutrino-neutrino collisions a possible source of the separation in Fig.~\ref{fig:Setup-I}; the exact contribution depends on the collision prescription used for that curve.

\section{Numerical settings for the performance benchmarks}
\label{app:performance-settings}

Tab.~\ref{tab:qcbe-performance-settings} gives the momentum grids and integration tolerances used for the timings in Tab.~\ref{tab:performance}.
The momentum coordinate is $y=ap$, where $p$ is physical momentum and the \qcbemethod scale factor is normalized by $aT_\gamma=1.00003$ at $T_\gamma=5\,\mev$.
Each grid has $N_p$ uniformly spaced points from $y_{\min}=0.01$ to $y_{\max}$, with composite trapezoidal quadrature.

\begin{table}[t!]
\centering
\begin{ruledtabular}
\begin{tabular}{lcccc}
Benchmark & $N_p$ & $y_{\max}$ & $\varepsilon_{\rm rel}$ & $\varepsilon_f$ \\
\hline
EM reheating & 101 & 40 & $10^{-6}$ & $10^{-8}$ \\
Hard neutrino injection & 151 & 300 & $10^{-8}$ & $10^{-12}$ \\
Active $1\leftrightarrow2$ processes & 51 & 40 & $10^{-6}$ & $10^{-8}$ \\
\end{tabular}
\end{ruledtabular}
\caption{Numerical settings for the \qcbemethod timings.
$N_p$ counts momentum points, $\varepsilon_{\rm rel}$ is the relative integration tolerance, and $\varepsilon_f$ is the absolute tolerance for the neutrino and resolved-$X$ occupations.
The absolute tolerance for the plasma variables and the comoving number of $X$ particles at rest is $10^{-8}$ in all three calculations.}
\label{tab:qcbe-performance-settings}
\end{table}

All three momentum-resolved calculations use RK23 and a decay cutoff at $20\tau_X$ after the initial time.
The grids and cutoff remain fixed during each timing run, and transport continues to $T_\gamma=0.024\,\mev$.
The preceding thermal evolution to $5\,\mev$ uses LSODA with relative tolerance $10^{-8}$ and absolute tolerance $10^{-12}$.
In this thermal stage, the resolved $X$ distribution has 240 logarithmically spaced physical momenta between $10^{-3}T_{\rm ini}$ and $25T_{\rm ini}$.

We normalize $N_{\rm eff}$ with the thermal Fermi-Dirac energy moment evaluated on the same momentum grid and with the same quadrature weights as the evolving distributions.
For the separate abundance comparisons, the transport is repeated with the same settings, and the BBN rates are evaluated using the spectra and plasma variables at every accepted integration node.
The BBN weak rates use shape-preserving cubic interpolation of the logarithm of positive neutrino occupations (log-PCHIP), with negative occupations replaced by zero.
The full nuclear network then determines $Y_P$ and D/H.

\end{document}